\documentclass[12pt]{JHEP3}
\pdfoutput=1
\usepackage{amssymb,amsfonts}
\usepackage{amsmath}
\usepackage{graphicx}
\usepackage{amsmath,epsfig}
\usepackage{amssymb,amsfonts}
\usepackage{latexsym}
\def\refeq#1{(\ref{#1})}
\def\hri#1#2{\href{http://arxiv.org/abs/#1}{[ArXiv:#1]#2}}
\def\hre#1#2{\href{http://arxiv.org/abs/#1/#2}{[ArXiv:#1/#2]}}

\def\pa{\partial}

\def\be{\begin{equation}}
\def\ee{\end{equation}}
\def\bea{\begin{eqnarray}}

\def\eea{\end{eqnarray}}
\def\sp{\;\;\;,\;\;\;}
\def\l{\lambda}
\def\lab{\label}

\def\e{\epsilon}
\def\o{\omega}
\def\le{\left}
\def\ri{\right}
\def\y{\psi}
\def\half{\frac12}
\def\q{\theta}
\def\cO{{\cal O}}

\def\d{\delta}
\def\6{\partial}
\def\de{\partial}

\def\ls{\ell_s}
\def\a{\alpha}
\def\b{\beta}
\def\lab{\label}
\def\parl{\parallel}
\def\tom{\tilde{\o}}

\def\r{x}

\def\dx{\delta X^\parl}
\def\dxy{\delta X^2}
\def\dxz{\delta X^3}

\title{\vskip -1cm Langevin diffusion of heavy quarks in non-conformal holographic backgrounds}
\author{Umut G\"ursoy$^1$, Elias Kiritsis$^{2,4}$, Liuba Mazzanti$^3$ and Francesco Nitti$^4$\\
$^1$\href{http://www1.phys.uu.nl/wwwitf}{Institute for Theoretical Physics, Utrecht University;
Leuvenlaan 4, 3584 CE Utrecht, The Netherlands.}\\
~\\
$^2$\href{http://hep.physics.uoc.gr/}{Crete Center for Theoretical Physics,\\Department of Physics, University of Crete
71003 Heraklion, Greece}\\
~\\
$^3$\href{http://www-fp.usc.es/~theory/}{Departamento de F\'{\i}sica de Part\'{\i}culas, Universidade
de Santiago de
Compostela\\and Instituto Galego de F\'{\i}sica de Altas
Enerx\'{\i}as (IGFAE)\\E-15782, Santiago de Compostela, Spain}\\
~\\
$^4$\href{http://www.apc.univ-paris7.fr}{APC, Universit\'e Paris 7,  B\^atiment Condorcet, F-75205, Paris Cedex 13, France,
 (UMR du CNRS 7164).}}
\preprint{\phantom{\hepth{yymmyyyy}}\\CCTP-2010-6\\ITP-UU-10/19\\SPIN-10/17}
\abstract{The Langevin diffusion process of a relativistic heavy quark in a non-conformal holographic setup is analyzed. The bulk geometry is a general,
five-dimensional asymptotically AdS black hole. The heavy quark is described by a trailing string attached to a flavor brane, moving at constant velocity.
{}From the equations describing linearized fluctuations of the string world-sheet, the correlation functions defining a generalized
Langevin process are constructed via the $AdS$/CFT prescription. In the local limit, analytic
expressions for the Langevin diffusion and friction coefficients
 are obtained in terms of the bulk string metric. Modified Einstein relations between these quantities are also derived. The spectral densities associated
to the Langevin correlators are analyzed, and simple analytic expressions
are obtained in the small and large frequency limits.
Finally, a numerical analysis of the jet-quenching parameter, and a
comparison to RHIC phenomenology are performed in the case of Improved Holographic QCD.
It is shown that the jet-quenching parameter is not enough to describe energy loss of very energetic
charm quarks and the full Langevin correlators are needed.}
\vspace{-1cm}
\keywords{\vspace{-1cm} AdS/CFT, Quark-Gluon Plasma,Langevin diffusion, heavy quarks}

\begin{document}

\section{Introduction and results}\label{INTRO}

RHIC experiments of heavy-ion collisions and related data on the deconfined phase of QCD,
\cite{rhic} have provided a window for string theory techniques to meet the real world.
The context is strong coupling dynamics near and above the deconfining transition in QCD.
String theory via the AdS/CFT correspondence has provided a framework in order to understand
strong coupling dynamics in the deconfined phase including the calculation of transport coefficients.
 Recent reviews on the progress in this direction are \cite{sonrev,csa,iancurev,gubserrev,ihqcdrev}.

Observables of particular importance are associated to heavy quarks. Heavy quarks may be produced in the Quark-Gluon Plasma (QGP) of the RHIC fireball
and are then travelling to the detectors while moving through the dense QGP. They can be
 tagged reasonably well and are therefore valuable probes of the
 dynamics in the plasma and in particular for the mechanism of energy loss.

A single heavy quark can be modeled in string theory by an open string. Its end-point is representing
the heavy quark while the string is trailing behind as the quark moves.
The large mass limit is important in order to neglect the non-trivial flavor dynamics associated with light quarks
 (although with improved techniques the light quarks may also eventually be addressed
reliably in the holographic context). As quarks are associated
with strings ending on flavor branes, a heavy quark ends on a
brane that is stretching in the UV part of the bulk geometry. The
motion of such a string, and the associated force acting on the
quark from the thermal medium, have been studied in detail with
several complementary methods, \cite{her,lrw,gub1}.
 In the simplest setup, the UV endpoint of a fundamental string is forced to move with constant velocity $v$ along a spacial direction.
 The equations of motion for the full string are solved and the radial profile of the trailing string is found as it moves in a bulk black-hole
 background representing the deconfined heat bath. The energy absorbed by the string is calculated and the drag force of the string is obtained.
 The picture remains roughly valid, while details change when conformal invariance is broken, \cite{lliu,transport}.

 An important improvement in this picture consists of the study of the stochastic nature of this system
  in analogy with the dynamics of heavy particles in a heat bath giving rise to Brownian motion.
  This involves a diffusive process, that was first considered in a holographic setting in
  \cite{tea}, by using the Schwinger-Keldysh formalism
 adapted to AdS/CFT in \cite{sonherzog}.

Subsequently, a study of the (quantum) fluctuations of the trailing string,
\cite{gubser,Casal} provided the information on the momentum broadening of
a heavy quark as it moves in the plasma. The stochastic motion was formulated as
a Langevin process, \cite{deboer,sonteaney} associated with the correlators
of the fluctuations of the string.

Many heavy quarks in experiments are relativistic. Therefore it is
necessary to study the associated relativistic Langevin evolution
of the trailing string, a feat accomplished in the ${\cal N}=4$
case in \cite{iancu}. The same type
of Langevin process was studied in \cite{guijosa} for the case of an accelerating quark  in the vacuum
 (rather than in a deconfined plasma),
 by analyzing the fluctuations of a trailing string in $AdS$ with
a non-uniformly moving endpoint.

On the experimental front, there have been several results from the RHIC experiments, \cite{ph0}-\cite{ph3}.
The experimental signatures are currently summarized by the $e^{\pm}$ spectra that originate in the semileptonic decays of charmed and bottom hadrons.
 From these spectra a modification factor $R^e_{AA}$ and an elliptic flow coefficient $v^e_2$ are extracted.
 They capture the effects of the medium to the propagation of the heavy quarks.
 The data exhibit a substantial elliptic flow, up to $v_2^e\simeq$ 10\%, and a high-$p_T$ suppression down to $R^e_{AA}\simeq 0.25$.
 These values are comparable to light hadrons. Radiative energy loss models
 based on pQCD, \cite{dainese} do not seem to explain well the experimental data, \cite{rapp}.
Elastic scattering energy loss plus non-perturbative interactions can on the other hand accommodate the data, \cite{rapp}.

In particular the Langevin approach has been applied to the study of the heavy quark energy loss by several groups, and the related physics is summarized in the
 review \cite{rapp}. The Langevin evolution used was relativistic and with symmetric diffusion coefficients.
 As there was no microscopic model to provide the proper
 fluctuation-dissipation relation, the Einstein equations used vary, and in all examples
  it was assumed that the equilibrium distribution is the J\"uttner-Boltzmann distribution.
  Moreover various combinations of friction forces were used, resonance models, pQCD, N=4 AdS/CFT and combinations.
 A further recent analysis was performed in \cite{lan} with similar conclusions.
 The associated relativistic and isotropic Langevin systems used have been introduced in the mathematical physics literature rather recently,
\cite{rel} (see \cite{rel1} for a review).

The purpose of the present paper is to study further the relativistic Langevin evolution of a heavy quark
 using holographic techniques in a general context, and going beyond conformally invariant backgrounds
characterized solely by $AdS$ geometries\footnote{A particular
 example in this class  was recently considered in \cite{hoyos}, which studied  the Langevin process in the non-conformal ${\cal N}=2^*$ background.}.
In this work we will study a large class of non-conformal
backgrounds captured by Einstein-dilaton gravity with a dilaton potential in 5 dimensions. In a series of recent works,
 such backgrounds were analyzed both qualitatively and quantitatively and have provided a rich variety
of holographic bulk dynamics. In particular, for a selected class of scalar potentials, they mimic the behavior of large-$N$ Yang Mills, \cite{ihqcd1}-\cite{gkmn3}.
This match can be quantitative, \cite{gkmn3}, agreeing very well both at zero and finite temperature with recent high-precision lattice data, \cite{panero}.
On the other hand, the analysis we carry out in the present paper is general, as it applies to any asymptotically $AdS$ background.

We therefore consider a fundamental string whose end-point lies in
the UV region of a bulk black-hole background of a non-conformal
holographic model. The string end-point is forced to move with
velocity $v$. Solving the Nambu-Goto equations of motion, the
classical profile of the trailing string can be found. The string
stretches inside the bulk until it becomes completely horizontal
at some value of radial coordinate $r_s$, given by $f(r_s)=v^2$
where $f(r)$ is the blackness function of the background. When the
quark is moving slowly, as $v\to 0$, the point $r_s$ approaches
the bulk black hole horizon.

The induced metric on the string world-sheet  has
the form of a two-dimensional black-hole metric with a horizon at
$r=r_s$ as first observed in \cite{Casal}.\footnote{This is a generic effect on strings and D-branes embedded in black-hole/black-brane backgrounds. It was first observed in
\cite{cli} where it was used to propose that a different speed of light is relevant for such branes. It is implicit or explicit
in many holographic computations using probe flavor branes, \cite{karch,mateos} and strings \cite{Casal}.}    This  black-hole is an important
ingredient of the dynamics of the system. In particular it is
crucial in the calculation of the thermal correlators using the
Schwinger-Keldysh formalism, as well as for the
fluctuation-dissipation relation. The world-sheet black hole has
an associated Hawking temperature $T_s$ that depends on several parameters: the
background temperature $T$,  the zero-temperature bulk scale
$\Lambda$\footnote{This appears as an integration constant in the
background geometry and corresponds to the dynamically generated
energy scale $\Lambda_{QCD}$ in the dual field theory.} and
the quark velocity $v$. It coincides with the temperature $T$ of the
heat bath only in the non-relativistic limit. In the conformal
case, one has $T_s= T_{\rm s,conf} = T(1-v^2)^{1\over 4}\leq T$.
The numerical analysis performed in Section 6 shows that, in our
non-conformal holographic model $T_s\leq T_{\rm s,conf}\leq T$. The
equality $T_s=T_{\rm s,conf}$, in the first relation is attained, for arbitrary $v$, in
the high T limit, as shown in figure \ref{fig Z} in section
\ref{numerics} and also in the ultra-relativistic limit,  $v\to 1$.

We next consider small fluctuations around the classical string profile.
They satisfy second-order radial equations that are
related to the associated thermal correlators by the holographic
prescription. It should be emphasized that such correlators are
thermal with temperature $T_s$ and not the temperature $T$ of the
heat bath\footnote{More precisely put, the Langevin correlators
that are obtained from the string fluctuations by the holographic
prescription obey a modified Einstein relation with temperature $T_s$ rather than
$T$.}. Moreover, they satisfy the fluctuation-dissipation relation
associated with the emergent temperature $T_s$. The fact that the
string fluctuations see a modified temperature crucially affects
the Einstein relation between the diffusion constants.

At the quadratic level of fluctuations, a relativistic
Langevin diffusion equation is obtained using the $AdS$/CFT prescription.
The diffusion constants and friction coefficients are
calculated analytically in terms of the bulk metric, for general
non-conformal backgrounds. The Einstein relation is now modified, due to the fact that the temperature is modified.
 Another important property is that the diffusion constants perpendicular and longitudinal to the motion
(denoted by $\kappa_\perp$ and $\kappa_\parl$) are different, a fact that was already observed in the conformal relativistic case, \cite{iancu}.
 This is persisting  here, and we are able to show, for general non-conformal backgrounds, that $\kappa_\parl \ge \kappa_\perp$,
 namely the longitudinal diffusion constant always dominates the transverse one.
Furthermore, both the diffusion constants and the friction term are momentum dependent,
as expected. This is in contrast to the conformal case.

The properties of this relativistic Langevin evolution differ substantially from rotationally invariant equations
that have been introduced recently in mathematical physics \cite{rel,rel1,lan}.
In particular, here the evolution is not rotationally symmetric, and the Einstein relation is different, because the fluctuation-dissipation relation is different.
This implies that the equilibrium configuration is not the standard rotationally-invariant J\"uttner-Boltzmann distribution.

The processes we discuss here are connected to  a general property of a class of statistical systems,
and provide a concrete and rather general solvable example thereof. Such systems, when in contact with a heat bath of temperature $T$, if stirred gently and continuously,
 end up in a stationary state that is thermal but with a temperature $T_s$, different from that of the heat bath.
 They satisfy a  fluctuation-dissipation relation involving the new temperature, \cite{kurchan}.
This phenomenon has been expected to occur in general, in situations with slow dynamics.
These include in particular glassy systems, as well as systems that are very gently stirred by external agents and reach stationarity.
Here, we have  a system that is stationary but strongly driven by the external source
 (an electric field on the flavor brane, that keeps the velocity large and constant as the quark is moving through the plasma).
Typically, $T_s\geq T$, \cite{kurchan} but in our case things are different.
In the conformal case $T_s= T_{s,conf} = T(1-v^2)^{1\over 4}\leq T$ and this seems
 to persist in several other cases, as already mentioned before.

The problem we are solving can be also cast in a different light. The following question
 has been asked since the beginning of the 20th century: what are the Lorentz-transformation
  properties of temperature and the associated stationary distribution? This question is still
   considered open, \cite{hanggi} with several conflicting answers. Our setup can be
   reconsidered as follows: the heavy quark is a ``thermometer" moving inside a heat
   bath of temperature $T$.
The way it measures temperature is via the fluctuation-dissipation theorem as argued
in generality in \cite{kurchan}. Therefore the temperature that it measures as it
moves in the thermal medium is $T_s$, which is velocity-dependent. This velocity
 dependence is simple in the conformal case, $T_s=T(1-v^2)^{1\over 4}$ but is
  rather complicated in the non-conformal case and is therefore system-
(and possibly thermometer-) dependent.

The Langevin correlators must be renormalized as they are divergent near the AdS boundary of the string world-sheet.
 We show that the only UV divergence is subtracted by a counterterm that renormalizes the (heavy) quark mass.
 The associated scheme dependence affects the real part of the correlators only.

The local Langevin equation arises when looking at the large-time limit of the fluctuations of the
 heat bath, i.e. at the small frequency modes. On the other hand, the holographic computation gives access to the full frequency
spectrum of the correlation functions driving the generalized Langevin dynamics. In this work we
compute holographically the full Langevin correlators and the associated spectral densities.
 In particular, we obtain analytic expressions (in terms of the bulk metric and dilaton profiles) in
the two opposite regimes of small and large frequencies $\omega$ (compared to
an appropriate temperature scale).

For small frequencies, an analytic expression for the spectral densities
is obtained using the {\em membrane paradigm} \cite{LiuIqbal}, which allows
to relate these quantities to the near-horizon values of the background functions.
 In the large-frequency regime, on the other hand, the spectral densities are obtained
via a modified WKB method, similar to the one followed in
\cite{Teaney} for {\em bulk} fluctuations in an $AdS$-Schwarzschild background. The high-frequency behavior is different,
depending on the mass of the probe quark. For finite mass, and for large $\omega$,
the spectral densities grows linearly with $\omega$, whereas in the limit when the quark mass becomes infinite this behavior changes to a cubic power-law.

Going beyond the zero-frequency limit is necessary when the
diffusion process happens on time scales comparable to, or smaller
than the auto-correlation time of the fluctuation propagators. More specifically,
 since these are thermal correlators at the temperature $T_s$, the large-time approximation breaks down over time-scales
shorter than $T_s^{-1}$. This condition puts a temperature-dependent upper bound on the momenta of the heavy quark,
above which the
diffusion process cannot be described by a simple {\em local} Langevin equation with white noise.
In this context, it is useful to have approximate expressions for
valid for large frequencies (for examples,
those we obtain with the WKB method) to model  the behavior of the system in the regime where the local Langevin approximation breaks down and the dynamics
becomes non-markovian (due to a non-trivial memory kernel)

The results described above apply to any five-dimensional holographic model
which admits asymptotically $AdS$ black-hole solutions. On the other hand,
it is  interesting to perform a {\em quantitative} comparison between
the diffusion constants calculated in concrete models, and characteristic
observables in heavy ion experiments that can simply be connected
to Langevin processes.
In the context of heavy-ion physics, the transverse diffusion constant
is directly related to the {\em jet-quenching parameter} $\hat{q}_\perp = 2 \kappa_\perp/v$. The latter is a convenient quantity
 to describe the observed phenomenon of {\em transverse momentum broadening}
of a heavy quark: this is the process by which  the transverse momentum of
 the heavy quark probe\footnote{Here, ``transverse'' refers to the initial quark trajectory,
 not to the direction of the colliding beams}, initially equal to zero,
 undergoes a stochastic diffusion process such that after a time
$t$ it acquires a dispersion $\Delta p_\perp^2 = \hat{q}_\perp v t $.

In the final part of this work we perform a quantitative analysis of both the full
Langevin correlators, and of the jet-quenching parameter, in a particular Einstein-dilaton model, namely
 Improved Holographic QCD \cite{ihqcd1,ihqcd2},
which agree quite well both qualitatively and quantitatively with the zero- and finite-temperature Yang-Mills
theory. In particular, we focus on the specific model which was put forward in \cite{gkmn3}, and displays a good quantitative match with the spectral
and thermodynamic properties of lattice Yang Mills theory.

The analysis is performed numerically, both with respect to the
background metric and to the solution of the fluctuation equations.
By a shooting technique, we determine the wave-functions describing
the world-sheet fluctuations, and obeying the appropriate retarded
boundary conditions. From the wave-functions, the holographic prescription
allows to determine the full Langevin retarded correlator, whose
imaginary part gives the associated spectral density.

Using the exact numerical evaluation we are able to test the different analytic results discussed above. In particular, we test the
validity of the WKB result for large frequency, and in various regimes
of quark mass and velocity. Unexpectedly, we find that the analytic
WKB formulae not only capture the large frequency regime, but are a very good approximation to the correlators at almost all frequencies.

The numerical evaluation of the diffusion constants may lead directly to a
comparison of the jet-quenching parameters between the holographic QCD model
and  data. We find that $\hat{q}_\perp$ displays a mild momentum dependence for large quark momenta, which
however differs from the one obtained holographically in the
conformal case. As the temperature rises, $\hat{q}_\perp$ increases significantly, approximately as $\sim T^3$. Interestingly,
it is found that for temperatures above $\sim 400~MeV$, the local description
of the diffusive process {\em breaks down} for charm quarks with momenta above $\sim 5-10~GeV$.
This is because the process occurs on time scales shorter than $1/T_s$.
This would imply that
 in order to describe heavy charm quark
diffusion in the ALICE experiment, one would need the full generalized
non-local Langevin equation, and the full frequency-dependent correlator,
rather than just its low-frequency limit captured by $\hat{q}_\perp$.
This would constitute an interesting testing ground for holographic
models, where the full correlators can be easily computed.
We also estimate the energies at which the energy loss mechanism described here, is not any more the dominant one and radiation becomes the dominant mechanism.
This is estimated by requiring that $r_s$ remains below the would-be position of the flavor brane, \cite{Casal,lrwr}.
We show that these limit do not substantially constrain this framework.

A direct quantitative comparison of the results of this paper with data is hampered by
 the fact that quark degrees of freedom in the plasma are not included in our analysis. We are compensating (partly at least)
 for this using the ``energy scheme" for comparison, however the recent results found in  \cite{bigazzi} (and reviewed in \cite{nunez})
suggest that even in that case we may be underestimating the result.

In summary the AdS/CFT calculation of the Langevin diffusion of heavy quarks has the following characteristics

\begin{itemize}

\item The diffusion coefficients are asymmetric. The longitudinal diffusion coefficients is always larger than the transverse one.
They both become large with increasing $\gamma$.

\item The Langevin correlators satisfy a thermal fluctuation-dissipation relation with a temperature $T_s$ that is typically smaller than then heat bath temperature $T$.
In the conformal case $T_s={T\over \sqrt{\gamma}}$, \cite{iancu}. The associated Einstein relations are non-standard, especially the longitudinal one.

\item The local (Markovian) Langevin diffusion breaks down at some energy scale. Beyond this scale, the full force correlators are needed.
This breakdown is expected to be relevant at LHC energies for the charm.

\end{itemize}

This paper is organized as follows.
Section 2 describes in detail the Langevin equation for a
relativistic heavy quark travelling in the quark-gluon plasma. In
this section we also review how to describe the Langevin
dynamics in the holographically dual geometry, in terms of
fluctuations of trailing strings. Section 3 presents the necessary
background for the holographic computation. In particular we
present the holographic dual geometry of our non-conformal
model, the relevant classical trailing string
solution, and the corresponding linear fluctuations. It is in this
section that we obtain the fluctuation equations in general
non-conformal black hole space-times, whose solutions enter the
construction of the Langevin propagators.

Sections 4 and 5 contain our main results. In Section 4 we
discuss the Langevin correlators and the associated spectral
densities, first in full generality, then in the various limits of
low- and high- frequency. In section 5 we specialize to the
low-frequency modes, which compute the long-time behavior of the diffusion and friction
coefficients of the local Langevin equation. We provide exact
analytic expressions for these quantities, in terms of the
background metric functions. We also discuss the non-relativistic
and ultra-relativistic limits, and derive the modified Einstein
relations. While the previous sections deal with a completely
general holographic dual, in Section 6 we provide a numerical
study of these results in a specific model, namely Improved Holographic QCD that was shown in \cite{gkmn3} to provide
a good quantitative description of the static properties of pure
Yang-Mills at zero and finite temperature. In particular, in this
section we compute the jet-quenching parameter arising from this
model, and discuss the results in light of RHIC data.

Several technical details are left to the Appendices.  In Appendix \ref{App GR} we discuss some subtleties in the
definition of the propagator, related
to boundary terms.
Appendix \ref{details kappa}  provides the details of the calculation of the diffusion constants;
In Appendix \ref{AppWKB} we give a detailed discussion of the WKB method that we use to
obtain the large-frequency limit of the spectral densities, as it is more involved than the conformal case. In Appendix \ref{AppN=4} we discuss
the Langevin correlators in the conformal, ${\cal N}=4$ case.

\section{Langevin Equation for a relativistic heavy quark}
\lab{LangevinSec}

In this section we review how the diffusion of a relativistic
heavy quark through the plasma is described by a generalized
Langevin equation. First we give the purely 4D picture. Then, in
subsection 4.2, we review the holographic description of the
Langevin process that appeared in the previous literature, for the
case of $AdS$-Schwarzschild black-holes. This will be extended to
general asymptotically $AdS$ geometries in Section 3.

\subsection{The Langevin equation in the boundary theory}

Consider a quark which, in a first approximation, experiences a
uniform motion across the plasma, with constant velocity $v$. Due
to the interactions with the strongly-coupled plasma, the actual
trajectory of the quark is expected to resemble Brownian motion. To
lowest order, the action for the external quark coupled to the
plasma can be assumed, classically, to be of the form:
\be\label{bound action} S[X(t)] = S_0 + \int d\tau X_\mu(\tau)
{\cal F}^\mu(\tau) \ee
where $S_0$ is the free quark action, and ${\cal F}(\tau)$ depends
only on the plasma degrees of freedom, and plays the role of a
driving force (the ``drag" force).

To obtain an equation for the quark trajectory one needs to trace
over the plasma degrees of freedom. If the interaction energies
are small compared with the quark kinetic energy (therefore for a very
heavy quark, and/or for ultra-relativistic propagation speeds),
tracing over the microscopic degrees of freedom of the plasma can
be performed in the semiclassical approximation, and the quark
motion can be described by a {\em classical} generalized Langevin
equation for the position $X^i(t)$, of the form:
\be\label{langeq} {\delta S_0 \over \delta X_i(t)} =
\int_{-\infty}^{+\infty} d\tau ~\theta (\tau) C^{ij}(\tau)
X_j(t-\tau) + \xi^i(t) , \qquad i=1,2,3
\ee
Here, $C^{ij}(t)$ is a {\em memory
kernel}, $\theta(\tau)$ is the Heaviside function and $\xi(t)$ is a Gaussian random variable with
time-correlation: \be\label{noise} \langle\xi^i(t)\xi^j(t')\rangle
= A^{ij} (t-t') \ee

The functions  $A^{ij} (t)$ and $C^{ij}(t)$ are determined
 by the symmetrized and anti-symmetrized real-time correlation functions
 of the forces ${\cal F}(t)$ over the statistical ensemble:
\be\label{correlators1} C^{ij}(t) = G_{asym}^{ij}(t) \equiv -i
\langle\left[{\cal F}^i(t), {\cal F}^j(0) \right]\rangle, \quad
A^{ij} (t) = G_{sym}^{ij}(t) \equiv -{i\over 2}\langle\left\{{\cal
F}^i(t), {\cal F}^j(0) \right\}\rangle . \ee

The results (\ref{langeq}) and (\ref{correlators1}) are very
general, and do not require any particular assumption about the
statistical ensemble that describes the medium (in particular, they
do not require thermal equilibrium). One way to arrive at equation
(\ref{langeq}) is using the double time formalism and the
Feynman-Vernon influence functional \cite{Feyver}. A clear and
detailed presentation can be found in \cite{kleinert}, chapter 18.

The retarded and advanced Green's function are defined by:
\be
G_R^{ij}(t) = \theta(t)C^{ij}(t), \qquad G_A^{ij}(t) =
-\theta(-t)C^{ij}(t),
\ee
 which lead to the relation
\be\label{comm} C^{ij}(t) = G_R^{ij}(t) - G_A^{ij}(t)
\ee
Notice
that the kernel entering the first term on the right in equation
(\ref{langeq}) is the retarded Green's function, $G_R^{ij}(t) =
\theta(t) C^{ij}(t)$.

It is customary to introduce a {\em spectral density}
$\rho^{ij}(\omega)$ as the Fourier transform of the
anti-symmetrized (retarded) correlator,
\be C^{ij}(t) =
-i\int_{-\infty}^{+\infty} d \omega\, \rho^{ij}(\omega)
e^{-i\omega t}, \qquad G_R^{ij}(\omega) = \int_{-\infty}^{+\infty}
d \omega' \, { \rho^{ij}(\omega') \over \omega - \omega' +
i\epsilon}. \ee

{}From equation (\ref{comm}) and the reality condition $G_A(t) =
G_R(-t)$, or in Fourier space, $G_A(\omega) = G_R^*(\omega)$, we
can relate the spectral density to the imaginary part of the
retarded correlator:
\be \rho^{ij}(\omega) = -{1\over \pi} {\rm Im}\,
G_R^{ij}(\omega) \ee

\paragraph{Local limit.}
Suppose the time-correlation functions  vanish for sufficiently large
separation, i.e. for times much larger than a certain correlation time $\tau_c$.
Then, in the limit $t\gg \tau_c$, equation (\ref{langeq}) becomes
a conventional {\em local} Langevin equation, with local friction and white
noise stochastic term. Indeed, in this regime the noise correlator can be
approximated by
\be A^{ij} (t-t') \approx \kappa^{ij}
\delta(t-t'), \qquad t-t' \gg \tau_c.
\ee
This equation defines the Langevin diffusion constants
$\kappa^{ij}$. Similarly, for the friction term, we define the function
$\gamma^{ij}(t)$ by the relation:
\be C^{ij}(t) = {d\over d t}
\gamma^{ij}(t) \ee
 so that the friction term can be approximated, for large times,
 as: \be\int_0^\infty d\tau C^{ij}(\tau)
X_j(t-\tau) \approx \left(\int_0^\infty d\tau\,
\gamma^{ij}(\tau)\right)\dot{X_j}(t), \qquad t\gg \tau_c. \ee

In this regime, equation (\ref{langeq})
becomes the local Langevin equation with white noise,
\be\label{langeq2} {\delta S_0 \over \delta X_i(t)} + \eta^{ij}
\dot{X}_j(t) = \xi^i(t), \qquad \langle\xi^i(t)\xi^j(t')\rangle =
\kappa^{ij} \delta(t-t'),
\ee
with the self-diffusion and
friction coefficients given by:
\be \label{coeff} \kappa^{ij} =
\lim_{\omega \to 0} G_{sym}^{ij}(\omega); \qquad \; \eta^{ij}
\equiv \int_0^\infty d\tau\, \gamma^{ij}(\tau) = -\lim_{\omega
\to 0} { {\rm Im}\, G_R^{ij} (\omega) \over \omega}. \ee

In the case of a system at equilibrium with a canonical ensemble
at temperature $T$, one has the following relation between the
Green's functions:
\be\label{thermal} G_{sym}(\omega) = -
\coth{\omega\over 2T}\, {\rm Im} \,G_R(\omega), \ee
 which using equation
(\ref{coeff}) leads to the Einstein relation $\kappa^{ij}= 2 T
\eta^{ij}$.
For such a thermal ensemble, the real-time correlators decay exponentially with a scale
set by the inverse temperature, therefore
the typical correlation time is $\tau_c \sim 1/T$.

 Determining and studying the Langevin correlators (\ref{correlators1}) , and
the diffusion constants (\ref{langeq2}) will be the main purpose
of the rest of this paper.

Next, we write down explicitly the classical part, $\delta S_0
/\delta X(t)$ of the Langevin equation, in order to arrive at an
equation describing momentum diffusion. We start with the kinetic
action for a free relativistic quark, \be
 S_0[X_\mu(\tau)] =- M_q \int d\tau \sqrt{{d X^\mu \over d\tau} {d X_\mu \over d\tau} }
\ee

 We choose the gauge $\tau = X^0$, and obtain
 \be
 \delta S_0
/\delta X^i(\tau) = d p_i/dt \sp {\rm with}\sp p_i \equiv M_q \dot{X}_i (1
- \dot{X}_i \dot{X}^i)^{-1/2}.
\ee
 Equation (\ref{langeq2}) becomes
the Langevin equation for momentum diffusion: \be\label{langeq3}
{d p^i \over d t} = - \eta_D^{ij}(\vec{p}^{~2}) p_j + \xi^i(t),
\ee where: \be\label{langeq3-bis} \eta_D^{ij}(\vec{p}^{~2}) =
{\eta^{ij} \over \gamma(\vec{p}^{~2}) M_q}, \qquad
\gamma(\vec{p}^{~2}) \equiv \sqrt{1 + \vec{p}^{~2}/M_q^2}. \ee

\paragraph{Linearized Langevin equations.}
For a generic quark trajectory, the Langevin equation
(\ref{langeq3}) is non-linear,
 due to the $p$-dependence implicit in $\eta^{ij}_D$.
To put it in a form which allows for the holographic treatment in
terms of the trailing string fluctuations, it is convenient to
derive from equation (\ref{langeq3}) a linearized Langevin equation
for the fluctuations in the position around a trajectory with
uniform velocity, $\vec{X}(t) = \vec{v}t + \delta \vec{X}$. To
this end, we separate the longitudinal and transverse components
of the velocity fluctuations:
\be \dot{\vec{X}}(t) = \left(v +
\delta \dot{X}^\parl(t)\right){\vec{v} \over v} + \delta
\dot{\vec{X}}^\perp. \ee

The corresponding linearized expression
of the momentum reads:
\be \vec{p} = M_q {\dot{\vec{X}}\over
\sqrt{1 - \dot{\vec{X}}\cdot \dot{\vec{X}}}} \simeq M_q
\left(\gamma + \gamma^3 v \delta \dot{X}^\parl \right)
\left(\vec{v} +\delta \dot{\vec{X}}\right) = \vec{p_0} + \delta
\vec{p},
\ee
where we introduced the zeroth-order Lorentz factor
$\gamma \equiv (1-v^2)^{-1/2}$. The zeroth-order term is
$\vec{p}_0 =\gamma M_q \vec{v}$, and the longitudinal and
transverse momentum fluctuations are given by:
\be \delta p^\parl =
\gamma M_q (1 + v^2 \gamma^2) \delta \dot{X}^\parl = \gamma^3 M_q
\delta \dot{X}^\parl, \qquad \delta p^\perp_i = \gamma M_q \delta
\dot{X}^\perp.
\ee
It is convenient to separate the longitudinal
and transverse components of the propagators, since as it will become
clear in the next section, the off-diagonal components vanish:
\be
G^{ij}(t) = G^\parl(t) {v^i v^j \over v^2} + G^\perp(t)
\left(\delta^{ij} - {v^i v^j \over v^2}\right)
\ee
and the
corresponding decompositions for $\eta^{ij}$ and $\kappa^{ij}$ from (\ref{coeff}).

Inserting these expressions in equation (\ref{langeq3}), we find to
zeroth order: \be\label{zeroth} {d p_0 \over d t} = -\eta^\parl_D
p_0 , \qquad p_0 \equiv \gamma M v \ee and to first order in
$\delta \vec{X}$ the relativistic Langevin equations for
position fluctuations:
\bea
&& \gamma^3 M_q\delta \ddot{X}^\parallel = - \eta^\parallel(v) \delta \dot{X}^\parallel + \xi^\parallel,
 \qquad \langle\xi^\parl(t) \xi^\parl(t') \rangle= \kappa^\parl \delta(t-t'), \label{langeq4-a}\\
&& \gamma M_q\delta \ddot{X}^\perp =- \eta^\perp(v) \delta
\dot{X}^\perp + \xi^\perp, \qquad \langle\xi^\perp(t)
\xi^\perp(t')\rangle = \kappa^\perp \delta(t-t')\label{langeq4-b}
\eea where the friction coefficients $\eta^{\parl,\perp}$ are
related to the coefficients $\eta_D^{ij}$ by \be\label{fric-p-x}
\eta^\perp =\gamma M_q \eta_D^\perp, \qquad \eta^\parl =
\gamma^3 M_q\left(\eta_D^\parl + p {\de \eta_D^\parl \over \de
p}\Big|_{p=\gamma M_q v}\right). \ee

As we shall see in the following sections, the holographic
prescription will directly compute the friction coefficients
$\eta^{ij}$ and the diffusion coefficients $\kappa^{ij}$ appearing
in equations (\ref{langeq4-a}-\ref{langeq4-b}).

\paragraph{Short-time solution: momentum broadening.}
For times shorter than the relaxation time $\tau_D~1/\eta_D$ we can treat
the quark as travelling at a constant velocity $v$ ( which is a
good approximation in the case of a very heavy quark). In this
regime\footnote{As we are relying on the local form of the Langevin process, equation (\ref{langeq3}), we must still require time separations much larger than the
auto-correlation time $\tau_c$. More explicitly, we consider time scales $t$ such
that $\tau_c\ll t \ll \tau_D$.
Therefore, consistency demands that $\tau_D \gg \tau_c$.}, one can write an approximate solution for equation
(\ref{langeq3}), which describes a Brownian-like diffusion for
momentum fluctuations.

 We start once again with equation (\ref{langeq3}), and linearize it (this time
staying in momentum space) around a uniform trajectory $\vec{p}
\simeq p_0\vec{v}/v + \delta\vec{p}$. In the longitudinal and
transverse directions we find the two equations: \be {d \delta
p^\perp \over dt} = - \eta^\perp_{D,0} \delta p^\perp + \xi^\perp,
\qquad {d \delta p^\parl \over dt} = - \eta^\parl_{D,0} p_0 +
\left[\eta^\parl_D + p\left(\de \eta^\parl_D \over \de p\right)
\right]_{p_0} \delta p^\parl + \xi^\parl \ee where
$\eta_{D,0}^{ij} \equiv \eta_D^{ij}(p_0)$. The solution to these
equations is straightforward: assuming initial conditions $\delta
\vec{p}(t=0) = 0$, it reads (notice that $p^\perp = \delta
p^\perp$): \bea
&& p^\perp(t) = \int_0^t dt'\, e^{\eta_{D,0}^\perp (t'-t)}\xi^\perp(t'), \\
&& p^\parl(t) = p_0 e^{-\eta_{D,0}^\parl t} + \int_0^t dt'\,
e^{\tilde{\eta}^\parl_{D,0} (t'-t)}\xi^\parl(t'), \qquad
\tilde{\eta}^\parl_{D,0}\equiv \left[\eta^\parl_D + p\left(\de
\eta^\parl_D \over \de p\right) \right]_{p_0}. \eea

{}From these solutions, we can compute the noise-average of the
transverse and longitudinal momentum fluctuations
 \bea
&&\langle (p^\perp)^2 \rangle = \int_0^t dt'\,\int_0^t dt''\, e^{\eta_{D,0}^\perp (t'+t''-2t)}\langle\xi^\perp(t')\xi^\perp(t'')\rangle \\
&&\langle (p^\parl-p_0)^2 \rangle = p_0^2\left(1
-e^{-\eta_{D,0}^\parl t}\right)^2 + \int_0^t dt'\,\int_0^t dt''\,
e^{\tilde{\eta}^\parl
(t'+t''-2t)}\langle\xi^\parl(t')\xi^\parl(t'')\rangle. \eea
Using
the fact that $\langle \xi(t') \xi(t'')\rangle = \kappa
\delta(t'-t'')$, and expanding to linear order in $t\eta_D \ll
1$, we arrive at the final result:
\be\label{broaden} \langle
(p^\perp)^2 \rangle = 2 \kappa^\perp t, \qquad \langle (\Delta
p^\parl)^2 \rangle = \kappa^\parl t. \ee The first equation
describes {\em transverse momentum broadening,} and it is
typically parametrized in terms of the {\em jet-quenching
parameter} $\hat{q}^\perp$, \be \hat{q}^\perp = {\langle (p^\perp)^2 \rangle
\over v t} = 2 {\kappa^\perp \over v} \ee

The following sections, namely \ref{correlators}, \ref{transport},
and \ref{numerics} will be devoted to the calculation, in a 5D
holographic setup, of the Langevin correlators
(\ref{correlators1}) and of the diffusion constants
$\kappa^\parallel$ and $\kappa^\perp$ appearing in equation
(\ref{langeq3}).

\subsection{The Langevin equation in the gravity dual picture}

As we have reviewed in the previous subsection, the memory kernel
and the noise time-correlation function that govern the
generalized Langevin equation (\ref{langeq})
 for an external quark, are given by appropriate real-time correlation functions
of the force operator ${\cal F}(t)$ over the ensemble that
describes the medium.

These correlation functions are precisely the kinds of objects one
can compute in the gravity dual picture: one needs to identify the
appropriate bulk field that couples to the boundary operator
${\cal F}$, then solve the bulk equations for this field with
appropriate boundary conditions.

As first discussed in \cite{her,lrw,gub1}, and as we will review in detail in Section 3,
a probe heavy quark propagating through the plasma is
described, in the gravity dual picture, by a probe string with an endpoint
attached to a flavor brane, and extending into the bulk.
 The string endpoint moves along the quark trajectory and the rest of the string
trails its endpoint extending in the holographic directions. The string
world-sheet is described by the embedding coordinates $X^A(\sigma,\tau)$ which,
in the static gauge $\tau=t,\sigma=r$, reduce to the spacial components $\vec{X}(r,t)$,
where $r$ is the non-compact holographic direction.

Using the trailing string picture for the heavy quark,
the identification of the appropriate bulk field is
straightforward: from equation (\ref{bound action}) it is clear that
the external source for the boundary field ${\cal F}_i(t)$ is
nothing but the quark position $X^i(t)$, i.e. the boundary value
for the string embedding $X^i(t,r)$. More precisely, for a heavy
quark that follows an approximately uniform trajectory $X^i(t) =
v^i t + \delta X^i(t)$ the boundary coupling is of the form
 \be
S_{coupling} = \int dt \delta X^i(t) {\cal F}_i(t), \qquad \delta
X^i(t) = \delta X(r_b,t) \ee
 where $\delta X(r_b,t)$ is the
boundary value of the {\em fluctuation in the trailing string} around the
classical profile. Therefore, the correlation functions (\ref{correlators1}) of
the force operators can be extracted, in the Gaussian approximation, by solving for the
bulk linear fluctuations around the trailing string
and using the appropriate holographic prescription.

This calculation was first performed in \cite{tea,gubser}, for the
$AdS$ case. In this case, the world-sheet fluctuations propagate
on a space-time with a metric of the form (\ref{H99}), with \be
 b(r) = \ell/r, \qquad f(r) = 1 - (\pi T r)^4,
\ee where $T$ is the black-hole temperature. For a quark velocity
$v$, the induced metric has a horizon at $0< r_s < r_h$, with
associated temperature $T_s = T /\sqrt{\gamma}$. The retarded
correlator for the longitudinal and transverse components of
$\vec{{\cal F}}$ was determined using the prescription of
\cite{SonStar}: \be\label{Green AdS}
\gamma^{-2}G_{R,AdS}^\parallel = G_{R,AdS}^\perp = - \Psi^*
{\cal G}^{rr}\de_r \Psi \big|_{Boundary}. \ee
 where $\Psi(r,\omega)$ is a solution to the fluctuation equation with
 unit normalization at the boundary and infalling boundary conditions at
the horizon, ${\cal G}^{rr} = (2\pi \ell_s^2)^{-1} H^{rr}$, with
$ H^{rr}$
 given by equation (\ref{Hab}) specialized to the $AdS$-Schwarzschild case.

In order to compute both terms entering the Langevin equation, one
needs also
 the symmetrized correlator, which gives the noise time-correlation function. In general,
the relation between the retarded and symmetrized correlator
depends on the statistical ensemble one is dealing with. For a
black hole (as in the case of the induced world-sheet metric), the
features of the ensemble can be obtained by connecting the mode
solutions along a Keldysh contour between the two boundaries of
the maximally extended Kruskal diagram. This corresponds to
obtaining a statistical ensemble by tracing over the degrees of
freedom of one of the causally disconnected regions. In the
context of $AdS$/CFT this idea was put forward in
\cite{sonherzog}, which also provides a justification of the
prescription (\ref{Green AdS}) for the retarded propagator.

We will not go into the details of this procedure, which can be
found in \cite{gubser,Casal}. The crucial point
is that the stationary statistical ensemble one obtains is a thermal
ensemble at the temperature $T_s=T/\sqrt{\gamma}$.
Therefore, one can compute $G_{sym}$ from the imaginary part of
$G_R$, as in equation (\ref{thermal}), with the substitution $T \to
T/\sqrt{\gamma}$.

Notice that the retarded Green's functions compute, through equations
(\ref{coeff}), the coefficients appearing in the Langevin
equations for the fluctuations $\delta X^i$, equation
(\ref{langeq4-a}-\ref{langeq4-b}). In particular, to extract the
coefficient $\eta_D^\parl$ one must divide $G^\parl(\omega)$ by an
extra factor $\gamma^2$ with respect to the corresponding result
for $\eta_D^\perp$.

Computing the zero-frequency limit of the retarded correlators,
the resulting Langevin diffusion coefficients are found to be
\cite{gubser,Casal}:
\be \gamma^{-2}\kappa^\parl= \kappa^\perp =
\pi {\ell^2\over\ell_s^2} \sqrt{\gamma}\, T^3, \ee
and the friction coefficients reproduce the classical drag force
calculation \cite{gub1},
\be
\eta^\parl_D = \eta^\perp_D = {\pi \over 2}{\ell^2\over \ell_s^2} {T^2\over M}.
\ee
The diffusion and friction coefficients indeed
satisfy an Einstein relation appropriate for the temperature $T_s$:
\be \kappa^\perp/\eta^\perp =
\kappa^\parl/\eta^\parl = 2 M T_s, \ee
where $\eta^\perp$ and $\eta^\parl$ are related to $\eta_D$ by equation (\ref{fric-p-x}).
Notice that, in the conformal case, $\eta_D$ is momentum-independent.

The approach taken in \cite{gubser} derives the Langevin
propagators by using the standard holographic prescription for
real-time correlators. An alternative procedure, giving the same
result, was adopted in \cite{deboer,sonteaney} for the
non-relativistic case, and later in \cite{iancu} for the general
relativistic case. These authors performed a direct derivation of
the Langevin equation: starting from the trailing string
fluctuations in the bulk, and integrating them out, they showed
explicitly that one arrives at equations like
(\ref{langeq4-a}-\ref{langeq4-b}) for the boundary fluctuations,
with coefficients given by the formulae previously found in
\cite{tea,gubser}. Furthermore, they showed that the same result
can be obtained by integrating out only a strip between the
world-sheet horizon $r_s$ and the stretched horizon $r_s +
\epsilon$: this gives a picture of the stochastic behavior of the
string fluctuations as originating from the world-sheet horizon.
In this work we will not follow explicitly this road, but rather
rely on the holographic computation of the Langevin correlators.
Nevertheless, by using the general formalism developed
\cite{LiuIqbal}
 we will compute directly the
transport coefficients.

\section{5D non-conformal backgrounds for Langevin holography}

In this section, we present the background material for the
holographic computation of the Langevin correlators that we carry
out in the Section 4. As a bulk geometry we take a general, five-dimensional,
asymptotically $AdS$ black-hole, dual to a non-conformal deconfined plasma.
These geometries arise generically as solutions in appropriate Einstein-dilaton
theories in five-dimensions \cite{gkmn1,gkmn2}.

A heavy external
quark moving through the plasma at temperature $T$ can be
described by a string whose endpoint at the boundary follows the
quark's trajectory \cite{her}-\cite{gub2}. The string extends into
the bulk, whose geometry is the dual black hole background with
appropriate temperature $T$. Once the motion of the endpoint at the boundary
is specified, one can find the trailing string solution through
the geodesic equation: the momentum flow along the string is dual
to the drag force experienced by the quark moving through the
plasma.

The fluctuations of the string world-sheet around the
geodesic solution are holographically dual to the stochastic forces
felt by the quark due to its interaction with the medium. Their
effect to leading (Gaussian) order is that of stochastic noise acting on the quark,
resulting in a Langevin-type diffusion with its associated transport
coefficients (diffusion constants).

\subsection{5D Einstein-Dilaton black holes}

We shall consider the dynamics of a probe string in a general 5D
black hole geometry, with a string frame metric: \be\label{metric}
ds^2 = b^2(r)\left[{dr^2\over f(r)} - f(r)dt^2 + dx^i dx_i
\right].
\ee
 We assume there is an asymptotically AdS region $r \to 0$
where
\be\label{UVmetricdil} \log b(r) \sim -\log {r\over \ell} +
{\rm subleading} , \qquad f(r) \sim 1 + O(r^4), \qquad r\to 0, \ee and a
horizon at $r=r_h$ where $f(r_h) = 0$, and $f'(r_h)$ and $b(r_h)$
remain finite. The black hole temperature is given by:
\be 4\pi T =
-f'(r_h). \ee

We make no particular assumptions on the subleading terms in equation
(\ref{UVmetricdil}). In case these subleading terms actually vanish as
$r\to 0$, then the metric is  $AdS$ in the usual
sense.

The black holes of the type (\ref{metric}) arise, in particular,
as solutions of a large class of 5-dimensional Einstein-dilaton
models, described by the Einstein-frame metric $g_{\mu\nu}^E$ and
a scalar field $\lambda$ with the action: \be\label{action2} S =
-M_p^3 N_c^2 \int \sqrt{-g^E} \left[R^E - {4\over3} {(\nabla
\lambda)^2\over \lambda^2} + V(\lambda)\right]. \ee In the
holographic interpretation of these models, the scalar $\lambda$
is dual to the running coupling $\l_t$ of the four-dimensional
gauge theory. This is the class of models we will have in mind,
although the results of this work apply to any 5D theory that
admits solutions such as (\ref{metric}).

For an appropriate choice of the potential $V(\l)$, the models with action (\ref{action2})
provide a good holographic dual to large-$N_c$ 4-dimensional pure
Yang-Mills theory, at zero and finite temperature
\cite{ihqcd1}-\cite{gkmn3}. The potential should have a regular
expansion as $\lambda \to 0$, with
\be V(\lambda) \sim {12\over \ell^2}(1
+ v_0 \lambda + \ldots)\,\,.
\ee
Furthermore, linear confinement in
the IR requires that, at large $\lambda$,  $V(\lambda)$ grows at
least as fast as $\lambda^{4/3}$. With these requirements,
\begin{enumerate}
\item The solutions in the Einstein frame are an asymptotically $AdS$ metric, with $AdS$ length $\ell$, and a non-trivial profile $\lambda(r)$;
\item There is a first order Hawking-page phase transition with a non-zero critical temperature $T_c$.
\end{enumerate}
For a short review of the main features of these models, the
reader is referred to  \cite{ihqcdrev}.

We will be interested in the fluctuations of a probe string in
the 5D black hole geometry for $T>T_c$ (corresponding to the
deconfined phase). The black hole solutions in the string frame
have the form (\ref{metric}), with string frame metric
$g_{\mu\nu} = \lambda^{4/3}g^E_{\mu\nu}$. In the UV region $r=0$,
we therefore have:
\be
\label{UVlimit} b(r) \sim {\ell\over r}
\lambda^{2/3}(r), \qquad f(r) \sim 1 - {{\cal C} \over \ell^3 }
\,r^4 , \qquad \lambda(r) \sim -{9\over 8 v_0 \log r}.
\ee
where the constant ${\cal C}$ depends on the thermodynamic
quantities that characterize the black hole, and it can be
expressed in terms of the temperature $T$ and entropy density $s$,
\cite{gkmn2}: \be\lab{CsT} {{\cal C}\over \ell^3} = {45\pi^2 \over
N_c^2} s\, T \ee

The expressions (\ref{UVlimit}) are corrected by terms of
${\cal O}(\lambda)$, which are negligible near the boundary.

\subsection{Classical trailing string and the drag force}

Before going into the details of the world-sheet fluctuations,
we review the calculation of the unperturbed trailing solution, that
was discussed in \cite{her,gub1} for pure AdS black holes, and generalized
in \cite{transport} for black holes in 5D Einstein-Dilaton theories.
In this subsection we review the setup and the results of \cite{transport}.

We consider an (external) heavy quark moving through an infinite volume of gluon plasma with a fixed velocity $v$ at a finite temperature $T$.
 In the dual picture, this is described by a classical ``trailing'' string
 with an endpoint on the UV boundary moving at constant velocity $v$.

 The world-sheet of the string is described by the Nambu-Goto action\footnote{Throughout the paper, we will
 denote 5D coordinates by $\mu,\nu\ldots$, world-sheet coordinates by $\alpha,\beta\ldots$, and
 boundary spatial coordinates by $i,j\ldots$. Indices $i,j\ldots$ in the boundary theory are raised
  and lowered with metric $\delta_{ij}$, so we will make no distinctions between upper and lower indices as far as boundary tensors are concerned.} ,
\begin{equation}\label{NGACTION}
S_{NG} = -\frac{1}{2\pi \ls^2}\int d^2\sigma \sqrt{-\det g_{\alpha\beta}} \;, \qquad g_{\alpha\beta}=
g_{\mu\nu} \partial_{\alpha}X^{\mu} \partial_{\beta}X^{\nu}, \qquad \left\{\begin{array}{c} \mu,\nu =0\ldots5\\
\a,\b= 0,1\end{array}\right.
\end{equation}
where $g_{\mu\nu}$ are the components of the bulk metric in the string frame\footnote{Unless otherwise
 stated, $b(r)$ will always denote the scale factor in the {\em string frame}. This is a slight change of
  notation with respect to our previous papers \cite{ihqcd1,ihqcd2,gkmn1,gkmn2,gkmn3}, where the same
  quantity was denoted $b_s(r)$, but it is justified since most expressions in this work are simpler in the string frame.},
\be\label{metric2}
ds^2= b^2(r)\left[{dr^2\over f(r)}\,-\,f(r)dt^2\,+\,dx^i dx^i\right] \;, \qquad i=1,2,3.
\ee
The ansatz for the classical trailing string is \cite{gub1},
\begin{equation}
\label{TRAILANSATZ}
 X^{1} = v t +\xi(z),\quad X^{2}=X^{3}=0 \;,
\end{equation}
and along with the gauge choice
\begin{equation}\label{TRAILGAUGE}
\xi^0=t, \quad \xi^1 =r\,
\end{equation}
leads to the induced metric:
\be \label{indu}
g_{\alpha\beta} = b^2(r) \left(\begin{array}{cc}v^2 - f(r)& v \xi'(r)\\ v \xi'(r) & f(r)^{-1} + \xi^{'2}\end{array}\right),
\ee
and the corresponding action:
\begin{equation}
S = -\frac{1}{2\pi \ls^2} \int dt dr ~ b^2(r)\sqrt{1 - {v^2\over f(r)}+ f(r) \xi^{'2}(r)} \;.
\end{equation}
Since $S$ does not depend on $\xi$ but only its derivative, the
conjugate momentum $\pi_\xi$ is conserved,
\be\label{pixi}
\pi_\xi = -{1\over 2\pi \ell_s^2}{ b^2(r) f(r) \xi'(r) \over \sqrt{1 - {v^2\over f(r)} + f(r) \xi^{'2}(r)}} = - {b^2(r_s) \sqrt{f(r_s)}\over 2\pi \ell_s^2}
\ee
where the final expression is obtained by evaluating it at the point $r=r_s$,
defined by
\be\label{rs}
f(r_s) = v^2.
\ee

For an infinitely massive  quark, the string endpoint is the boundary, $r=0$.
For a quark of finite mass $M_Q$, the endpoint should be located  at a position 
 $r_Q$ in the interior, as discussed in detail in \cite{transport}. This
puts an upper bound on the quark velocity $v$, since the trailing string
picture fails when $r_s< r_Q$. At this point, the flavor brane dynamics
should become important.

\paragraph{The Drag Force} The {\em drag force} on the quark can be
determined by calculating the momentum that is lost
by flowing from the string to the horizon, which results in:
\be\label{b2}
 F_{\rm drag} = \pi_\xi = - \frac{v\, b^2(r_s)}{2\pi \ls^2},
\ee
where we have replaced $f(r_s)$ by $v^2$ in the last equality.

One defines the momentum {\em friction coefficient} $\eta_D$ as the characteristic attenuation constant for the momentum of a quark of mass $M_q$:
\be
F_{\rm drag}= {dp\over dt}\equiv-{\eta_D}p,\sp p={M_q v \gamma}
\label{b10}\ee
where $\gamma = (1-v^2)^{-1/2}$ is the relativistic contraction factor. With this definition we obtain:
\be\label{frictioncoeff}
\eta_D = {1\over \gamma M_q} { b^2(r_s) \over 2\pi \ell_s^2 }.
\ee
 In the conformal case, $\eta_D$ is independent of $p$,
\be\label{b10-2}
\eta^{conf}_D={ \pi \sqrt{\l}~T^2 \over 2M_q},
\ee
where $\lambda=(\ell/\ell_s)^4$ is the fixed 't Hooft coupling of ${\cal N}=4$ sYM.
This is not anymore so in the general case, where $\eta_D$
is momentum dependent.

\paragraph{The world-sheet black hole.} The coordinate value $r=r_s$ is
 a horizon for the induced world-sheet
metric. In order to ascertain this, we can invert equation (\ref{pixi}) to obtain
$\xi'(r)$ in the form:
\be
\xi'(r)={C\over f(r)}\sqrt{f(r)-v^2\over b^4(r)f(r)-C^2}\sp C \equiv v \, b^2(r_s).
\label{H5}\ee
 We may now change coordinates to diagonalize the
induced metric, by means of the reparametrization:
\be
t=\tau+\zeta(r), \qquad \zeta'={v\xi'\over
f-v^2}={Cv\over f(r)\sqrt{(f(r)-v^2)( b^4f-C^2)}}. \label{H14} \ee
In these coordinates the induced metric is
\be
ds^2=b^2\left[-(f(r)-v^2)d\tau^2+{b^4\over
(b^4f-C^2)}dr^2\right].
\label{H15}\end{equation}
The coefficient of $d\tau^2$ vanishes at $r_s$, so this point corresponds to a world-sheet black-hole horizon.
Since $f(r)$ runs between $0$ and $1$ as $0<r<r_h$, by definition (\ref{rs})
the world-sheet horizon is
always outside the bulk black hole horizon, and it coincides with it
only in the limit $v\to 0$. In the opposite limit, $v\to 1$, $r_s$
asymptotes to the boundary $r=0$.

The Hawking temperature associated to the black hole metric (\ref{H15}) is found as usual, by expanding around $r=r_s$ and demanding regularity of the Euclidean geometry. The resulting temperature is:
\be \label{Ts}
 T_{s} \equiv {1\over 4\pi}\sqrt{f(r_s)f'(r_s)\left[{4b'(r_s)\over b(r_s)}+{f'(r_s)\over f(r_s)}\right]}.
\ee

In the conformal limit, where the dilaton is constant and the background solution reduces to $AdS$-Schwarzschild, the world-sheet temperature and horizon position are
simply given by:
\be\label{conformal}
T_s^{conf} = {T\over \sqrt{\gamma}}, \qquad r_s^{conf} = {1\over \pi \sqrt{\gamma}\, T}.
\ee

More generally, in the ultra-relativistic limit $v\simeq 1$, one can express $r_s$ in terms of thermodynamic quantities. In this limit $r_s$ approaches the boundary $r=0$, and in this region the geometry approaches that of $AdS$-Schwarzschild, equations (\ref{UVlimit},\ref{CsT}). Therefore, from the definition (\ref{rs}) we obtain:
\be\label{rslim}
 r_s \simeq {1\over \sqrt{\gamma}}\le( \frac{4N_c^2}{45\pi^2
 sT }\ri)^\frac14
, \qquad v\to 1 \ee

\subsection{Fluctuations of the trailing string}
We now proceed to study the quadratic fluctuations around the
classical trailing string solution reviewed in the previous section.
This analysis was performed in the $AdS$ black hole background in
\cite{gubser}. Here, we extend it to the general 5D background (\ref{metric}).

We continue to work in the static gauge $\xi^0=t$, $\xi^1=r$, but we will allow for a more general ansatz for the embedding coordinates:
\be
X^1=vt+\xi(r)+\dx(r,t)\sp X^2=\dxy(r,t)\sp X^3=\dxz(r,t).
\label{H4}
\ee
We will treat the quantities $\delta X^i$ as perturbations around the
background solution (\ref{TRAILANSATZ}).

The Nambu-Goto action (\ref{NGACTION}) is now given by:
\be
S_{NG}=-{1\over 2\pi \ell_s^2}\int dtdr\sqrt{\hat g_{rt}^2-
\hat g_{tt}\hat g_{rr}}
\label{H2}\ee
where
\be
\hat g_{tt}=b^2(-f+\dot X^i\dot X^i)\sp \hat g_{rr}=b^2
\left({1\over f}+{X^i}'{X^i}'\right)
\sp \hat g_{rt}=b^2~{X^i}'\dot X^i
\label{H3}\ee
where a dot and a prime represent derivatives w.r.t. $t$ and $r$
respectively.

Expanding the Nambu-Goto action in $\delta X^i$ around the classical solution
 (\ref{H5}) we obtain, to quadratic order:
\be
S_2=-{1\over 2\pi \ell_s^2}\int dtdr~ G^{\a\b}\left[{1\over 2}
\partial_{\a}\dx\partial_{\b}\dx+ {Z^2\over 2} \sum_{i=2}^3
\partial_{\a}\delta X^i\partial_{\b}\delta X^i\right],
\label{H9}\ee
where
\be
G^{\a\b} = {b^2\over Z^3}\left( \begin{array}{cc}
-{Z^2f+v^2\over f^2}~~ & ~~{v\xi'} \\
{v\xi'}~~ & {f-v^2}\end{array}\right)\;,
\label{H99}\ee
and we have defined:
\be
Z\equiv b^2\sqrt{f-v^2\over b^4f-C^2}.
\label{H10}\ee
Note that $\det(G^{\a\b})=-b^4/Z^4$ and that in the
${\cal N}=4$ case $Z=\sqrt{1-v^2}$ is a constant.

In terms of the induced world-sheet metric (\ref{indu}), we obtain
\be
G^{\a\b}=Z^{-1} b^4 g^{\a\b}, \qquad \sqrt{-\det g}=b^2 Z\;.
\label{H12}
\ee
We may therefore rewrite the action as
\be
S_2=-{1\over 2\pi \ell_s^2}\int dtdr {b^2\over
2} \sqrt{-\det g}~g^{\a\b}\left[{1\over Z^2}\partial_{\a}\dx\partial_{\b}\dx + \sum_{i=2}^3 \partial_{\a}\delta X^i\partial_{\b}\delta X^i\right]
\label{H13}\ee
To simplify the action, we change
coordinates to diagonalize the induced metric,
as in the previous subsection. By a reparametrization of the world-sheet
time coordinate as in (\ref{H14}), the new induced metric is
(\ref{H15}), and the action read:
 \be S_2=-{1\over
2\pi \ell_s^2}\int d\tau dr ~{1\over 2} H^{\a\b}\left[{1\over Z^2}
\partial_{\a}\dx\partial_{\b}\dx + \sum_{i=2}^3 \partial_{\a}
\delta X^i\partial_{\b}\delta X^i\right]
\label{H16}\ee with \be \label{Hab}
 H^{\a\b} = \left( \begin{array}{cc}
-{b^4\over \sqrt{(f-v^2)(b^4f-C^2)}}~~ & ~~0\\
0~~ & \sqrt{(f-v^2)(b^4f-C^2)}\end{array}\right)\;,
\ee
Equations (\ref{H9}) and (\ref{H16}) show that the longitudinal fluctuations
(i.e. those parallel to the direction of the unperturbed trailing string
 motion), namely $\dx$, and the transverse fluctuation $\dxy$ and $\dxz$,
have different kinetic terms, as the effective two-dimensional metrics
they are sensitive to differ by a factor $Z^2$. From now on, we will denote the longitudinal fluctuations as $\delta X^{\parallel}$ and the transverse
fluctuations as $\delta X^{\perp}$.

{}From equation (\ref{H15}) one can immediately derive the field
equations satisfied by the fluctuations: \be
\partial_{\a} \,Z^{-2}\,H^{\a\b}\partial_{\b}\delta X^{\parallel}=0\sp \partial_{\a}\,H^{\a\b}\partial_{\b}\delta X^{\perp}=0.
\label{H18}\ee

For a harmonic ansatz of the form
$\delta X^i(r,\tau)=e^{i\omega \tau}\delta X^i(r,\omega)$,
equations (\ref{H18}) become:

\be
\partial_r\left[ \sqrt{(f-v^2)(b^4f-C^2)}\,\,\partial_r\left(\delta X^{\perp}\right)\right]+{\omega^2b^4\over \sqrt{(f-v^2)(b^4f-C^2)}}\,\delta X^{\perp}=0
\label{H19}\ee
\be
\partial_r\left[{1\over Z^2}\sqrt{(f-v^2)(b^4f-C^2)}\,\,\partial_r\left(\delta X^{\parallel}\right)\right]+{\omega^2b^4\over Z^2\sqrt{(f-v^2)(b^4f-C^2)}}\delta X^{\parallel}=0
\label{H20}\ee
 In the next sections we will compute the Langevin
correlation functions from these fluctuation equations and extract
the diffusion constants and the spectral densities from them.

We note however that the diffusion constants can also be read-off
directly from the quadratic action (\ref{H16}) by using the method
of the {\em membrane paradigm} as explained in section {\ref{mempar}.

\section{Holographic computation of Langevin correlators }\label{correlators}

\subsection{The Green's functions}

{}From the discussion in the previous section, it emerges that in a
4D theory with a 5D gravity dual we can compute the Langevin
correlators holographically, from the classical solutions for the
fluctuations of the trailing string. As we have observed in section 3,
these fluctuations behave as free fields propagating on a 2D
black-hole background, whose metric is essentially the induced
metric on the bulk trailing string, equation~(\ref{H15}). The asymptotic
form and the causal structure of
 this black hole are exactly the same as the one for the
trailing string embedded in an $AdS$ black hole.
As a consequence, the results of \cite{gubser,Casal,iancu}
discussed in the previous sections immediately
generalize to the more general metric (\ref{metric}): following the Keldysh contour
in the extended Kruskal diagram of the black hole, one finds a
thermal spectrum of transverse
and longitudinal fluctuations with effective temperature $T_s$, given in equation (\ref{Ts}).

In this ensemble, the symmetrized and retarded Green's functions obey the  relation
(\ref{thermal}), with $T = T_s$. Therefore, one can obtain both the memory kernel and the noise correlator
entering equations (\ref{langeq}-\ref{noise}) from the knowledge of the retarded Green's function.

{}From the structure of the action for the fluctuations, equation
(\ref{H13}), one can observe that there are essentially two types of
retarded correlators, $G_R^\parl (\omega)$ and
$G_R^\perp(\omega)$, for the longitudinal and transverse
fluctuations. We introduce a notation similar to \cite{gubser} and
define
\be \label{Gab} {\cal G}^{\a\b}_\perp \equiv {1 \over 2 \pi
\ell_s^2} H^{\a\b}, \qquad {\cal G}^{\a\b}_\parallel \equiv {1 \over 2
\pi \ell_s^2} {H^{\a\b} \over Z^2}, \ee
where $H^{\a\b}$ is defined in equation (\ref{Hab}).

The holographic prescription for the retarded correlator,
 computed with the diagonal induced metric \refeq{H15} is given by:
\be\label{full GR}
G_R(\omega) = - \left[\Psi_R^*(r,\omega) {\cal G}^{rr} \pa_r \Psi_R(r,\omega)\right]_{\rm boundary} \;.
\ee
Here $\Psi_R(r,\omega)$ denotes collectively the fluctuations
$\delta X^\parallel$, $\delta X^\perp$, solutions of equations (\ref{H19}-\ref{H20}) with the
appropriate boundary conditions, i.e. unit normalization at the boundary, and infalling
conditions at the world-sheet horizon (as we discuss more extensively below) and the factor
$ {\cal G}^{rr} $ is the appropriate one from equation (\ref{Gab})

The expression in equation~\refeq{full GR} must be evaluated at the boundary of the trailing
string world-sheet. In the case of
an infinitely massive quark, the string is attached at the $AdS$ boundary at $r=0$ (when needed, in order to keep quantities
finite, we introduce a cut-off boundary at $r=\epsilon$).
In case we want to keep the quark mass finite, the trailing string is attached to a point $r_Q$, which
is determined by demanding that the free energy of a static string in the $T=0$ background
gives the mass of the quark:
\be\label{mass cutoff}
M_q = \frac{1}{2 \pi \ell_s^2} \int_{r_Q}^{r_*} b(r)^2 dr\;
\ee
(here $r_*$ is the point at which $b(r)$ reaches its minimum, see the discussion in \cite{transport}).

Next, we discuss in greater detail the boundary conditions for the $\Psi_{R}$'s.
The solutions to the fluctuation equations \refeq{H19}--\refeq{H20} share the same asymptotics both for the transverse and longitudinal components
(since $Z(r)$ asymptotes to a constant both at the horizon and at the boundary,
where the equations have singular points). At
the world-sheet horizon $r\to r_s$ equations \refeq{H19}--\refeq{H20} both take the form
\be
\partial_r^2 \Psi + {1 \over |r-r_s|} \partial_r \Psi + \left( \frac{\omega}{4\pi T_s |r-r_s|} \right)^2 \delta \Psi = 0 \;,
\ee
 so that the solutions near the horizon behave as
\be
\Psi (r,\omega) \sim (r_s-r)^{\pm {i \omega\over 4 \pi T_{s}}}+\cdots \;
\label{H23}\ee
The $+$ sign in the exponent corresponds to a wave which is outgoing with respect to the world-sheet horizon, while the $-$ sign characterizes an
in--falling wave.

Near the boundary $r\to0$ both transverse and longitudinal fluctuations have to solve the following equation:
\be\label{eqUV}
\partial_r^2 \Psi(r,\omega) - \left({ 2 \over r } - {4\over 3}{\l'\over \l}\right)\partial_r \Psi(r,\omega) + \gamma^2 \omega^2 \Psi(r,\omega) = 0 \;,
\ee
As long as $r \l'/\l\ll 1$ \footnote{The condition $r \l'/\l\ll 1$ is realized
in particular in the case of logarithmic running: in that case, for small $r$, $r\l'/\l \sim \l << 1$. In the case where $\l$ is dual to a relevant operator it is also valid a fortiori since $r\l'/\l \sim \tilde \Delta r$, with $\tilde \Delta=min(\Delta,4-\Delta)$.}
the two independent solutions are a normalizable mode and a non-normalizable mode,
\be\label{psiUV1}
\Psi \sim C_s + C_v r^3 \l^{-4/3}.
\ee

According to the standard prescription \cite{SonStar}, the
appropriate boundary conditions for the wave functions $\Psi_R$ in the expression ~\refeq{full GR} for
the retarded correlator are the
in--falling behavior at the world-sheet horizon with the condition $\Psi_{R}(r)=1$ at the boundary:
\bea
&& \Psi_R(r_b,\omega) = 1 \qquad r_b =
\left\{\begin{array}{c} 0 \quad {M_q \to \infty}\\ r_Q \quad M_q \,\,\,{\rm finite} \end{array}\right. \label{psibound}\\
&& \Psi_R(r,\omega) \simeq \Psi_h \,\,(r_s-r)^{-{i \omega\over 4 \pi T_{s}}} \qquad r\sim r_s. \label{psihor}
\eea
where $\Psi_h$ is a constant.

Given the wave-function, obeying the near-boundary and near-horizon asymptotics specified by equations (\ref{psibound})-(\ref{psihor}), we can
extract the propagator from equation. (\ref{full GR}). Below,
 we  separately discuss the features of the real and imaginary parts
of the retarded Green's functions, and the associated spectral densities.

\paragraph{Real part  of the  retarded correlators.}
The real part of the correlator (\ref{full GR})
suffers from ambiguities related to the possibility of adding
boundary counterterms to the action (\ref{NGACTION}). This was discussed e.g. in
\cite{gubser-bv} in the context of the calculation of 4-dimensional
transport coefficients.

The ambiguities in the propagator are, as usual, associated to UV-divergences
in the on-shell action, that arise when we try to evaluate it in the
full $AdS$ space-time. To obtain a finite result, we must consider the action
on a  regularized  space-time, with boundary  at $r=\epsilon$ rather than $r=0$. Then, once the divergences in the limit $\e\to 0$  are identified, one can
add counterterms to subtract them and obtain a finite limit as $\e \to 0$.

The essence of holographic renormalization is that these counterterms
are {\em local, covariant boundary terms}.
 As we show in appendix \ref{App GR}, we  only need a single boundary
counterterm to regularize the action, and this is given  by the boundary-covariant point-particle action:
\be\label{count4}
S_{count} = \Delta M(\e) \int dt\, \sqrt{\dot{X}^\mu\dot{X}_\mu}
\ee

The same UV divergences
appear in the real part of the propagator,
if we try to compute it naively from  equation (\ref{full GR}). In fact, the expression  (\ref{full GR}) is nothing but the unrenormalized  on-shell action, as can be easily observed by integrating equation (\ref{H16})  by parts and using the field equations (\ref{H18}). Therefore, as a consequence of  the analysis of the on-shell action carried out  in  Appendix \ref{App GR}
 the divergent parts of the transverse and longitudinal
Green's functions  are:
\be\label{propdiv}
\Big({\rm Re}~G_R^\perp\Big)^{(div)} = {\l^{4/3}(\e)\over \e} {\ell^2 \over 2\pi \ell_s^2}\, \gamma \,\o^2, \qquad \left({\rm Re}~G_R^\parl\right)^{(div)} = {\l^{4/3}(\e)\over \e}{\ell^2 \over 2\pi \ell_s^2}\, \gamma^3 \,\o^2.
\ee

This result can also be explicitly derived from the
explicit form of  the wave-functions, close to the boundary. As we
will show in Subsection \ref{WKB},  (see eq. (\ref{psibndry}),  and for  more a more  detailed derivation, Appendix \ref{AppWKB}) the solution of eq. (\ref{eqUV}) is,
\be\label{psiUV2}
\Psi_{UV}(r) = \left[\cos(\gamma \o r) + (\gamma \o r) \sin  (\gamma \o r)\right] +  C_v \l^{-4/3}(r)\left[ (\gamma \o r )\cos  (\gamma \o r) - \sin  (\gamma \o r)\right]
\ee

This solution  generalizes eq. (\ref{psiUV1})  for any finite $\o$, and
it is valid  in the near-boundary region, i.e. for $\gamma \omega r \ll 1$,
and $\l(r)\ll1$.
The value of the coefficient $C_s$ of the leading term is fixed to ensure
unit normalization at $r=\epsilon$ (we are keeping in mind that we
will take the  $\e \to 0$ limit at the end).

Evaluating the real part of  (\ref{full GR}) at $r=\epsilon$ with  the wave-function given by (\ref{psiUV2}), we find the following divergent term for the  transverse and longitudinal components:
\be\label{ReG WKB}
{\rm Re}~G_R^\perp \simeq \gamma^{-2} {\rm Re}~ G_R^\parl \simeq {\l^{4/3}(\e)\over \e}  \gamma \o^2  \left[1  + {1\over 2}\e^2 \gamma^2 \o^2 + O(\e^4) \right] {\ell^2 \over 2\pi \ell_s^2}, \qquad \epsilon\to 0
\ee
The divergence is purely in the $\o^2$ term, and the coefficient agrees with
the result we found from the on-shell action  (\ref{propdiv}).
 Notice that the second term in equation (\ref{psiUV2}),
proportional to $C_v$, starts at $O(\e^3\o^3)$, so it does not contribute to the divergent part of the propagator.

To eliminate the divergence, and obtain a finite result, we must add the
contribution  from the boundary counterterm. However, different results
can arise due to different choices for  the {\em finite} contributions
included in the counterterm, i.e. different subtraction schemes.
As discussed  in appendix \ref{App GR}, in our case this ambiguity reduces to a
term of the form $\delta G_R (\o)= \delta m\, \o^2$, which can be reabsorbed
in the renormalization of the quark mass. We are going to use a {\em minimal} scheme, and fix the coefficient of the counterterm action (\ref{count4})  to be:
\be
\Delta M(\e) = - {\l^{4/3}(\e)\over \e} {\ell^2 \over 2\pi \ell_s^2}.
\ee
This choice exactly subtracts  the  divergences (\ref{propdiv}) (see Appendix \ref{App GR}), and moreover removes all (finite
or infinite)  O($\o^2$)  terms in the large-$\o$ behavior of the
propagator.

Once we subtract the divergence in the  minimal scheme, and we take the limit $\e \to 0$,  the
 right hand side of equation (\ref{ReG WKB})  vanishes. This means that, in  this scheme,  ${\rm Re}~G_R(\o) \to 0$ as $\omega \to \infty$. Moreover,
as the  only ambiguity in ${\rm Re}~G_R$ is proportional to $\o^2$ (see Appendix \ref{App GR}), we conclude that {\em in any other  scheme} the real
part of the Langevin Green's function  grows as $\o^2$ for large $\o$.

As a final remark, the previous discussion only applies if we consider
the quark mass to be infinite. In the case of a finite mass,
the trailing string is attached at a radial point $r_Q > 0$,
the cut-off is physical, and the result is not divergent. However,
one should still specify the finite  boundary term included in the
action in order to arrive at an unambiguous result.

\paragraph{Imaginary part, and the symmetric correlators.}

Unlike the real part, the imaginary part of the retarded correlator
does not suffer from  ambiguities. One of the reasons is that it is
proportional to a conserved quantity, which can be shown to be finite at the horizon.
In fact,  we can write ${\rm Im}~G_R$ in the form:
\be\label{current}
{\rm Im}~G_R(\omega) = - {1 \over 2 i} {\cal G}^{rr} \Psi_R^* \overleftrightarrow{\pa_r} \Psi_R \equiv - J^r \;.
\ee
Here $J^r$ is a conserved current--- this follows directly from the equations for the fluctuations, equations
 \refeq{H18}--- hence the imaginary part of the retarded correlator can be analytically evaluated at any $r$, not necessarily at the boundary. It is convenient to evaluate it at the horizon.
{}From the definitions
 (\ref{Gab}) and (\ref{Hab}), we find that, in the near-horizon limit :
\be
{\cal G}^{rr}_\perp \simeq Z^2(r_s) {\cal G}^{rr}_\parl \simeq (4\pi T_s) b(r_s)^2 (r_s -r), \qquad r\to r_s.
\ee
 Inserting this
expression in equation (\ref{current}), and using the expression (\ref{psihor}) for $\Psi(r)$, we find:
\be\label{IMGR}
 {\rm Im}~G_R^\perp(\omega) = -{b^2(r_s) \over 2\pi \ell_s^2} |\Psi_h^\perp(\omega)|^2\, \omega, \qquad {\rm Im}~G_R^\parl(\omega)= -{b^2(r_s) \over 2\pi \ell_s^2 Z^2(r_s)} |\Psi_h^\parl(\omega)|^2 \, \omega,
\ee
where $\Psi_h$ is the coefficient of the in-falling wave-function, see equation (\ref{psihor}).

{}From  the imaginary part of $G_R(\omega)$  we can immediately extract the  symmetrized correlator
 $G_{sym}(\omega)$, i.e. the generalized Langevin noise time-correlation function:   as discussed at the beginning of this section,
due to the thermal nature of the world-sheet fluctuations,  $G_{sym}(\omega)$ is related
 to  ${\rm Im}~G_R(\o)$  by the following equation, which generalizes the analogous equations in \cite{gubser,Casal,iancu}:
\be\label{Gsym}
G_{sym}(\omega) = -\coth \left( { \omega \over 2 T_s} \right) {\rm Im}~G_R(\omega) \;.
\ee

\paragraph{The spectral densities.}

The spectral densities associated with the Langevin dynamics are
defined by
%%%%%%%%%%%%%%%%%%%%%%%%%%%%%%%%%
\bea\lab{sd1}
&& \rho^a (\omega)= -{1\over \pi }{\rm Im}~G_R^a(\omega), \\
&& \rho_{sym}^a(\omega) = -\frac{1}{\pi} G_{sym}^a(\omega) =
-\coth \left( { \omega \over 2 T_s} \right) \rho^a(\omega) ,
\eea
%%%%%%%%%%%%%%%%%%%%%%%%%%%%%%%%%
where $ a=\perp,\parl$ , and we have used equation (\ref{Gsym}) in the second line. The imaginary part of
$G_R$ is given by the flux for the perpendicular and the parallel
components by the formula \refeq{current}.
%%%%%%%%%%%%%%%%%%%%%%%%%%%%%%%%%
We will give an analytical estimation of the large-frequency
behavior of the spectral densities in Subsection \ref{WKB}, using a WKB method; in Section \ref{numerics} we will  use
numerical methods to obtain the full functional dependence of $\rho(\omega)$ in a concrete model.

\subsection{The membrane paradigm}
\lab{mempar}

Here we introduce an alternative method to calculate both the
diffusion constants and the spectral densities that goes under the
name of {\em the membrane paradigm} \cite{LiuIqbal}. We apply this
method to obtain the spectral densities in the next section and to
obtain the diffusion constants, in section \ref{memdiff}.

In \cite{LiuIqbal}, it was established that the transport
coefficients associated with generic massless fluctuations can be
read off directly from their effective coupling in the action,
evaluated at the horizon. For an
arbitrary massless fluctuation $\phi$ with an action
\begin{equation}\label{IL1}
 S_2 = -\half \int d^{d}x dr \sqrt{-\det g}\, Q(r)\, g^{\a\b} \6_{\a}
 \phi \6_{\b} \phi,
\end{equation}
the transport coefficient {\em associated with the retarded
Green's function} is given by,
\begin{equation}\label{IL2}
 \chi_R = - \lim_{k_{\mu}\to 0} \frac{{\rm Im}\,\, G_R(\o,\vec{k})}{\o} = \lim_{k_{\mu}\to 0} Q(r_s) \sqrt{\frac{-\det
 g}{g_{rr}g_{tt}}}\bigg|_{r_s}
\end{equation}
where $Q$ is the effective coupling of the fluctuation defined in
(\ref{IL1}). We refer the reader to \cite{LiuIqbal} for a
derivation of this formula, noting that here we apply the same
idea to the {\em world-sheet black-hole} rather than {\em the bulk
black-hole} as in \cite{LiuIqbal}.

In our case, comparison of (\ref{IL1}) with (\ref{H13}) yields,
\begin{equation}\label{Qs}
 Q_{\parl} = \frac{1}{\pi\ell_s^2}\frac{b^2}{Z^2},
 \qquad Q_{\perp} = \frac{1}{\pi\ell_s^2} b^2.
\end{equation}
Once we know these effective couplings, we can immediately write
down the diffusion constants. Therefore, the method provides a very
efficient and fast way of computing the latter. This is done in
section \ref{memdiff}.

\subsubsection{A differential equation for spectral densities}

Although the method of the membrane paradigm is most effective in
the low frequency limit, where one can read off transport
coefficients directly from the $Q$'s, it is still a convenient
method for arbitrary $\o$ where one has to use the flow equations
\cite{LiuIqbal}. We define the canonical momentum associated with
$\phi$ in (\ref{IL1}) with respect to foliations in $r$, as
\begin{equation}\label{canmom}
 \Pi(r,\o,\vec{k}) = Q(r) \sqrt{-\det g} g^{rr} \6_r
 \phi(r,\o,\vec{k}),
\end{equation}
where we performed the Fourier transform on the 4D space-time. We
also define the ``$r$-dependent" response function
\begin{equation}\label{chibar}
\bar{\chi} = i \frac{\Pi(\r,\o,\vec{k})}{\o \phi(r,\o,\vec{k})}.
\end{equation}
Using the general AdS/CFT relation for the retarded Green
function,
%%%%%%%%%%%%%%%%%%%%%%%%%%%%%%
\be\lab{corfunc} G_R(\o,\vec{k}) = \lim_{r\to 0}
\frac{\Pi(\r,\o,\vec{k})}{\phi(r,\o,\vec{k})}, \ee
%%%%%%%%%%%%%%%%%%%%%%%%%%%%%%
we find that,
%%%%%%%%%%%%%%%%%%%%%%%%%%%%%%
\be\lab{chiflux} \rho(\o,\vec{k}) = -\frac{1}{\pi} {\rm
Im}~G_R(\o,\vec{k}) = \frac{\o}{\pi} \lim_{r\to 0} {\rm Re}
\bar{\chi}.\ee
%%%%%%%%%%%%%%%%%%%%%%%%%%%%%%
{}From the equations of motion, one derives a first-order equation
for $\bar{\chi}$, \cite{LiuIqbal}\footnote{In the rest of this section, we set
$\vec{k}=0$.} :
\begin{equation}\label{eqchibar}
 \6_r \bar{\chi} = i\o \sqrt{\frac{g_{rr}}{g_{tt}}} \le( \frac{1}{Q(r)} {\bar{\chi}}^2 - Q(r)
 \ri).
\end{equation}
Here, the effective coupling $Q$ is given by (\ref{Qs}) and,
\begin{equation}\label{metfunc}
\sqrt{\frac{g_{rr}}{g_{tt}}} = \frac{Z(r)}{f(r)-v^2}.
\end{equation}
For regularity at $r=r_s$ one should require that $\bar{\chi} \to
Q(r_s)$ as $r\to r_s$. One solves (\ref{eqchibar}) with this boundary condition
at the horizon, and determines $\bar{\chi}$ on the boundary
$r=r_b$. The spectral density associated with the symmetric
Green's function is given by,
\begin{equation}\label{yeter1}
 \rho_{sym}(\o) = \frac{\o}{\pi} \coth \left( { \omega \over 2 T_s}
 \right)Re \bar{\chi}(r_b,\o).
\end{equation}
As we show in section \ref{WKB},  $\rho_{sym}$ is divergent
as $\o\to\infty$. This is similar to the familiar short-distance divergences
of correlators of quantum field theory.

\subsection{Universal results for the spectral densities}
\lab{membrane}

The membrane paradigm allows us to obtain interesting relations
concerning the spectral densities in certain limits. We first note
that, by employing the equation of motion for $\phi$ that follows
from (\ref{IL1}) and (\ref{eqchibar}) one can show that
$|\phi|^2\rm{Re}~\bar{\chi}$ is independent of $r$:
%%%%%%%%%%%%%%%%%%%%%%%%%%%%%%
\be\lab{const1} |\phi(r)|^2{\rm Re}~\bar{\chi}(r) = N(\o). \ee
%%%%%%%%%%%%%%%%%%%%%%%%%%%%%%
Through equation~(\ref{chiflux}), this is of course equivalent to the
fact that the imaginary part of the Green function is proportional
to the conserved flux, hence it is constant.

We may evaluate
(\ref{const1}) in the two limits $r=r_b$ and $r=r_s$. We first
evaluate it at the horizon: the in-falling condition is,
%%%%%%%%%%%%%%%%%%%%%%%%%%%%%%%
\be\lab{infl} \phi(r,\o)\to C_h(r_s,\o)(1-r/r_s)^{-\frac{i\o}{4\pi
T_s }}, \qquad r\to r_s.\ee
%%%%%%%%%%%%%%%%%%%%%%%%%%%%%%%
Therefore, $|\phi(r_s)|^2 = |C_h|^2$. On the other hand, by the boundary
condition of (\ref{eqchibar}), $\bar{\chi}(r_s) = Q(r_s)$.
Therefore we obtain, $N(\o) = C_h(\o,r_s) Q(r_s)$.

Secondly, we evaluate (\ref{const1}) at the boundary $r_b$ using
(\ref{chiflux}) and we obtain $N(\o) = \pi \rho(\o)/\o$. Hence we
have,
%%%%%%%%%%%%%%%%%%%%%%%%%%%%%%%
\be\lab{const2} \rho(\o) = \frac{\o}{\pi} Q(r_s)
|C_h|^2(r_s,\o).\ee
%%%%%%%%%%%%%%%%%%%%%%%%%%%%%%%
Now, we consider the special limit $r_s\to r_b$ while keeping $\o$
finite, or more precisely we consider $\omega r_s \ll 1$.
If we think of $\omega$ as being fixed, this limit can be attained in two
ways:
\begin{itemize}
\item either by sending $T\to \infty$, so that the black hole horizon, and
consequently the world-sheet horizon, are pushed to the $AdS$ boundary;
\item or by keeping $T$ finite, and sending $v\to1$. In this case the black
hole temperature is fixed, but the string is ultra-relativistic.
\end{itemize}
Therefore, the regime $\omega r_s \ll 1$ corresponds to a UV limit for
the {\em background} quantities $T$ and $v$.

On the other hand, for a given fixed $\omega$, as $r_s$ approaches the boundary $C_h(\omega)$ becomes
independent of $\o$ and approaches unity. This is because of the
unit normalization of $\phi$ at the boundary, $\phi(r_b) = 1$.
Therefore, (\ref{const2}) simplifies and one obtains,
%%%%%%%%%%%%%%%%%%%%%%%%%%%%%%%
\be\lab{const3} \rho(\o) \simeq \frac{\o}{\pi} Q(r_s), \qquad \omega r_s \ll 1\ee
%%%%%%%%%%%%%%%%%%%%%%%%%%%%%%%
We can make this expression more explicit by taking the near-boundary approximate expressions, valid for $r_s \sim 0$, in (\ref{Qs}): we
approximate the metric functions as in equation (\ref{UVlimit}),
and consequently, from equation (\ref{H10}), $Z(r_s)\simeq \gamma^{-1} $.
These approximations are valid up to terms of ${\cal O}(\lambda)$, which
signal the departure from conformality of the UV geometry.
Using these approximations we find:
%%%%%%%%%%%%%%%%%%%%%%%%%%%%%%%
\be\lab{Qperplim} Q_{\perp}(r_s) \, \simeq \,\gamma^{-2}Q_\parl(r_s) \,\simeq\, \frac{\ell^2}{\ell_s^2}
\frac{\l^{\frac43}(r_s)}{\pi r_s^2} \Big[1+\cO(\l(r_s))\Big]. \ee
%%%%%%%%%%%%%%%%%%%%%%%%%%%%%%%

Finally, $r_s$ can also
be expressed purely in terms of the boundary quantities in the same
limit, using the near-boundary approximation (\ref{rslim}).

The conclusion of the previous discussion is that, {\em in the high energy limit one obtains
universal expressions for the spectral densities, where they
become linear in frequency.}

 This universal behavior is to be expected based on the consideration
that, for fixed $\omega$, both limits $v\to 1$ and $T\to \infty$ correspond
to the limit $r_s \to 0$. In this regime the equation governing the string fluctuation
become essentially the equation one finds in pure $AdS$, close to the boundary. 
This case is discussed explicitly in Appendix \ref{AppN=4}, from which one concludes (see equation (\ref{fluctu conf}))
 that the wave-functions depend on $\omega$ and $T_s$ only
 through the combination $\omega/T_s \propto \omega r_s$. Thus,
for fixed $\omega$ and small $r_s$, i.e. for $\omega r_s\ll 1$, the spectral density is approximated by the linear term in the expansion in  $\omega r_s$.

\subsection{The WKB approximation at large frequency}\lab{WKB}

The WKB approximation can be used to obtain the high-frequency
behavior of the Langevin spectral densities. Here we summarize the
method and present the results. The detailed derivation is given in
appendix \ref{AppWKB}.

By a rescaling of the wave-function, the fluctuation equations (\ref{H19})--(\ref{H20}) can be put in a Schr\"odinger-like form,
and the large $\o$ solution can be obtained by
an adaptation of the WKB method. This method has been applied to the case
of shear perturbations in AdS-Schwarzschild black-hole in
\cite{Teaney}.

The fluctuation equations \refeq{H19}--\refeq{H20}
can be put in the Schr\"odinger form
%%%%%%%%%%%%%%%%%%%%%%%%%%%%%%
\be\label{schro1} -\y''+V_s(r)\y=0,\qquad V_s(r) =
-\frac{\o^2 b^4}{R^2} + \half \big(\log{\cal R}\big)'' + \frac14
\big(\log {\cal R}\big)^{'2}. \ee
%%%%%%%%%%%%%%%%%%%%%%%%%%%%%%
defining the wave function $\psi = {\cal R}^{\half} \Psi$ and $R = \sqrt{(f-v^2)(b^4f-C^2)}$, where
%%%%%%%%%%%%%%%%%%%%%%%%%%%%%%
\be
\Psi = \left\{\begin{array}{l} \delta X^\perp \\ \delta X^\parl \end{array}\right. , \quad
{\cal R} = \left\{\begin{array}{l} R \\ R/Z^2\end{array}\right.\quad.
\ee

For large frequency (compared to $r_s^{-1}$), we can approximate the potential over essentially the entire range of $r$ by the expression: \be
\label{SchroWKB} V_s(r) \simeq -\frac{\o^2 b^4}{R^2},\qquad \o r_s \gg 1,\quad r_{tp}\ll r < r_s, \ee where $r_{tp}$ denotes the classical turning
point, $V(r_{tp}) =0$. The range of $r$ for which equation (\ref{SchroWKB}) is valid is the classically allowed region. For large $\omega r_s$, the
turning point is approximately $r_{tp} \simeq \sqrt{2}/(\gamma \omega)\ll r_s$.  Therefore, the classically allowed region covers almost all the range
$0<r<r_s$, but for a small region close to the boundary (that includes the turning point) where the approximation (\ref{SchroWKB}) breaks down.
Finally, and most importantly, for $\omega r_s \gg 1$ there is always an overlap region in which the boundary asymptotics of the Schr\"odinger
equation and equation (\ref{SchroWKB}) are both valid , since $r_{tp}\ll r_s$. This will allow the matching of the WKB solution and the boundary
solution.

In order to obtain a large-$\o$ analytic approximation for $\rho(\omega)$, it is necessary to specify whether we consider the quark mass as infinite, or
we are working with a finite but large mass $M_q$.

\paragraph{Infinitely massive quarks.}
In this case the endpoint of the string is attached to the $AdS$ boundary
$r=0$, and this is where we should set the normalization of the wave-functions.

The WKB computation proceeds along the steps detailed in appendix \ref{AppWKB}. The WKB solution is matched with the horizon and boundary asymptotics, for
the wave function, determining the coefficients $\Psi_h$ appearing in equation \refeq{IMGR}

\be\lab{psih}
\Psi_h(\omega) = \omega {i \ell \gamma \l_{tp}^{2/3} \over b(r_s)}\left\{\begin{array}{ll} 1, & \quad \perp \\ \gamma Z(r_s), & \quad \parl
\end{array}\right.\quad.
\ee

Inserting these expressions in equation (\ref{IMGR}) and using the definition (\ref{sd1}), we determine $\rho (\omega)$, in the limit $\o\gg 1/r_s$:

\be\lab{rhoperpWKB} \rho_{\perp}(\o) \simeq \gamma^{-2}\rho_\parl(\o) \simeq \frac{\o^3}{2\pi^2}
\frac{\ell^2}{\ell_s^2}~\gamma^2~\l_{tp}^{\frac43}(\o) \ee
%%%%%%%%%%%%%%%%%%%%%%%%%%%%%%%%%
Here $\l_{tp}=\l(r_{tp})$ where $r_{tp} \simeq \sqrt{2}/(\o\gamma)$ is the classical turning point, as discussed in  Appendix \ref{AppWKB}. For very
large $\o$, the dilaton profile is can be approximated as in equation (\ref{UVlimit}),
%%%%%%%%%%%%%%%%%%%%%%%%%%%%%%%%%
\be\lab{UVhooft} \l_{tp} \simeq b_0^{-1}\log^{-1}\le[\frac{\o\gamma}{\sqrt{2}\Lambda}\ri]. \ee

\paragraph{Finite mass quarks.}

The computation for the finite mass quarks follows the same steps as for the infinitely massive case, except that the boundary normalization of the
wave function $\Psi(r)=1$ has to be imposed at the cutoff $r=r_Q$ (determined by equation~\refeq{mass cutoff}), rather than at the proper boundary
$r=0$. As $M_Q$ becomes large (with respect to the UV scale $\Lambda$), $r_Q \sim 1/M_Q$.

The presence of a finite cutoff at $r_Q$ implies some subtleties in the matching of the WKB solution with the boundary solution, as explained in
Appendix \ref{AppWKB}.

To give an explicit result, we must distinguish two regimes which, for fixed $\o/M_q$, correspond to small and large velocities.

\begin{itemize}
\item {\bf Small velocities.}
If $r_Q \ll 1/\gamma \omega$ the finite mass spectral densities at
large frequencies behave like the infinite mass ones, except for $(\gamma \omega r_Q)^2$ corrections. More explicitly, we obtain
\bea\lab{rhoperpWKB small cutoff}
\rho_{\perp}(\o) &\simeq& \gamma^{-2}\rho_\parl(\o) \nonumber\\&\simeq&
\frac{\o^3}{\pi^2} \frac{\ell^2}{\ell_s^2}~\gamma^2~\l_{tp}^{\frac43}(\o)
\Bigg[ 1 + \left( \gamma \omega r_Q \right)^2
+ \left[ \left( {\ell^2 \lambda_{tp}^{4 \over 3} \over \gamma r_Q^3} \int^{r_Q} { dr' \over {\cal R}(r') } \right)^2 - 1 \right] \nonumber\\
&& \cdot \left( \sin(\gamma \omega r_Q) - \gamma \omega r_Q \cos(\gamma \omega
r_Q) \right)^2 \Bigg]^{-1}.
\eea
These expressions are valid in the regime where $\o r_s \gg 1 $, but at the same time $\gamma \omega r_Q < 1 $, i.e. for small velocities, given a fixed (large) quark mass and a given frequency. On the other hand, for fixed frequency and velocity but for $M_q\to \infty$,
$r_Q \to 0 $, the r.h.s. asymptotes  to unity, and we recover the infinite quark mass expressions (\ref{rhoperpWKB}).

\item {\bf Large velocities.} Analogously, in the limit where $r_Q \gg r_{tp} \simeq \sqrt2/\gamma \omega$, the spectral functions read
\be\lab{rhoperpWKB large cutoff}
\rho_{\perp}(\o) \simeq \gamma^{-2}\rho_\parl \simeq \frac{\o^3}{\pi^2\ell_s^2}\gamma^3 r_Q^2 {\cal R}_Q~ \bigg[ 1 + \left( \gamma \omega r_Q \right)^2 \bigg]^{-1}.
\ee
Hence, the difference with respect to the infinite mass result in this case is that $\lambda_{tp}^{4/3}$--- which is $\omega$--dependent ---is replaced by
$\lambda_Q^{4/3}$ ---which, instead, is $\omega$--independent.
\end{itemize}

We note that the large $\omega $ behavior for finite mass, both in the large and in the small cutoff regimes, changes with respect to the infinite
mass case and becomes linear rather than cubic, due to the extra $(\gamma \omega r_Q)^2$ term in equations~\refeq{rhoperpWKB small
cutoff}--\refeq{rhoperpWKB large cutoff}. This extra term comes from the fact that the solution has a subleading linear dependence on $r$, which is
negligible in the infinite mass case, but enters the expression of the spectral function, $\rho \sim \Psi^* \Psi'$, in the finite mass case, giving it
a quadratic dependence on $r_Q$.

\section{Langevin diffusion constants}\label{transport}

The correlators and spectral densities discussed in the previous section are the building
blocks of the generalized Langevin equation, (\ref{langeq}). Now, we will focus on the
long-time limit, discussed in Section 2, in which equation (\ref{langeq}) reduces to the local
 form (\ref{langeq2}), in which only the $\omega$-independent friction and diffusion coefficients, $\eta$ and $\kappa$, appear.
  They are given in equation (\ref{coeff}) in terms of the zero-frequency
limit of the Langevin Green's functions.

Therefore, in this Section we consider the zero-frequency limit of the Green's functions
constructed in Section \ref{correlators}. This will allow us to give the analytic results for the
diffusion constants, both from the direct evaluation of the correlators, and using the membrane paradigm.

\subsection{Diffusion constants via the retarded correlator}

The diffusion constant is defined in terms of the symmetric correlator $G_{sym}$ (see equation (\ref{coeff}):
\be\label{kappa}
\kappa = \lim_{\omega\to0} G_{sym} = - 2 T_s \lim_{\o \to 0} {{\rm Im} G_R(\o)\over \o},
\ee
where in the second equality we have used the $\o\to0$ limit of equation \refeq{Gsym}.

The small frequency limit of the symmetric correlator can be evaluated analytically since we can determine the boundary--to--bulk wave function in
this limit and discard higher orders in the evaluation of \refeq{kappa}. More precisely, we write the small frequency limit of the horizon
asymptotics of the $\Psi_R$'s. Given the in--falling boundary condition (\ref{psihor}) we obtain
\be
\Psi_R (r,\omega)= \Psi_h (r_s-r)^{-{i \omega\over 4 \pi T_{s}}} \simeq \Psi_h \left[ 1- {i \omega \over 4 \pi T_s } \log |r-r_s| + \dots\right]
\label{H26}\ee

This solution can be connected to the boundary asymptotics by the exact solution of the fluctuation equations \refeq{H19}--\refeq{H20} at $\omega=0$
which reads $\Psi_R(r,0) \equiv 1$ once we impose the appropriate boundary condition $\Psi_R(r_b) = 1$ (see Appendix \ref{details kappa} for details). On the other
hand, equation~\refeq{H26} reduces to $\Psi_R = \Psi_h$ in the strict $\omega = 0$ limit. Therefore, the near-horizon solution at small frequencies is entirely
determined by equation~\refeq{H26} and the match to the boundary solution which yields $\Psi_h = 1$ (both for transverse and longitudinal modes).

Furthermore, expanding for $\omega \ll 1$, we may also show (see Appendix \ref{details kappa}) that the solution for all values of $r$
and small frequencies is given by
\be\label{wave small omega}
\Psi_R = \left(r_s-r\right)^{-{i\omega\over 4 \pi T_{s}}} \left[ 1 + {\cal O}\left(\omega\right) \right] \;.
\ee

Now we may substitute the solution \refeq{wave small omega} in the expression for $\kappa$, using \refeq{kappa} and evaluating the current at
the horizon.

\paragraph{Infinitely Massive Quarks.}

For infinite mass quarks, the boundary is located at $r=0$ and the appropriate boundary--to--bulk wave function $\Psi_R$ is given by equation\refeq{wave small
omega} at small frequencies.

To compute the diffusion constants \refeq{kappa}, we evaluate $J^r$,
defined in equation (\ref{current}), at the radius value $r = r_s - \epsilon$ and then let $\epsilon \to 0$ (since $J^r$ is conserved it
can be evaluated at any radius and not necessarily at the boundary, where subleading ${\cal O}(\omega)$ terms in \refeq{wave small omega} would
contribute). This allows to neglect the sub-leading terms in the solution \refeq{wave small omega}. For the longitudinal component we also need
the near-horizon limit of $Z(r)$, which can be easily obtained from equation (\ref{H10}):
\begin{equation}\label{Zhor}
 Z^2\to \frac{1}{16\pi^2}\frac{f'(r_s)^2}{T_s^2}, \qquad r\to r_s.
\end{equation}

 Using the explicit expressions (\ref{IMGR}) in equation (\ref{kappa}), with $\Psi_h=1$, we obtain to the following results:

\bea\label{kappa perp}
\kappa_\perp &=& {1 \over \pi \ell_s^2} b^2(r_s) T_s \\
\kappa_\parallel &=& {16 \pi \over \ell_s^2} {b^2(r_s) \over {f^{'2}}(r_s)} T_s^3 \;. \label{kappa par}
\eea

We note that $\kappa_\perp$ and $\kappa_\parallel$ are simply related by $Z^2 \kappa_\parallel = \kappa_\perp$, as it can be read off from
equation\refeq{kappa}, using equation\refeq{current} and \refeq{Gab} and the fact that the small frequency behavior of the wave function is the same for both
transverse and longitudinal directions.

In the conformal limit $b(r)=\ell/r$, $f(r)=1-(\pi T r)^4$ and $T_s=T/\sqrt\gamma$, $r_s=1/(\pi T \sqrt\gamma)$, we recover the results of \cite{gubser,Casal}
for the holographic ${\cal N}=4$ SYM:
\bea
\kappa_{\perp{\cal N}=4} &=& {\pi} \sqrt{\lambda_{{\cal N}=4}} \gamma^{1/2} T^3 \label{kconf1}\\
\kappa_{\parallel{\cal N}=4} &=& {\pi} \sqrt{\lambda_{{\cal N}=4}} \gamma^{5/2} T^3 \;. \label{kconf2}
\eea
where $(\ell/\ell_s)^{4} = \lambda_{{\cal N}=4}$ is identified with the ${\cal N}=4$ 't Hooft coupling, in the $AdS_5$ background.

\paragraph{Finite Mass Quarks}

For massive quarks, the appropriate boundary condition should be $\Psi_{R}(r_Q)=1$, where $r_Q$ is the UV cutoff determined by the
value of the quark mass $M_q$, using equation \refeq{mass cutoff}. In this case, equation \refeq{wave small omega} gets modified and reads
\be\label{wave mass small omega}
\Psi_R = \left({r_s - r \over r_s - r_Q}\right)^{-{i\omega\over 4 \pi T_{s}}} \left[ 1 + {\cal O}\left(\omega\right) \right] \;.
\ee

Nevertheless, as in the conformal case \cite{iancu}, the results for
$\kappa_\perp$ and $\kappa_\parallel$ remain unchanged, since they
are independent of $r_Q$. Indeed $r_Q$ enters in the wave function,
as equation~\refeq{wave mass small omega} shows, and cancels out in the $\o \to 0$ limit as we take the product $\Psi_{R}^*\Psi_{R}$ in \refeq{current}.

\subsection{The jet-quenching parameter}

As discussed in Section \ref{LangevinSec}, the jet-quenching
parameters can be defined in terms of the diffusion constants as:
\be\label{qhat22} \hat{q}^\perp = 2 {\kappa^\perp\over v}, \qquad
\hat{q}^\parl = {\kappa^\parl\over v}. \ee
The first parameter
defines the transverse momentum broadening of heavy quark probes.

There is also a different definition of the jet-quenching
parameter, which is related to the perturbative relation between
this quantity and an appropriate limit of a Wilson loop joining
two light-like lines in Yang-Mills theory (see e.g.
\cite{salgado}).
This was the basis of a different holographic calculation of
$\hat{q}$, that was carried out in \cite{lrw} in the conformal
case, and in \cite{transport} for the general backgrounds
(\ref{metric}), which gives: \be \hat{q}_{WL} = {\sqrt{2}\over \pi
\ell_s^2}\left(\int_0^{r_s} {dr \over b^2
\sqrt{f(1-f)}}\right)^{-1}, \ee where the subscript $WL$ is
introduced to distinguish this definition of the jet-quenching
parameter from the original definition (\ref{qhat22}). As in the
conformal case, this result differs from the result obtained via
the Langevin equation. The reasons for this were analyzed in \cite{lrwr}.

\subsection{The diffusion constants via the membrane paradigm}
\lab{memdiff}

The method of the membrane paradigm that we introduced in section
\ref{mempar} allows us to read off the diffusion constants {\em
directly from the action (\ref{H13}) with no need to derive and
solve for the fluctuation equations as in the previous section}.
The diffusion constants are defined in terms of the {\em symmetric
Green's functions} by (\ref{kappa}). Employing the relation
(\ref{thermal}) between the retarded and the symmetric Green's
functions and the basic formula of the membrane paradigm, equation
(\ref{IL2}), we arrive at,
\begin{equation}\label{kappas}
 \kappa^a = 2T_s \chi_{R}^a = T_s Q^a(r_s),
\end{equation}
where $a = \{ \perp,\parl\}$. In the second equation above we used the fact that the metric dependence in (\ref{IL2}) drops out in 2d.

{}From the expressions (\ref{Qs}), and using the near-horizon limit of the function $Z$ from (\ref{Zhor}), we find, unsurprisingly, the same result as equation
 (\ref{kappa perp},\ref{kappa par}).

We note that these results establish one of the very few examples
of trivial flow as defined in \cite{LiuIqbal}, in the sense that the effective
couplings $Q$, determined on the horizon membrane, stay unchanged
through the flow from the horizon to the boundary. Therefore the
dual field theory quantities i.e. the diffusion constants, which
should be evaluated on the boundary, can also be computed directly
on the horizon due to trivial flow. The other basic example of
trivial flow is the shear viscosity $\eta/s$. The reason for
trivial flow is that there is no mass term in the fluctuation equations
(\ref{H18}), because the geometry is flat on the domain-wall
directions. For the same
 reason one expects trivial flow for any transport coefficient that stem from string fluctuations on a generic
 domain wall background, as long as the fluctuations do not involve the radial direction.

\subsection{A universal inequality: $\kappa_\parl \ge \kappa_\perp$}
\lab{unineq}

{}From the expressions (\ref{kappa perp},\ref{kappa par}), one derives the
ratio,
%%%%%%%%%%%%%%%%%%%%%%%%%%%%%%%%%%%%%%%%%
\be\lab{kaprat} \frac{\kappa_\parl}{\kappa_\perp} = \le(\frac{4\pi
T_s }{f'(r_s)}\ri)^2 = 1 + 4v^2~\frac{b'(r_s)}{f'(r_s)b(r_s)},
\ee
where in the last equation, we used the definition of the
world-sheet temperature $T_s$ in (\ref{Ts}).

We note that the second term on the RHS of (\ref{kaprat}) is
always positive definite in the deconfined phase $T>T_c$. This
can be seen as follows: First of all, $f'(r)b(r)$ is a negative
definite quantity at any $r$. This follows from the general
relation, see e.g. \cite{gkmn2}.
%%%%%%%%%%%%%%%%%%%%%%%%%%%%%%%%%%%%%%%%%
\be\lab{fprime} f'(r) = - \frac{s T}{M_p^3N_c^2} b_E(r)^{-3}, \ee
%%%%%%%%%%%%%%%%%%%%%%%%%%%%%%%%%%%%%%%%%
where $s$ is the entropy density and $b_E$ is the Einstein frame
scale factor. The left hand side is manifestly negative definite.

Secondly, the quantity $b'(r_s)$ is also negative-definite in the
deconfined phase. This follows from the fact that, in the type of
geometries that confines color, the string frame scale factor
$b(r)$ at zero temperature always possesses a minimum at some point $r=r_*$. Hence
$b'(r)<0$ for $r<r_*$ and $b'(r)>0$ for $r>r_*$. Moreover, we can argue that, for
$T>T_c$, the location of the bulk horizon $r_h$ should be closer to
the boundary than $r_*$, i.e. $r_h<r_*$, otherwise the Wilson loop
would have linear behavior, as a result of saturation of the
corresponding string at $r_*$. Since, the world-sheet horizon is
always smaller than the bulk-horizon, it follows that, in the
de-confined phase:
%%%%%%%%%%%%%%%%%%%%%%%%%%%%%%%%%%%%%%%%%
\be\lab{dephase} r_s<r_h<r_*, \qquad for\,\,\, T>T_c.\ee
%%%%%%%%%%%%%%%%%%%%%%%%%%%%%%%%%%%%%%%%%
Therefore $b'(r_s)$ should be negative-definite, and the entire second
term on the RHS of (\ref{kaprat}) should be positive-definite.
{\em Therefore, we arrive at the universal result that }
\be
\kappa_\parl
\ge \kappa_\perp~~~ {\rm for}~~ T>T_c
\ee
Equality is attained for $v\to
0$. We check by a numerical computation in section \ref{numerics} that this inequality is obeyed in the  particular background used in that section.

\subsection{A generalized Einstein relation}

We may derive a generalized Einstein relation by relating the
diffusion constant (\ref{kappa perp},\ref{kappa par}) with the
friction coefficient $\eta_D$ in (\ref{langeq3}).

On the one hand, we have found holographically the relation
 between the diffusion coefficients $\kappa^{ij}$ and
the friction coefficients $\eta^{ij}$ for the Langevin equations in position space (\ref{langeq4-a}-\ref{langeq4-b}). From equation (\ref{Gsym}) and the definitions (\ref{coeff}) we arrive at:
\be
\kappa^{ij} = 2 T_s \eta^{ij}
\ee
On the other hand, we can relate $\eta^{ij}$ to the {\em momentum} diffusion coefficients $\eta_{D}^{ij}$, by equations (\ref{fric-p-x}). Therefore we find:
\bea\label{kappa-1}
&&\kappa^\perp = 2 T_s \gamma M_q \eta_D^\perp, \\
&& \kappa^\parl = 2 T_s \gamma^3 M_q \left(\eta_D^\parl + p {\de \eta_D^\parl \over \de
p}\Big|_{p=\gamma M_q v}\right). \label{kappa-2}
\eea
These relations lead to an important consistency condition: notice that, by equation (\ref{zeroth}), the coefficient $\eta_D^{\parl}$ must coincide with the zeroth order drag
coefficient (\ref{frictioncoeff}), calculated via the classical trailing string solution,
\be\label{fric6}
\eta_D^\parl = \eta_D = {1\over \gamma M_q}{b(r_s)^2\over 2\pi \ell_s^2 },
\ee
Therefore, consistency requires that, inserting the expression (\ref{fric6}) for $\eta_D^\parl$ in equation (\ref{kappa-2}), the resulting expression for $\kappa^\parl$ agree with equation (\ref{kappa par}). This is indeed the case: using the explicit expression (\ref{fric6}),
and the definition $f(r_s) = v^2$,
we find:
\be
\left( \eta_D + p {\de \eta_D \over \de p}\Big|_{p=\gamma M_q v}\right) ={1\over \gamma^2 M_q}{b(r_s)^2 \over2\pi \ell_s^2}\left(1 + 4 v^2 {b'(r_s) \over b(r_s) f'(r_s)}\right)
\ee
Inserting this expression in the right hand side of equation
(\ref{kappa-2}), and using
the identity (\ref{kaprat}), the resulting expression exactly agrees with
 $\kappa^\parl$ obtained from the Langevin correlator, (\ref{kappa par}).

Finally, from equation~(\ref{kappa-1}) and the explicit expression (\ref{kappa perp}),
we find that $\eta_D^\perp$ {\em also} equals the drag force coefficient (\ref{fric6}). This implies that, at the end of the day, the friction term in the momentum diffusion equation is isotropic,
\be
\eta_D^{ij} = {1\over \gamma M_q} {b^2(r_s) \over 2\pi \ell_s^2} \delta^{ij} \equiv {1\over \tau} \delta^{ij}.
\ee
The last equality defines the momentum diffusion time $\tau = 1/\eta_D$.

We arrive at the {\em generalized Einstein relation:}
% %%%%%%%%%%%%%%%%%%%%%%%%%%%%%%%%%%%%%%%%%
 \be\lab{GER} \tau \kappa_\perp = 2M_q \gamma T_s.\ee
% %%%%%%%%%%%%%%%%%%%%%%%%%%%%%%%%%%%%%%%%%
 where $M_q$ is the quark mass and $T_s$ is the emergent ``world-sheet
 temperature''. (\ref{GER}) can be viewed as a generalization of the usual
non-relativistic Einstein relation which has the form:
%%%%%%%%%%%%%%%%%%%%%%%%%%%%%%%%%%%%%%%%%
\be\lab{usGER} \tau \kappa = 2M_q T.\ee
%%%%%%%%%%%%%%%%%%%%%%%%%%%%%%%%%%%%%%%%%
The modified Einstein relations (\ref{kappa-1}) and (\ref{kappa-2}) had already been found in \cite{hoyos} in
the particular  case of the holographic dual of the ${\cal N}=2^*$ theory, and the results of that
analysis fit consistently in the general framework we are considering here.

We note that in the conformal limit, the world-sheet temperature
is related to the bulk temperature as
%%%%%%%%%%%%%%%%%%%%%%%%%%%%%%%%%%%%%%%%%
\be\lab{confWST} T_s = \frac{T}{\sqrt{\gamma}}, \qquad
\rm{conformal},\ee
%%%%%%%%%%%%%%%%%%%%%%%%%%%%%%%%%%%%%%%%%
and (\ref{GER}) becomes,
%%%%%%%%%%%%%%%%%%%%%%%%%%%%%%%%%%%%%%%%%
\be\lab{GERlim} \tau \kappa_\perp = 2M_q T \sqrt{\gamma}.\ee
%%%%%%%%%%%%%%%%%%%%%%%%%%%%%%%%%%%%%%%%%
This is quite different from (\ref{usGER}) and reduces to it only
in the non-relativistic limit $\gamma \to 1$.

The generalized relation in (\ref{GER}) is defined in terms of a set of physical boundary quantities, and the geometric
quantity $T_s$. In a sense, $T_s$ is the temperature that is $read$ by the quark as it moves through the medium.
$T_s$ provides the answer to the following interesting question: what is the temperature read by a thermometer moving with speed $v$ inside a strongly
coupled QGP of temperature $T$.

Note that in the conformal case the answer is universal, and transforms simply with boosts. In the non-conformal case, associated with the Einstein-dilaton system the answer is dynamics dependent.

\subsection{Special limits of the diffusion constants}

In this section we study the diffusion constants (\ref{kappa
perp}) and (\ref{kappa par}) in the extreme relativistic and
non-relativistic limits and express these quantities in terms of
thermodynamic functions.

\subsubsection{Non-relativistic limit}\lab{nrelsec}

As $f(r_s)=v^2$, in the non-relativistic limit $v\to 0$, the
world-sheet horizon approaches the bulk horizon: $r_s\to r_h$.
Using the near-horizon expressions for the metric functions, one
also finds from (\ref{Ts}) that $T_s\to T$ in this limit. Finally,
we use the expression that relates the entropy density with the
scale factor at the horizon, \cite{gkmn2},
 \be\lab{ent} s =
\frac{b^3_E(r_h)}{4G_5} = 4\pi M_p^3N_c^2 b^3_E(r_h), \ee
 to obtain
\be\lab{kappev0} \kappa_{\perp} \to
\frac{2}{\pi}\le(\frac{45\pi}{4}\ri)^{\frac23}\frac{\ell^2}{\ell_s^2}
\le(\frac{s}{N_c^2}\ri)^{\frac23}\l_h^{\frac43}T, \quad v\to 0,
\ee where $\l_h$ is the horizon value of $\l$. On the other hand,
the ${\cal N}=4$ result becomes $\kappa_{\perp{\cal N}=4}\to\pi
\sqrt{\lambda_{{\cal N}=4}} T^3$. Hence the ratio becomes,
\be\lab{kappev0norm} \frac{\kappa_{\perp}}{\kappa_{\perp{\cal
N}=4}} \to
\frac{2}{\pi^2}\le(\frac{45\pi}{4}\ri)^{\frac23}\frac{\ell^2}{\ell_s^2}
\frac{1}{\sqrt{\lambda_{{\cal N}=4}}}
\le(\frac{s}{N_c^2T^3}\ri)^{\frac23} \l_h^{\frac43}, \qquad v\to
0. \ee A similar analysis for the parallel component yields,
\be\lab{kappav0} \kappa_{\parl} \to
\frac{2}{\pi}\le(\frac{45\pi}{4}\ri)^{\frac23}\frac{\ell^2}{\ell_s^2}
\le(\frac{s}{N_c^2}\ri)^{\frac23}T \l_h^{\frac43}, \qquad v\to 0.
\ee We note that this is exactly the same as (\ref{kappev0}). This
is what one expects from the physical perspective. In the
non-relativistic limit, the main source of momentum broadening is
due to thermal fluctuations in the plasma, that itself is
isotropic. Similarly the ratio of QCD and ${\cal N}=4$ results
also become the same as in (\ref{kappev0norm}).

\subsubsection{Ultra-relativistic limit}\lab{ulrelsec}

We consider the opposite limit $v\to 1$. Here the expression
$f(r_s)=v^2$ tells us that $r_s\to 0$ hence the world-sheet
horizon approaches the boundary. Using the near-boundary
expression for $f(r)$ in equation (\ref{UVlimit}),
and the near-horizon expressions for
the metric functions, we find that, \be\lab{Tsv1}
 4\pi T_s\to r_s (4{\cal C}/\ell^3)^{1/2} \le(1+ \cO(\log^{-1}(r_s))\ri).
\ee
where the constant ${\cal C}$ is given in equation (\ref{CsT}). Upon substitution in (\ref{kappa perp}), we finally obtain,
\be\lab{kappev1} \kappa_{\perp} \to
\frac{(45\pi^2)^{\frac34}}{\sqrt{2}\pi^2}\frac{\ell^2}{\ell_s^2}\frac{(sT/N_c^2)^{\frac34}}{(1-v^2)^{\frac14}}
\le(-\frac{b_0}{4}\log(1-v^2) \ri)^{-\frac43}, \qquad v\to 1.
\ee
We observe that the result diverges in the extreme relativistic limit $v=1$.
However, one obtains a finite expression by considering the ratio with the
${\cal N}=4$ result: \be\lab{kappev1norm}
\frac{\kappa_{\perp}}{\kappa_{\perp{\cal N}=4}} \to
\frac{(45\pi^2)^{\frac34}}{\sqrt{2}\pi^3}\frac{\ell^2}{\ell_s^2}\le(\frac{s}{N_c^2T^3}\ri)^{\frac34}\frac{\le(-\frac{b_0}{4}\log(1-v^2)
\ri)^{-\frac43}}{\sqrt{\lambda_{{\cal N}=4}}},\qquad v\to 1. \ee

Similarly, for the parallel component one finds, \be\lab{kappav1}
\kappa_{\parl} \to
\frac{(45\pi^2)^{\frac34}}{\sqrt{2}\pi^2}\frac{\ell^2}{\ell_s^2}\frac{(sT/N_c^2)^{\frac34}}{(1-v^2)^{\frac54}}\le(-\frac{b_0}{4}\log(1-v^2)
\ri)^{-\frac43}, \qquad v\to 1. \ee Again this is divergent as
$v\to 1$, but the ratio with ${\cal N}$=4 result again remains finite in
this limit: \be\lab{kappav1norm}
\frac{\kappa_{\parl}}{\kappa_{\parl{\cal N}=4}} \to
\frac{(45\pi^2)^{\frac34}}{\sqrt{2}\pi^3}\frac{\ell^2}{\ell_s^2}\le(\frac{s}{N_c^2T^3}\ri)^{\frac34}\frac{\le(-\frac{b_0}{4}\log(1-v^2)
\ri)^{-\frac43}}{\sqrt{\lambda_{{\cal N}=4}}},\qquad v\to 1. \ee

We observe that the parallel and
perpendicular components of the diffusion constants asymptote essentially to the conformal result both in the $v\to 0$ and in the $v\to 1$ limits.
 modulo logs in the second case and the appropriate adjustment of relevant parameters like temperature, entropy etc.

We should warn the reader that, as $v\to 1$, we expect the break down of the treatment based on the local Langevin equation: in fact, for very large $v$ the world-sheet temperature drops to zero,
and the auto-correlation time of the Langevin Green's functions, $\tau_c \sim T_s^{-1}$, diverges. On the other hand, the
relaxation time $\tau_D = \eta_D{-1}$ stays approximately constant. Therefore, eventually the
relation $\tau_D \gg \tau_c$, necessary for the local treatment, will break down at large enough $v$.
In the next subsection we give a detailed discussion of this validity condition.

Another important caveat, when considering the extreme relativistic limit of our results, lies in the fact that they are sensitive to the UV region of the geometry, and as discussed at length in previous work (see e.g. \cite{k}), the details of the gravity theory we are using are not fully reliable in this limit.

 \subsection{Time scales and validity of the local approximation} \label{Sec valid}

The results of this section so far were obtained based on two separate approximations concerning the
time scales involved. On the one hand, we assume we are in a
short-time approximation, compared to the typical relaxation time.
 This means
that, the quark velocity $v$ can be assumed to be constant only
within time scales that are much shorter than the relaxation time
$\tau_D = 1/\eta_D$. On the other hand, the analysis based on the
local Langevin equation relies on a {\em long-time} approximation,
this time compared to the typical time scales entering the
Langevin correlators, and determined by the inverse temperature
that the quark ``feels" as it travels through the plasma.
According to our previous discussions this is given by
$\tau_c=1/T_s$ where $T_s$ is the world-sheet temperature
(\ref{Ts}). Therefore our analysis in terms of diffusion constants
will be valid only for time scales $t$ such that $\tau_c \ll t \ll
\tau_D$. Existence of time intervals satisfying this condition
requires that: \be\lab{validity} \frac{1}{\eta_D} \gg
\frac{1}{T_s}. \ee Since both $\eta_D$ and $T_s$ depend
non-trivially on the quark momentum, this condition translates in
an upper bound on the quark momentum (or velocity), above which
the local treatment breaks down\footnote{Note that this condition
is different than the condition given in \cite{tea} for the
classical non-relativistic Langevin dynamics that is $1/\eta\gg
1/T$. As the thermal behavior of the Green's functions are set by
$T_s$ rather than $T$, it should be this effective temperature
that enters in the validity condition (\ref{validity}).}.

 The relaxation time $1/\eta_D$ is given by (\ref{frictioncoeff}). We can write equation (\ref{validity}) more explicitly as
\be\label{validity-11}
M_q \gamma \gg {\ell^2 \over 2\pi \ell_s^2} {b^2(r_s)\over T_s}
\ee

We can read this condition as
a lower bound on the quark mass. It is most
 restrictive in the UV (large $v$), where it reads,
\be\lab{validity2} M_q\gg \sqrt{\gamma} \left(\ell\over \ell_s\right)^2
\frac{\l(r_s)^{\frac43}}{2} \le(\frac{\cal
C}{\ell^3}\ri)^{\frac14}, \ee
 where ${\cal C}$ is defined in
(\ref{CsT}) and in the UV region $\l(r_s)$ is approximately given
by, \be\lab{lsUV} \l(r_s)\equiv \lambda_s \approx -\frac{1}{b_0
\log\le[\Lambda \ell^3/ ({\cal C} \sqrt{\gamma})\ri]}. \ee We
observe that $\l(r_s)$ vanishes in the extreme relativistic limit
because of the dependence on $\gamma$\footnote{In fact, the
numerical studies in the next section shows that the other factor
in the log dominates over the $\gamma$ dependence and $\l$ can be
treated as a constant except in the extreme relativistic limit
$v=1$.}. However, this logarithmic dependence is milder when
compared to the explicit dependence on $\gamma$ in
(\ref{validity2}).

Alternatively, equation (\ref{validity}) can be read as a condition on the quark velocity,
or momentum. For fixed quark mass, and for $v\to 1$, $r_s$ approaches the $AdS$
boundary, and the right hand side of equation (\ref{validity-11}) scales
approximately as $~\gamma^{3/2}$.
Explicitly we find
\be\lab{validity3}
\sqrt{\gamma} \ll 2M_q \left(\ell_s\over \ell\right)^2\l_s^{-\frac43} \le(\frac{\cal C}{\ell^3}\ri)^{-\frac14}.
\ee
Consequently the condition (\ref{validity}) puts an upper bound to the quark velocity.

To obtain an estimate of the upper bound
on momentum, in the right hand side of equation (\ref{validity-11}) we can approximate the scale factor as $b(r) \simeq \l^{2/3}(r) (\ell/r)^2$, and replace the quantities $r_s$ and $T_s$ by by the corresponding conformal expressions, equations (\ref{conformal}). We arrive at the bound (for an
ultra-relativistic quark):
\be\label{bound-p}
p \ll {1\over 4} \left(\ell_s\over \ell\right)^4 {M^3_q \over T^2} \l_s^{-8/3}.
\ee
For $v$ close to unity, the dependence on $v$ in $\l(r_s)$ is very mild and the
right hand side can be considered as a constant, depending only by the quark mass,
temperature and value of the (holographic) coupling. We will give a numerical estimate
of these quantities in the next section, in Improved Holographic QCD.

\section{Improved Holographic QCD and comparison with data}\lab{numerics}

In the previous sections we obtained general results for the correlators of world-sheet fluctuations, and for the Langevin diffusion constants, valid
in any 5D Einstein-Dilaton theory admitting asymptotically $AdS$ black hole solutions.
 Here, we will study in detail these results in Improved Holographic QCD, with the potential
 proposed in \cite{gkmn3}, whose thermodynamic properties are in good agreement with lattice YM thermodynamics
\cite{panero} as well as the $T=0$ spectra of glueballs obtained
on the lattice.

We take the potential to be, as in \cite{gkmn3,transport}:
\be\lab{dilpot} V(\l) = {12\over \ell^2} \left\{ 1 + V_0 \l + V_1
\l^{4/3} \left[\log \left(1 + V_2 \l^{4/3} + V_3 \l^2\right)
\right]^{1/2} \right\}. \ee

The coefficients $V_i$ entering the potential, are fixed as follows
(for a detailed discussion see \cite{gkmn3,transport}):
\be
V_0 = {8\over 9} \beta_0, \quad V_1 = 14, \quad V_2 = \beta_0^4 \left({23 + 36\,
\beta_1/\beta_0^2 \over 81 V_1 }\right)^2, \quad V_3 = 170.
\ee
where $\beta_{0}$ and $\beta_1$ are the first two pure Yang-Mills the beta-function coefficients,
\be
\beta_0 = {22\over 3 (4 \pi)^2}, \qquad \beta_1 = {51\over 121} \beta_0^2.
\ee
The coefficients $V_0$ and $V_2$ are fixed to match the perturbative YM $\beta$-function, whereas $V_1$ and $V_3$ are fixed phenomenologically
by comparing the equation of state of the model with that of YM on the lattice.

The coefficient $\ell$ is the scale of the asymptotic $AdS_5$
space-time at $r=0$ and it sets the energy scale in the field
theory. All observables defined holographically using the metric
in the Einstein frame are measured in units of $\ell$. For a given
class of black hole solutions with fixed UV asymptotics, the value
of $\ell$ can be set by matching the mass of the lowest glueball
excitation \cite{gkmn3}.

It may seem that there is an extra scale associated to these models, with respect to 4D Yang-Mills,
where there is a {\em single scale}, i.e. the quantity $\Lambda_{QCD}$ setting the scale of conformal symmetry breaking in the UV. In our model, the analog
of the QCD scale emerges as an integration constant that labels different solutions (distinguished
by different UV boundary conditions) of the {\em same} theory, with $\ell$ fixed. Explicitly, it controls the UV asymptotics of the field $\lambda(r)$,
given in equation (\ref{UVlimit}), as $\lambda(r) = -(b_0 \log r \Lambda)^{-1} + O(\log^{-2}r \Lambda) $. Therefore, it may appear we have two independent scales,
$\ell$ and $\Lambda$.
However,
as shown in \cite{ihqcd2}, physical observables depend on $\Lambda$ only via an overall scaling.
Therefore, we can choose an arbitrary value for the dimensionless parameter $\ell \Lambda$, and
subsequently fix the value of $\ell$ to  match some reference energy
scale, as explained in the previous paragraph.

The quantities that are computed by probe strings depend on
another scale, independent of $\ell$, namely the fundamental
string scale $\ell_s$. In string-derived models, the ratio
$\ell/\ell_s$ is known. In phenomenological models on the other
hand it must be adjusted to fit observation. For example,
 the ratio $\ell/\ell_s$ can be fixed by comparing the confining string tension of the holographic model (controlled by the Nambu-Goto action (\ref{NGACTION}) , hence by $\ell_s$) to the lattice value $\sigma_c = (440 MeV)^2$.
One finds: \be\label{ells} {\ell^2\over 2\pi \ell_s^2} = 6.5. \ee This
fixes the ubiquitous overall coefficient entering in the diffusion
constants, in equations  (\ref{kappa perp}-\ref{kappa par}).

With the potential given by (\ref{dilpot}), we numerically solve Einstein's equation for the metric and dilaton functions $b(r),f(r),\lambda(r)$, to
obtain black hole solutions of different temperatures $T$, but obeying fixed UV boundary conditions (for a detailed discussion of the solution
procedure, see Appendix A of \cite{gkmn3}). Once the solutions are given, for each temperature we determine the position of the world-sheet horizon
$r_s$ as a function of velocity $v$, by numerically solving the equation $f(r_s) = v^2$.

{}From the metric coefficients evaluated at $r_s$ we can obtain the
world-sheet temperature $T_s$, through equation (\ref{Ts}). The ratio $T_s/T$
is plotted as a function of $v$, and for various bulk temperatures,
in Figure \ref{fig Z} (a). We observe that $T_s < T$ for all velocities,
and as the bulk temperature increases this ratio approaches the $AdS$-Schwarzschild curve $(T_s/T)_{AdS} = (1-v^2)^{1/2}$.

\FIGURE[h!]{
\includegraphics[width=6cm]{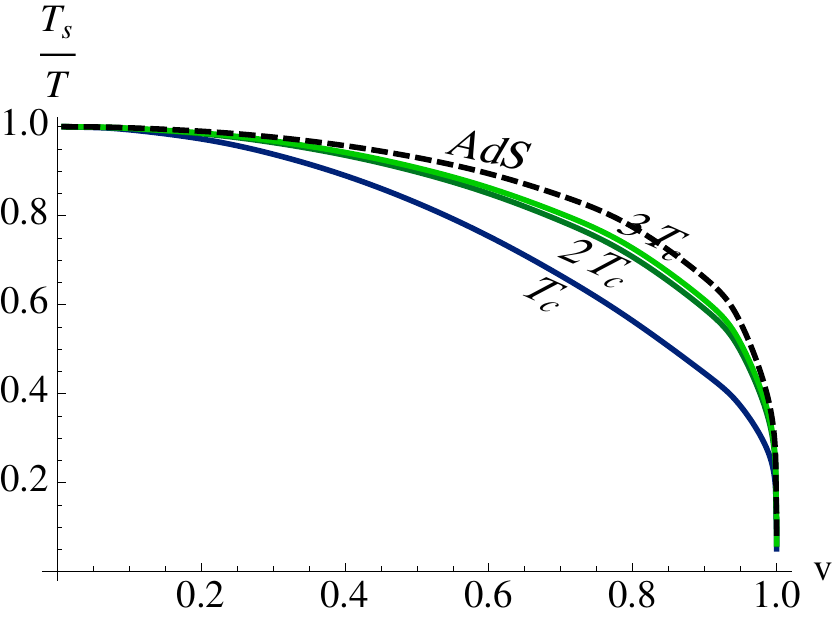}\hspace{1cm}
\includegraphics[width=7cm]{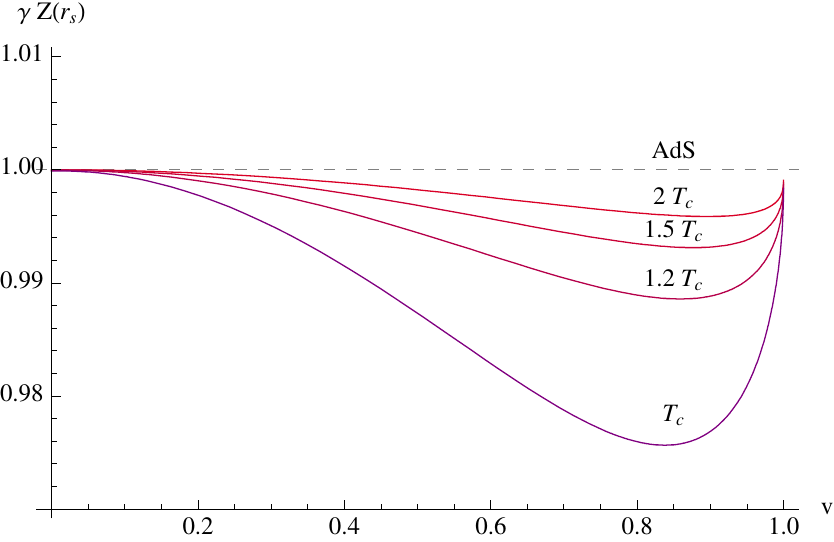}\\
(a) \hspace{7cm} (b)
\caption{(a) The ratio of the world-sheet temperature to the bulk black hole
temperature, as a function of velocity, for different values of the bulk temperature. The dashed line indicates the $AdS$-Schwarzschild curve, $T_s = T/\sqrt{\gamma}$.
(b) The function $ \gamma Z(r_s)$ as a function of velocity (with $Z$ defined as in equation~\refeq{H10} and $\gamma\equiv 1/\sqrt{1-v^2}$), computed numerically varying the velocity, at different temperatures. The dashed line represents the conformal limit, in which $\gamma Z = 1$ exactly.}\label{fig Z}}
%\end{center}
%\end{figure}

Another interesting quantity that provides an indication of how much the backgrounds deviate from the conformal case, is the function $Z(r)$, defined
in equation (\ref{H10}). For $AdS$-Schwarzschild,  this function is exactly constant, $Z(r) = 1/\gamma$. In Figure \ref{fig Z} (b) we portray the behavior
of $\gamma Z(r_s)$ as a function of velocity, for different bulk temperatures.
We observe that as $v\to 0, 1$ this quantity asymptotes to unity, as can also be seen
analytically from equation (\ref{H10}) by taking the limits $r_s \to 0$, $r_s\to r_h$. Again, as the bulk temperature increases, we move closer to the $AdS$-Schwarzschild behavior, represented by the dashed line in the graph.

At this point, we note that Figure \ref{fig Z} also provides a confirmation of the universal inequality derived in section \ref{unineq}. The function $Z(r)$ of equation (\ref{H10}) at the world-sheet horizon is given by $Z(r_s) = f'(r_s)/4\pi T_s$. Then, from equation (\ref{kaprat}) we see that $\kappa_\parl/\kappa_\perp = Z(r_s)^{-2}$. On the other hand, the numerical computation shown in figure \ref{fig Z} implies that,
$$Z(r_s)< \frac{1}{\gamma} \quad \Rightarrow \quad  Z(r_s)^{-2}>\gamma^2 >1.$$
Therefore the numerics confirm that the inequality $\kappa_\parl/\kappa_\perp>1 $ is satisfied.

Knowing the numerical black-hole solutions and the values of $r_s$, we can immediately compute the diffusion constants, (\ref{kappa perp}-\ref{kappa par}). The results are discussed in subsections \ref{qhat1}, \ref{qhat2}.
On the other hand, in order to compute the full Langevin correlators, we additionally need to solve the world-sheet fluctuation equations. This is analyzed in the next subsection.

\subsection{Correlators and spectral functions}

The retarded correlator of the trailing string fluctuations is given by equation (\ref{full GR}) as a function of the frequency $\omega$, where the
wave-functions $\Psi_{R}^{\perp,\parallel}(r,\omega)$ are the eigenmodes of the world-sheet fluctuations. From the
full retarded propagator, one can further obtain the symmetric one through equation (\ref{Gsym}), and the spectral density through equation (\ref{sd1}).

To compute the Green's function (\ref{full GR}) numerically, one must solve the linear fluctuation equations (\ref{H19}-\ref{H20}), with
infalling conditions (\ref{H26}) at the world-sheet horizon $r_s$, and unit normalization at the UV boundary.

%\begin{figure}[h!]%\begin{center}
%\centering
\FIGURE[h!]{
\begin{tabular}{cc}
\includegraphics[width=7cm]{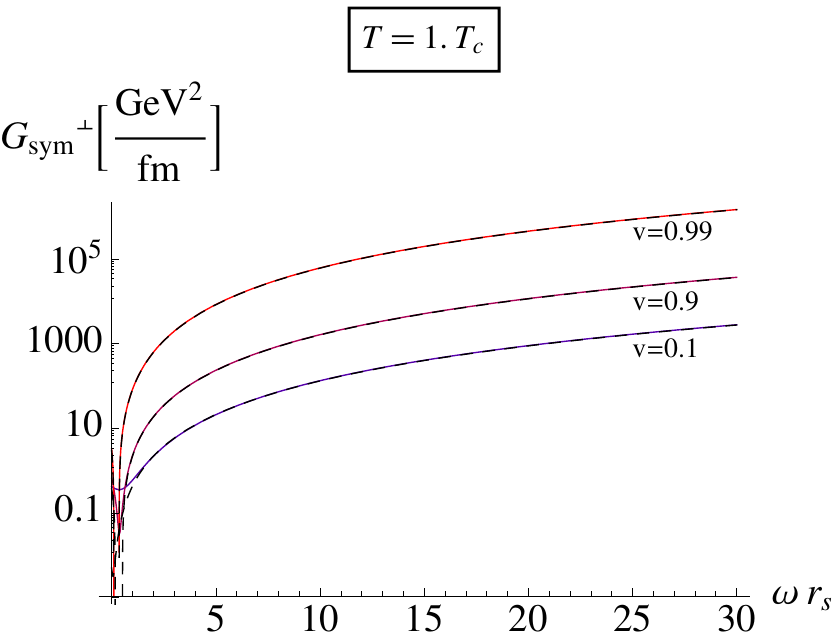} &
\includegraphics[width=7cm]{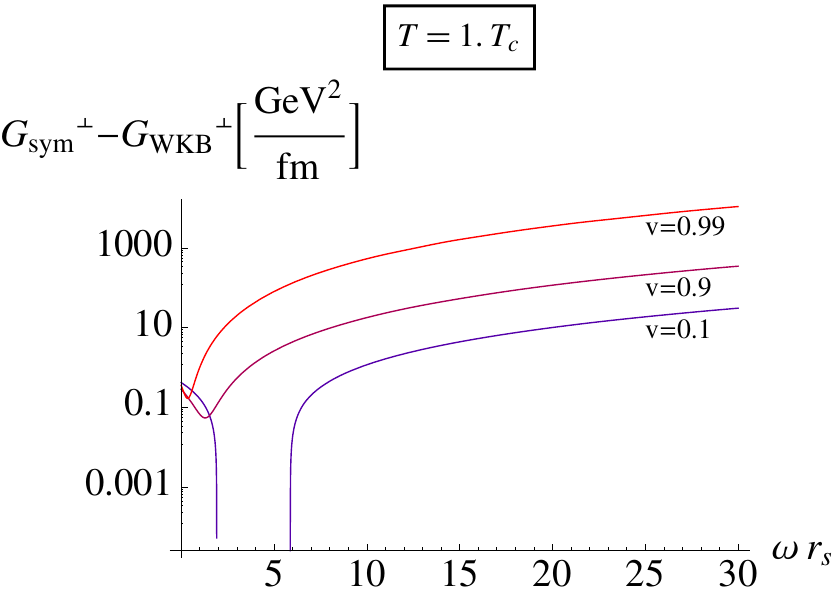} \\
\includegraphics[width=7cm]{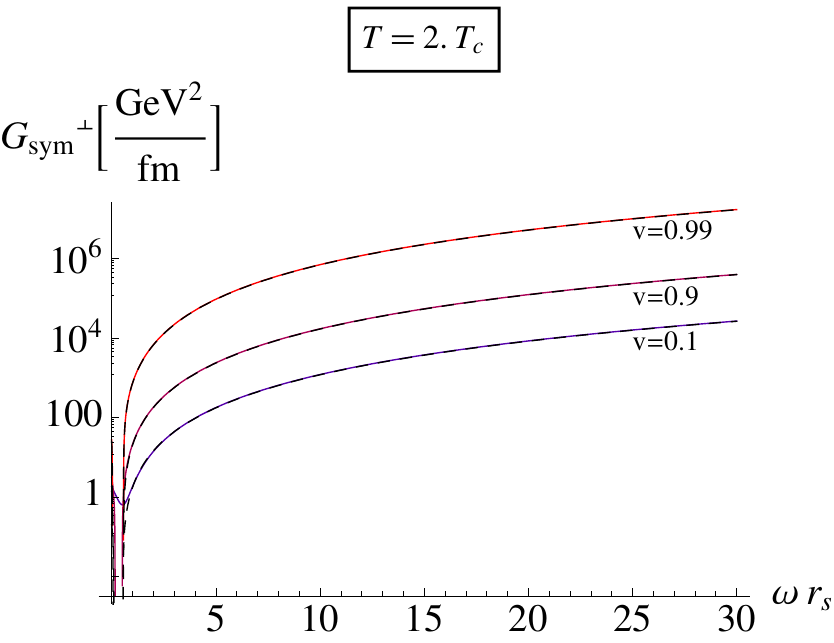} &
\includegraphics[width=7cm]{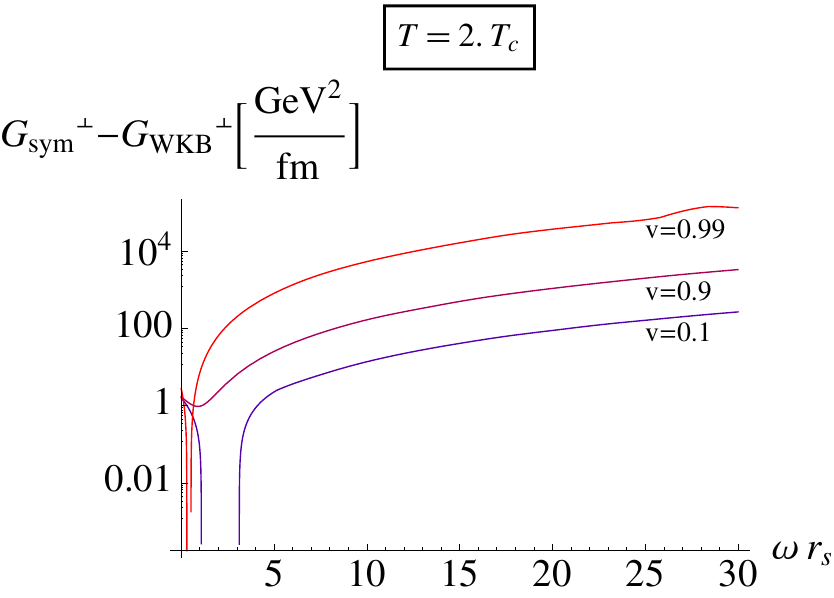} \\
\includegraphics[width=7cm]{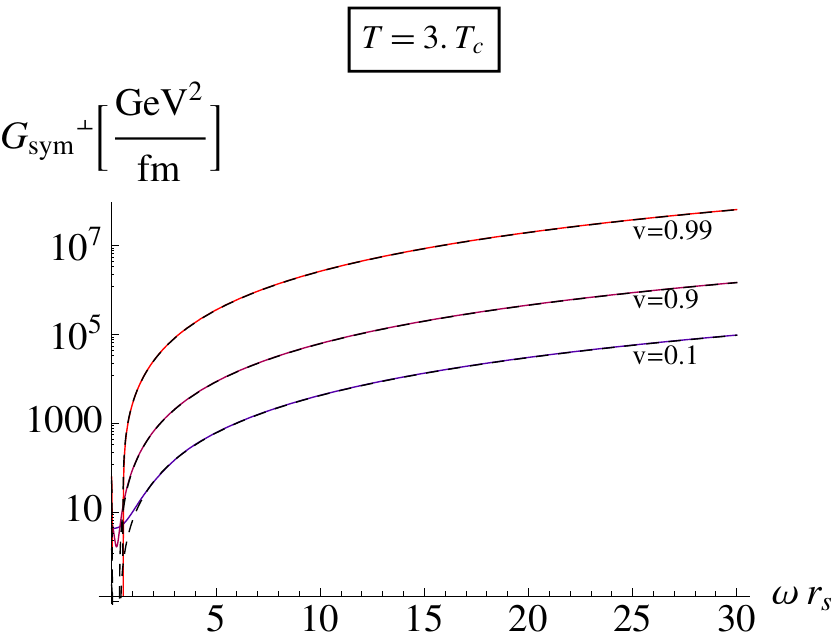} &
\includegraphics[width=7cm]{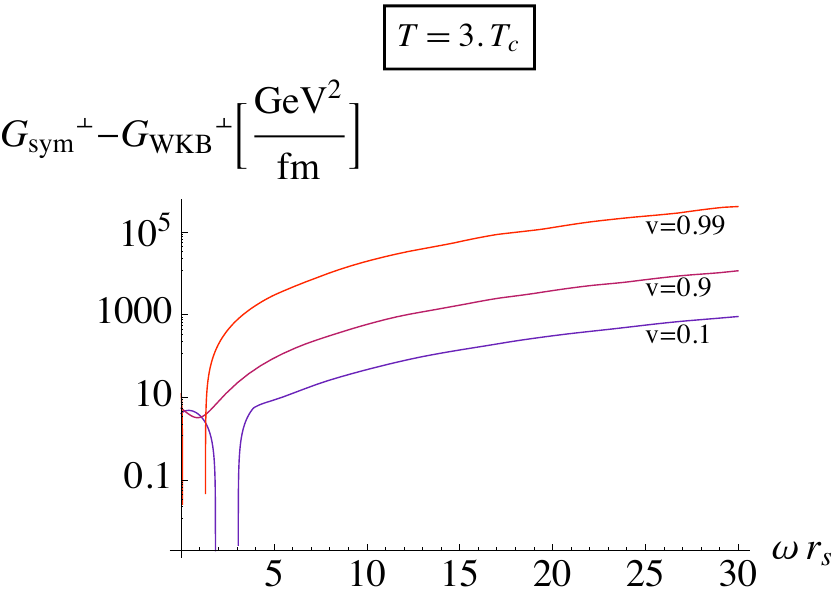}\\
(a) & (b)
\end{tabular}
%\end{center}
\caption{ (a) The symmetric correlator of the $\perp$ modes by the
numerical evaluation (solid line) and by the large-frequency WKB
computation of Section \ref{WKB} (dashed line), both in the $M_q
\to \infty$ limit. (b) Difference of the numerical and WKB
results. We show in each plot the curves corresponding to the
velocities $v=0.1,0.9,0.99$ and different plots for the
temperatures $T=T_c,2T_c,3T_c$. \label{fig corr WKB}}}
%\end{figure}
\FIGURE[h!]{
%\begin{center}
\begin{tabular}{cc}
\includegraphics[width=7cm]{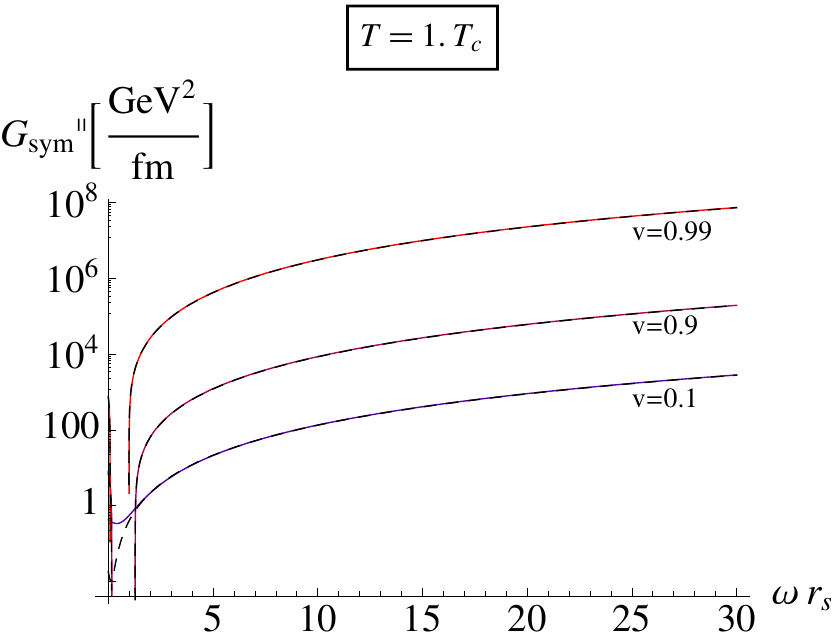} &
\includegraphics[width=7cm]{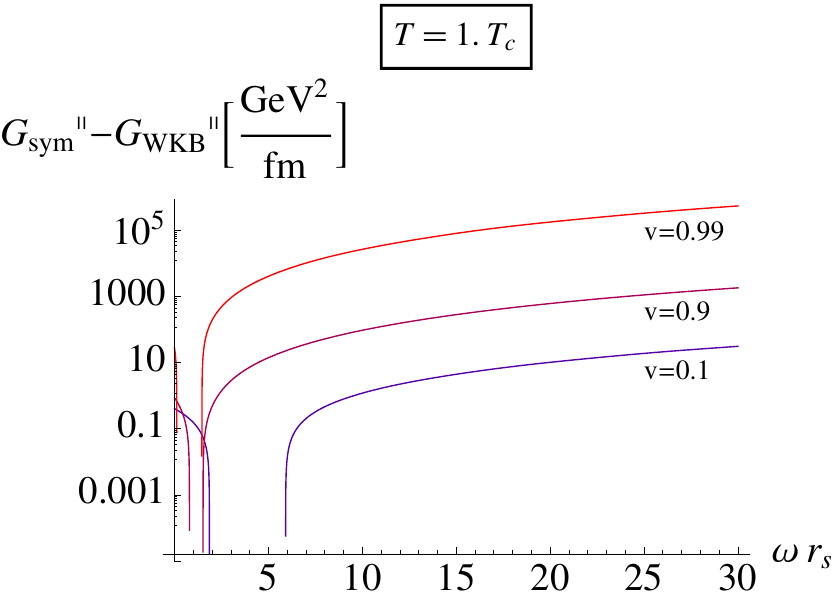} \\
\includegraphics[width=7cm]{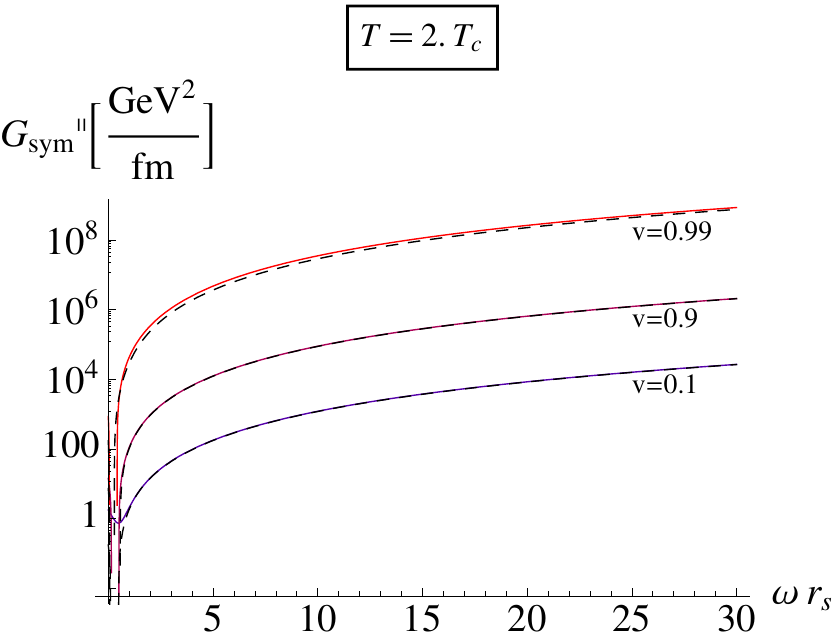} &
\includegraphics[width=7cm]{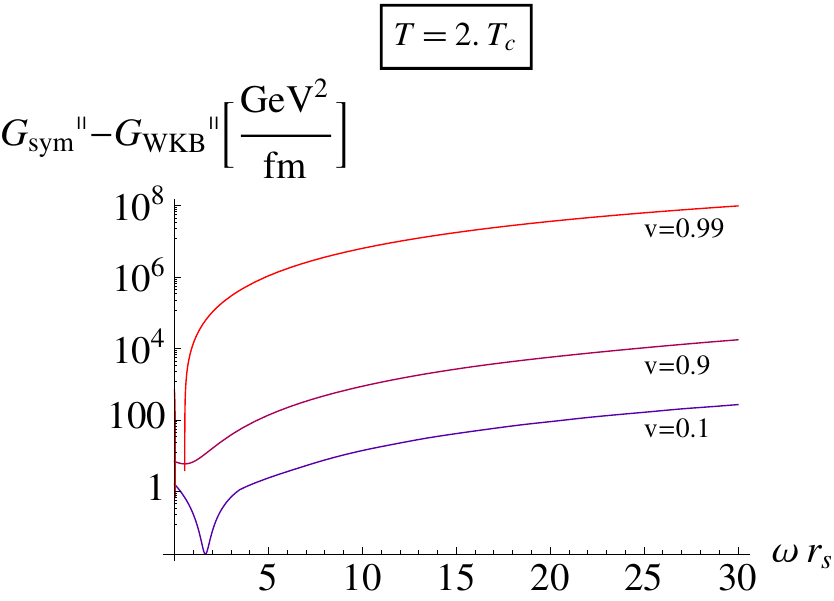} \\
\includegraphics[width=7cm]{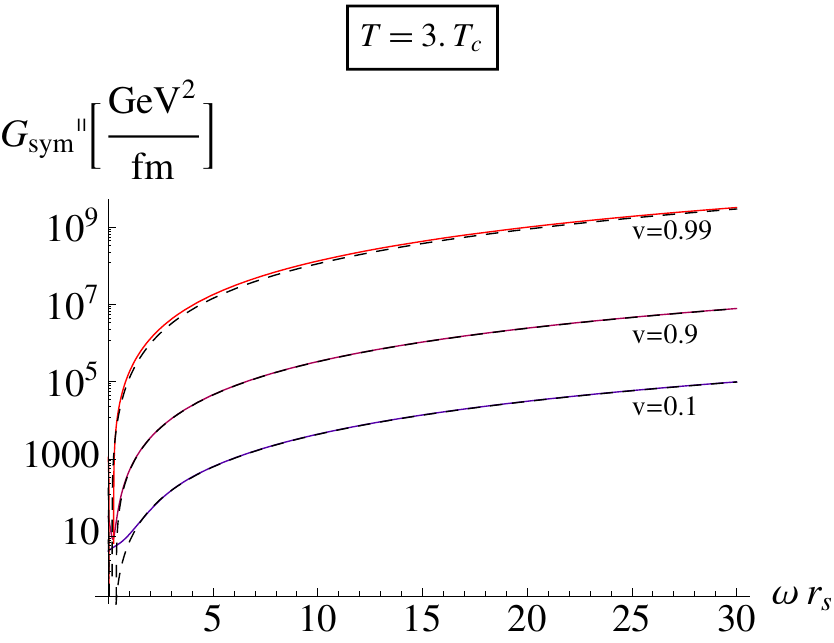} &
\includegraphics[width=7cm]{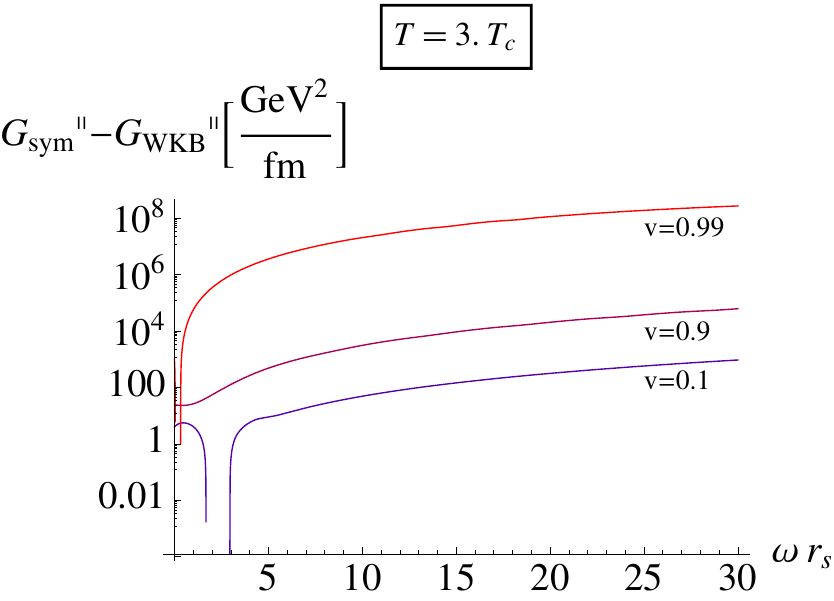}\\
(a) & (b)
\end{tabular}
%\end{center}
\caption{
(a) The numerical result for the symmetric correlator of the $\parl$ modes (solid line) together with the large frequency result from the WKB computation of Section \ref{WKB} (dashed line), for $M_q =\infty$. (b) Difference
between the exact and WKB results.
As for the $\perp$ modes, we show in each plot, $T=T_c,2T_c,3T_c$, the curves corresponding to the velocities $v=0.1,0.9,0.99$.}\label{fig corr WKB
parl}}
\FIGURE[h!]{%\begin{center}
\begin{tabular}{cc}
\includegraphics[width=7cm]{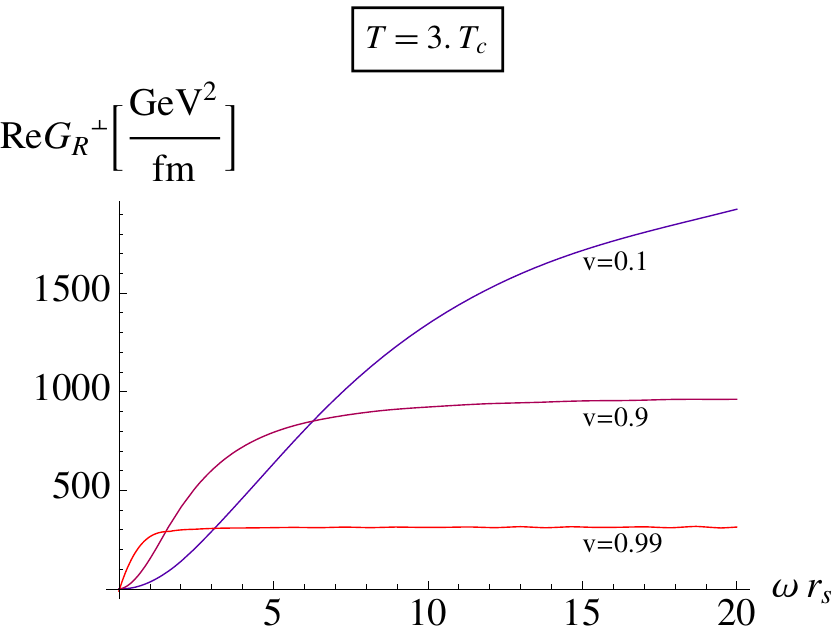} &
\includegraphics[width=7cm]{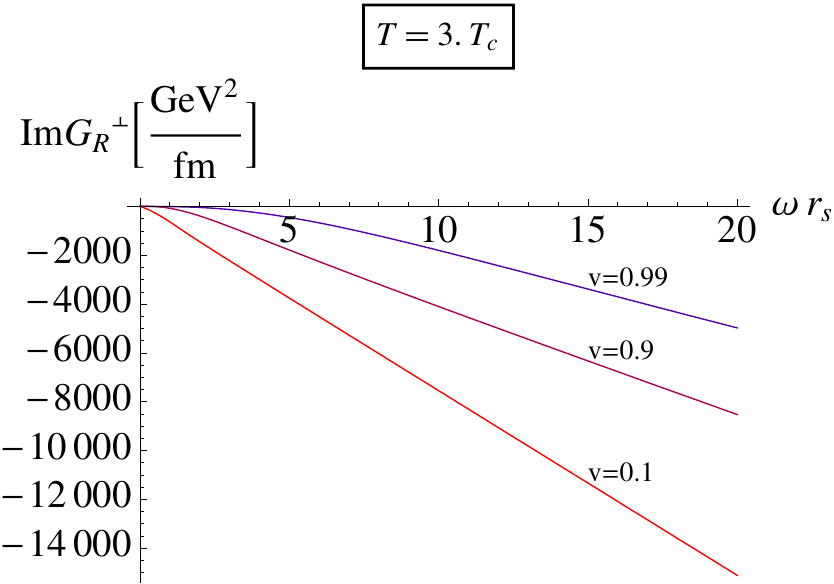} \\
(a) & (b) \\
& \\
\includegraphics[width=7cm]{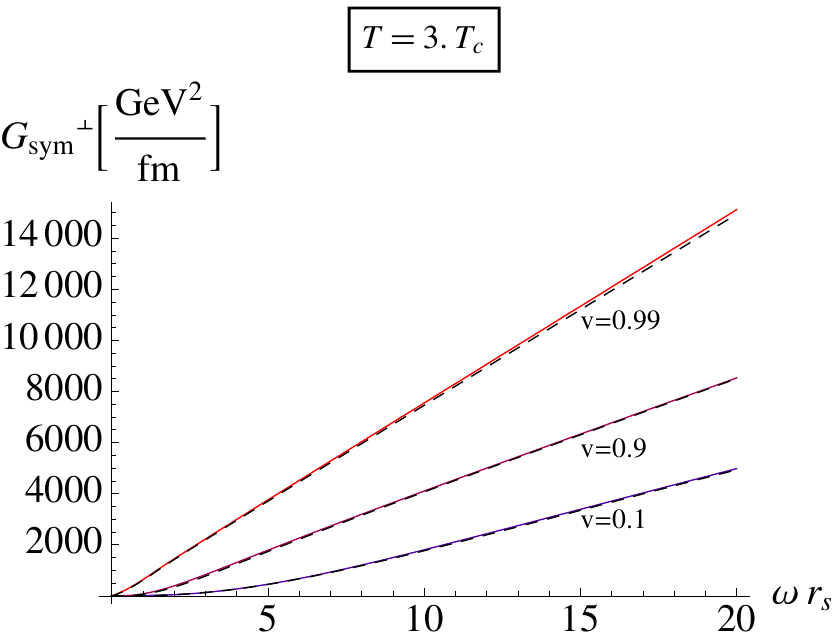} &
\includegraphics[width=7cm]{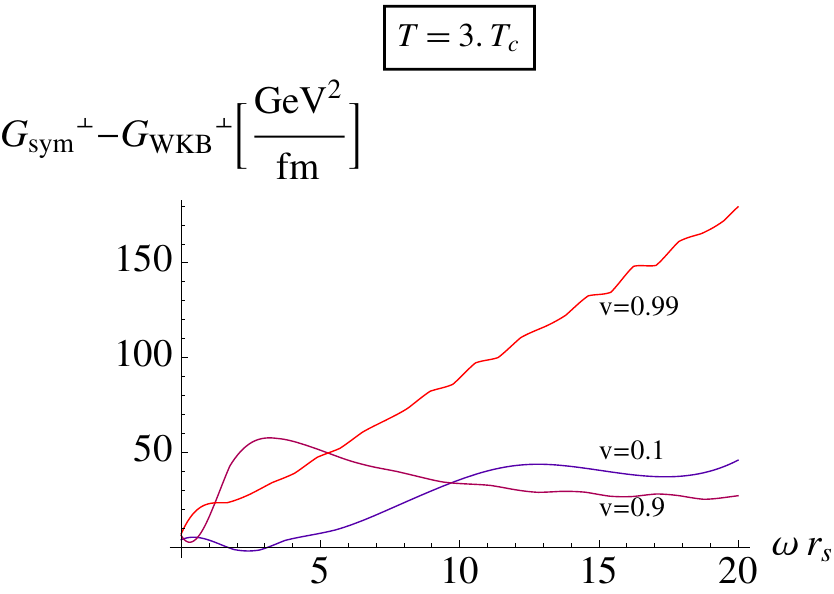}\\
(c) & (d)
\end{tabular}
%\end{center}
\caption{ The retarded correlator ((a) real and (b) imaginary
part) for finite but large quark mass, calculated numerically. (c)
The symmetric correlator for finite quark mass, from numerical
evaluation (solid lines) and WKB result of Section~\ref{WKB}
(dashed lines). (d) The difference between the numerical and WKB
result. In each plot we represent the curves corresponding to the
velocities $v=0.1,0.9,0.99$. We have taken $T=3T_c$  and
$M_Q=M_{Charm}$.}\label{fig mass WKB}}
The numerical computation makes use of the shooting technique from
the world-sheet horizon (specifying the in--falling initial
conditions for the wave function and its derivative). Once the
full solution is obtained, we normalize it dividing by the value
of the solution at the boundary, in order for the wave function to
obey the required boundary conditions. The results of the
numerical analysis are shown in figures \ref{fig corr WKB} through
\ref{fig mass WKB}, and we will discuss them below in more detail.

As discussed in Section \ref{correlators}, the real part of the correlator is UV-divergent and therefore
 it is very sensitive to the cut-off used in the numerical integration. Even if one subtracts the
  divergent term $\sim\gamma \o^2 /\epsilon$, one should be very careful  to extract  the limit $\e \to 0$ from
the numerics, and eliminate all terms which grow as higher powers of $\omega$,
but whose coefficient would vanish at $\e=0$. For example, after
subtracting the divergence, the numerical calculation will be dominated by the subleading
term in eq. (\ref{ReG WKB}), which at finite $\epsilon$ grows as $\o^4$,
but which is absent when the cut-off is  removed.

The imaginary part of the propagator on the other hand does not present these problems, and can be obtained in a clean way from the numerical computation.
 Since the imaginary part of the
retarded correlator is related in a simple way to the symmetric
correlator through equation~\refeq{Gsym}, we chose to only show
plots of the latter, which are discussed below.

In Figure~\ref{fig corr WKB} and Figure~\ref{fig corr WKB parl} the symmetric correlator corresponding to a quark with infinite mass is shown, and
it is compared to
the WKB result in \refeq{rhoperpWKB}, for the transverse and longitudinal modes.
 From these plots we observe that the WKB
result is a very good approximation to the spectral densities even
at low frequency, which is {\em a priori} unexpected. In
particular, from comparison with the WKB result, we learn that the
symmetric correlators scale with a {\em cubic power-law} at large
frequency. The difference w.r.t. the WKB result is small compared
to the value of the correlator (the apparent discontinuity in some
of the curves is an artifact due to the logarithmic scale--- it is
a small bump if plotted on a linear scale).

Figure~\ref{fig mass WKB} shows the result for the finite mass
correlators and their comparison to the WKB approximation for
large frequencies, using the results of section~\ref{WKB} and
evaluating the factor \refeq{WKB mass factor}. The correlators
indeed display the linear behavior for large $\omega$ that was
derived analytically in section~\ref{WKB}.

The numeric computation in this case is performed by normalizing
the wave function at the cutoff $r_Q$. The value of $r_Q$ is
determined by the quark mass (using equation~\refeq{mass cutoff}),
rather than being given by the regulated boundary $\epsilon$. For
the charm and bottom quarks, we take $M_{charm}  = 1.5$ GeV and
$M_{bottom} = 4.5$ GeV.\footnote{These values are subject to
renormalization in the plasma due to interactions with the media.
Therefore, they become temperature dependent. However, as we show
in \cite{transport}, this temperature dependence is very mild in our
holographic model, within the relevant temperature range.}

The correlators are also evaluated at $r_Q$, following the formula
\refeq{full GR}. We use the results obtained in \cite{transport}
for the cutoff, yielding $r_Q \simeq 1.4$ for the bottom quark and
$r_Q \simeq 7.5$ for the charm quark. We chose to show in
Figure~\ref{fig mass WKB}, as an example, the results for the
transverse mode of a charm quark at $T=3T_c$, for different
velocities.

\subsection{The jet-quenching parameters} \label{qhat1}

The diffusion constants $\kappa_{\perp}$ and $\kappa_{\parallel}$
are computed directly, as a function of temperature and velocity,
by evaluating equations (\ref{kappa perp}-\ref{kappa par}) at the
world-sheet horizon $r_s$ specified by equation (\ref{rs}).

To give an idea of the effect of the running of the scalar field, and of the breaking of conformal invariance in our model, in Figure \ref{fig kappa}
we show the ratios of $\kappa_\perp$ and $\kappa_\parallel$ to the corresponding quantities obtained in the $AdS$ black hole background, equations
(\ref{kconf1}-\ref{kconf2}) representing strongly coupled ${\cal N}$=4 SYM. The ratios $\kappa/\kappa_{conf}$ are shown as a function of velocity, at
different temperatures. The conformal results are obtained by fixing $\ell_s$ by its $AdS$/CFT relation to the fixed coupling of
 ${\cal N}=4$ SYM, $(\ell/\ell_s)^4 = \l$. We take as $\lambda = 5.5$ as in \cite{gubserrev}.
\FIGURE[h!]{
\begin{tabular}{cc}
\includegraphics[width=7cm]{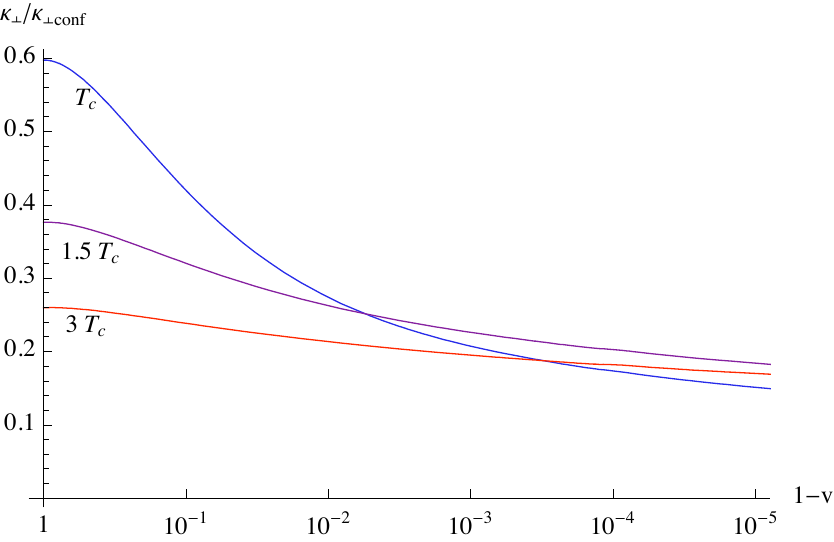} & \includegraphics[width=7cm]{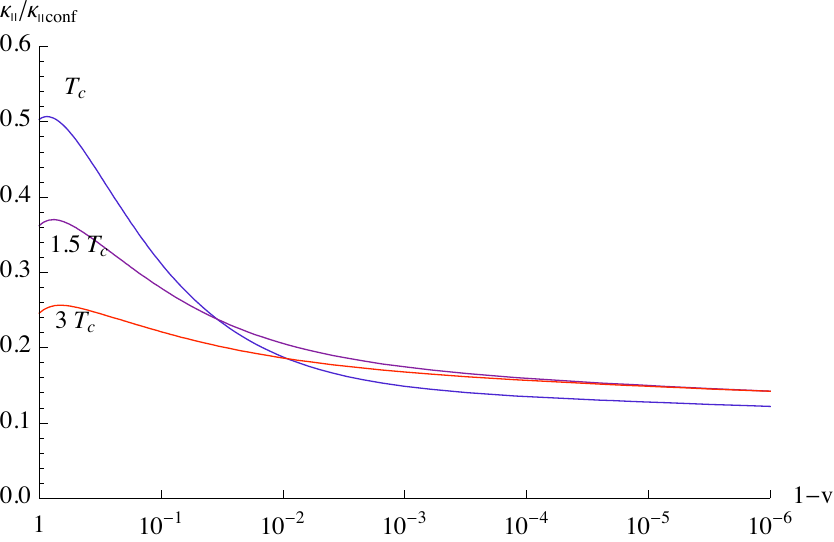} \\
\end{tabular}
\caption{The ratio of the diffusion coefficients $\kappa_\perp$ and $\kappa_\parallel$ to the corresponding value in the holographic conformal ${\cal
N}=4$ theory (with $\lambda_{{\cal N}=4}=5.5$) are plotted as a function of the velocity $v$ (in logarithmic horizontal scale) from equations \refeq{kappa
perp}-\refeq{kappa par}. The results are evaluated at different temperatures $T=T_c,1.5T_c,3T_c$ in the deconfined phase of the non-conformal model. }
\label{fig kappa}}

{}From Figure \ref{kappa} we observe that, apart from an overall
normalization, the non-conformality in this particular model
significantly affects the diffusion constants only for
temperatures close to $T_c$, and for velocities that are not too
large. Indeed, if we choose $\lambda\sim 0.5$ instead of $\lambda
= 5.5$ in the conformal case (just to make the overall magnitudes
similar in the comparison), then our result agrees with the
conformal result within the 10\% level, in the range $v\gtrsim
0.6$ and for $T\gtrsim 1.5 T_c$.

In the rest of this section we will focus on the {\em jet-quenching parameters} $\hat{q}$, which are related to the diffusion constants by
\be\label{qperp}
\hat{q}^\perp = 2 {\kappa^\perp \over v} ,\qquad \hat{q}^\parl = {\kappa^\parl \over v} \;.
\ee
The Langevin dynamics defines the two independent parameters, $\hat{q}_{\perp}$ and $\hat{q}_\parallel$. The first  controls
the transverse momentum broadening of a heavy quark probe moving through the
plasma, and it is the one usually quoted in relation to experimental results.
In this Subsection we present the result for the dimensionless quantities $\hat{q}/T_c^3$, since they do not depend on how we fix the overall energy scale
in the holographic QCD model. We will translate the result to physical units
in the next Subsection.

 In
figure \ref{fig q} we plot the two jet-quenching parameters (in units
of the critical  temperature $T_c$) as a function of the velocity, for different values of the temperature $T$.
The behavior for small $v$ is dominated by the $1/v$ factor in the
definitions (\ref{qperp}).

We note that the difference between the longitudinal and
transverse modes is due to the function $Z$ defined in
equation~\refeq{H10}, which reduces to $ f'(r_s)/4\pi T_s$ when
evaluated at the world--sheet horizon. More precisely, the
relation between the diffusion constants is $\kappa_\perp =
\kappa_\parallel Z^2(r_s)$. In sections \ref{nrelsec} and
\ref{ulrelsec} we showed that $\gamma Z(r_s) \to 1$, both when
$v\to 0$ and $v \to 1$. This is also apparent in the numerical
result shown in Figure \ref{fig Z} (b).

\FIGURE[h!]{
%\begin{center}
\begin{tabular}{cc}
\includegraphics[width=7cm]{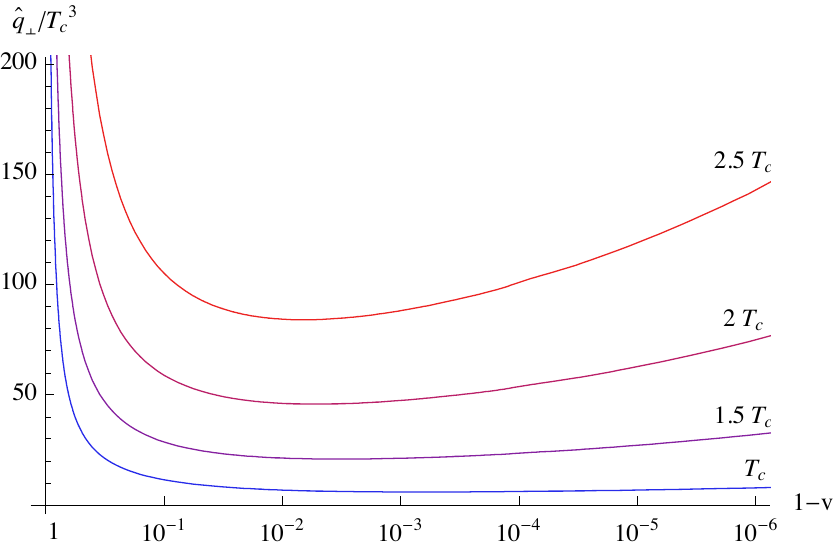} & \includegraphics[width=7cm]{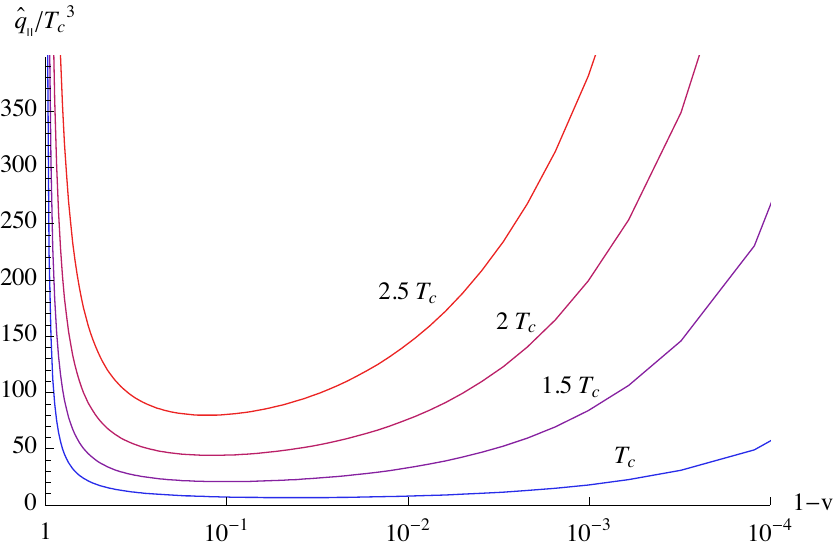} \\
\end{tabular}
%\end{center}
\caption{The jet-quenching parameters $\hat q_\perp$ and $\hat q_\parallel$
obtained from the diffusion constants \refeq{kappa perp}-\refeq{kappa par},
 normalized to the critical temperature $T_c$,  are plotted as a function of the
velocity $v$ (in a logarithmic horizontal scale). The results are evaluated at different temperatures.
}\label{fig q}}
%\end{figure}

\begin{figure}[h!]\begin{center}\begin{tabular}{cc}
\includegraphics[width=7cm]{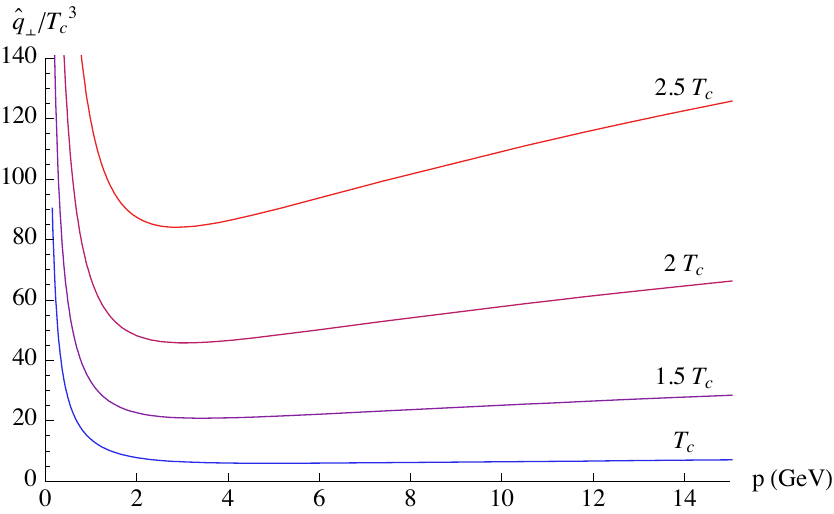} & \includegraphics[width=7cm]{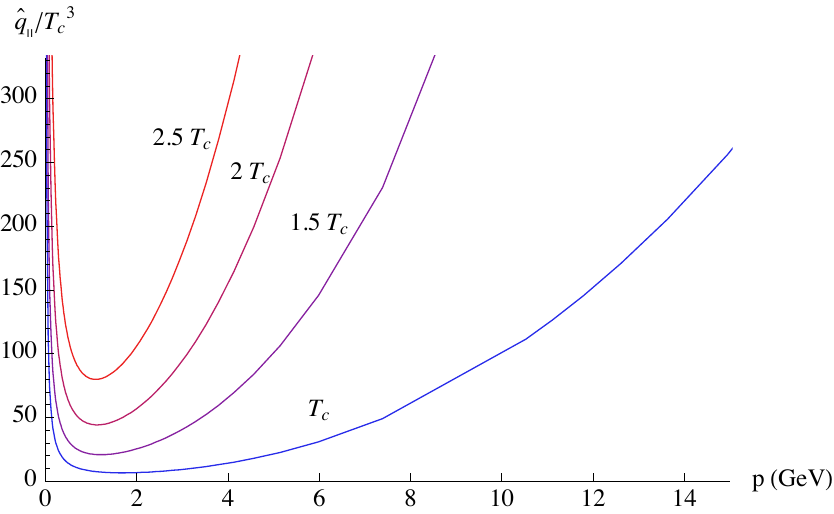} \\
$\hat{q}_\perp$ charm & $\hat{q}_\parallel$ charm\\
\includegraphics[width=7cm]{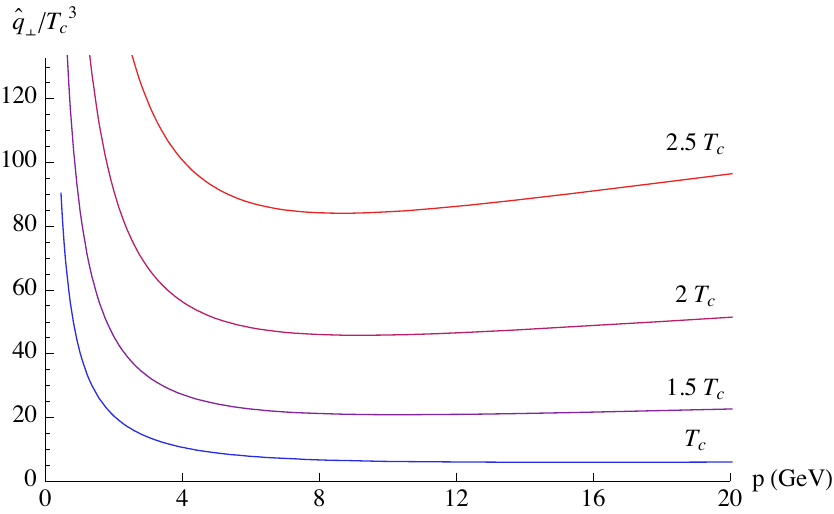} & \includegraphics[width=7cm]{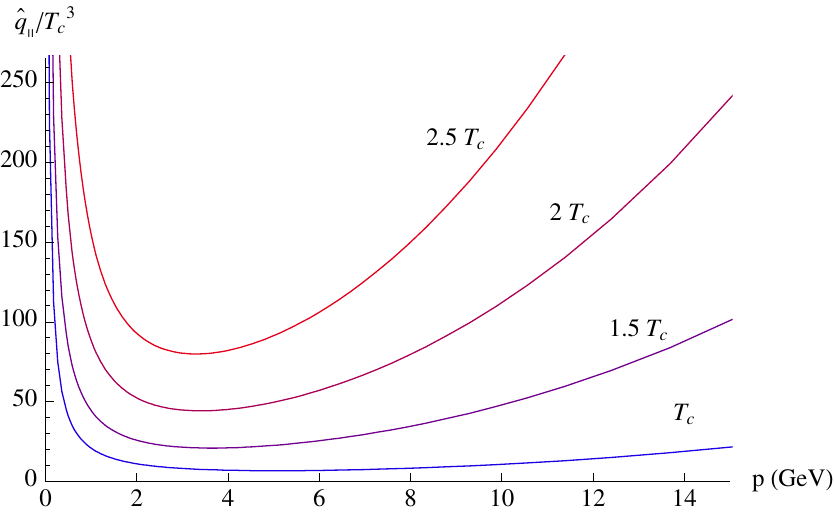} \\
$\hat{q}_\perp$ bottom & $\hat{q}_\parallel$ bottom
\end{tabular}\end{center}
\caption{The quantities $\hat q_\perp/T_c^3$ and $\hat q_\parallel/T_c^3$ plotted as a function of the quark momentum $p$. The plots for the charm and the bottom quark differ by a scaling of the horizontal direction.}
\label{fig q p}
\end{figure}

It is instructive to translate the quark velocity $v$ on the
horizontal axis of Fig. \ref{fig q} into momentum, $p = M_q v
\gamma$, where $M_q$ is the quark mass. Taking $M_{charm} = 1.5\,
GeV$, $M_{bottom} = 4.5\, GeV$, the resulting plots are shown in
Fig. \ref{fig q p}\footnote{As shown in \cite{transport}, thermal
corrections to its mass are negligible in our set-up.}. From these
plots, we observe that $\hat{q}_{\perp}$ is almost constant over a
wide momentum range, for a relativistic heavy quark. This is not
so for $\hat{q}_{\parallel}$, the difference being due to the
extra factor of $Z^2(r_s) \sim \gamma^{-2}$ in the latter.

{}From Figure \ref{fig q p} we observe that, for a fixed momentum,
$\hat{q}$ increases with temperature, as can also be inferred from
the analytic expressions \refeq{kappa perp}-\refeq{kappa par}.
This behavior is shown more clearly in Fig. \ref{fig q T}, that
displays $\hat{q}$ as a function of temperature (in units of the
critical temperature $T_c$) for different quark momenta. This is
the $\hat{q}_\perp \propto T^3$ behavior predicted by both the
relativistic approximation (\ref{kappev1}) and the
non-relativistic one (\ref{kappev0}), once we use the fact that $s
\propto T^3$ approximately.
\begin{figure}[h!]\begin{center}\begin{tabular}{cc}
\includegraphics[width=7cm]{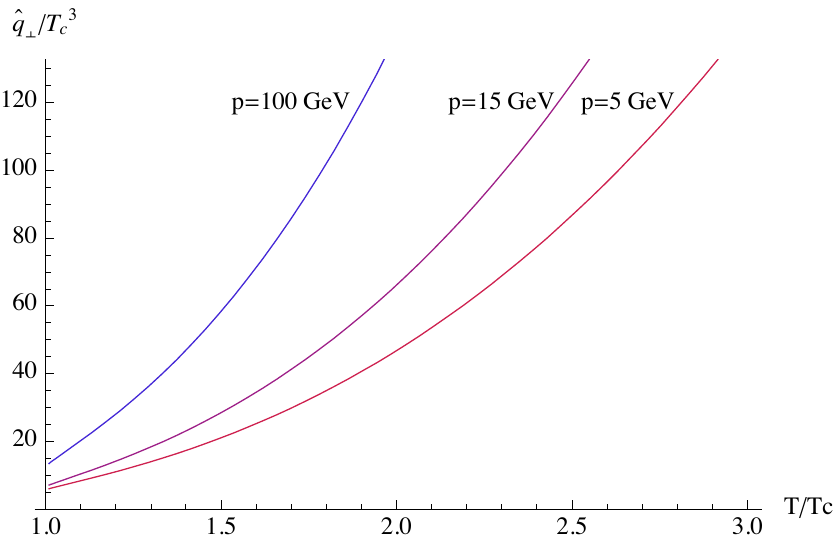} & \includegraphics[width=7cm]{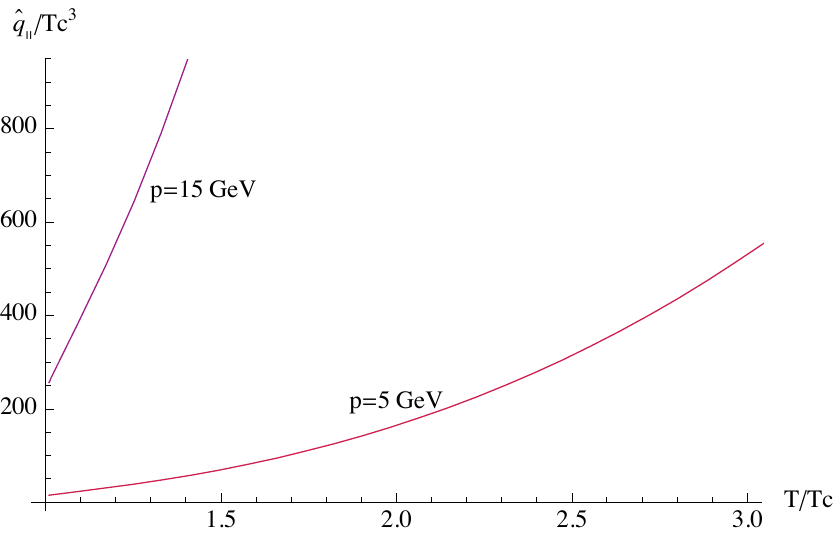} \\
$\hat{q}_\perp$ charm & $\hat{q}_\parallel$ charm\\
\includegraphics[width=7cm]{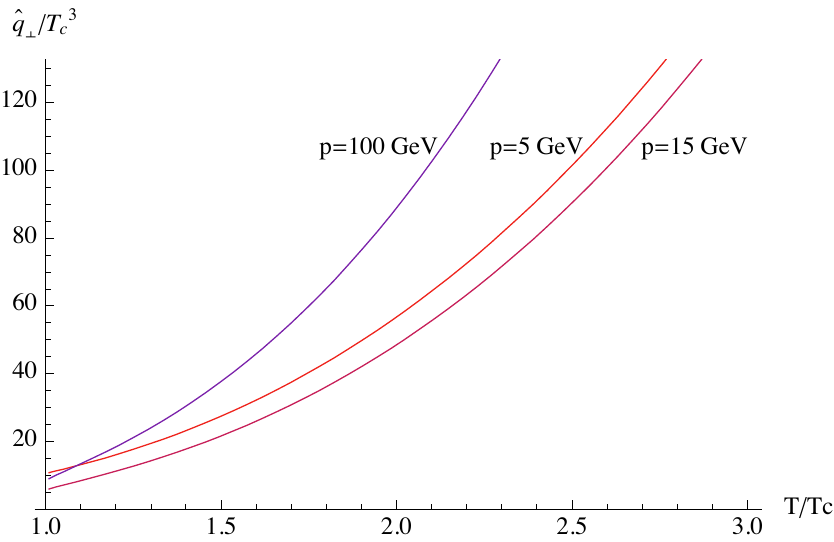} & \includegraphics[width=7cm]{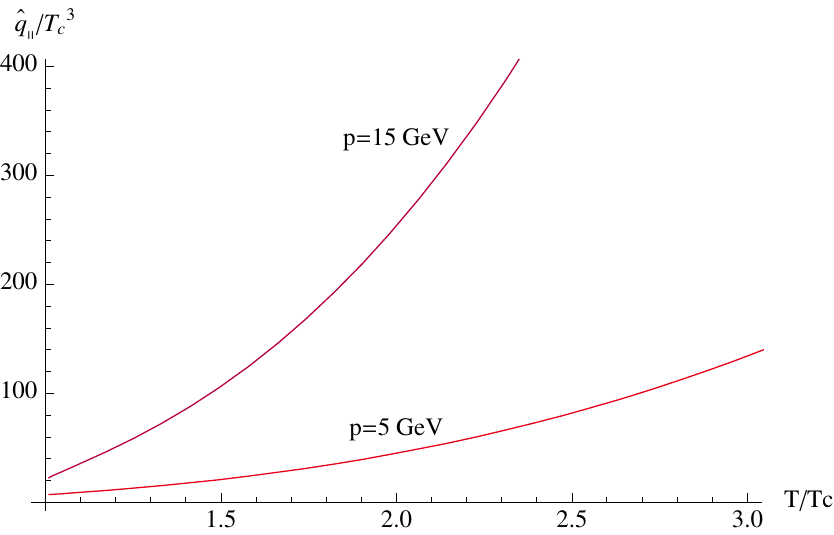} \\
$\hat{q}_\perp$ bottom & $\hat{q}_\parallel$ bottom
\end{tabular}\end{center}
\caption{The jet-quenching parameters $\hat q_\perp$ and $\hat q_\parallel$ plotted as a function of $T/T_c$, for different quark momenta.}
\label{fig q T}
\end{figure}

It is important to keep in mind that, for a finite quark mass, the trailing
string description breaks down when the world-sheet horizon coordinate
$r_h$  becomes smaller  than $r_Q$, the point where the trailing string
is attached (this is infinite for an infinite quark mass). This translates
into an upper bound on the velocity, or  momentum, beyond which the setup
breaks down, and the string  dynamics on the flavor branes becomes important.
To estimate this bound, we can use the $AdS_5$ relations for the  quark mass
and world-sheet horizon: for a heavy quark $r_Q$  is approximately
$r_Q \simeq (\ell^2/2\pi \ell_s^2) \l^{4/3}(r_Q)\, M_Q^{-1}$, and for large momentum $p$,  $r_s\simeq \gamma^{-1/4}(p) (\pi T)^{-1} $, with $gamma(p) = \sqrt{1+p^2/M_q^2}$. This results in the approximate bound:
\be\label{bound-p1}
p < M_Q \left({M_Q \over \pi T}\right)^2 \left({2\pi \ell_s^2 \over \ell^2}\right)^2 \l^{-8/3}(r_Q)
\ee
Notice that the
bound is stronger for higher temperatures and lower quark masses.

We estimated numerically  the bound on $p$  in the model we are using. The
results are displayed in Figure \ref{fig pmax}. From this figure we see that
the the bound is easily satisfied for both the Charm and Bottom quarks in
the RHIC and LHC regimes, if we use $T_c\sim 200\, MeV$

\begin{figure}[h!]\begin{center}
\includegraphics[width=7cm]{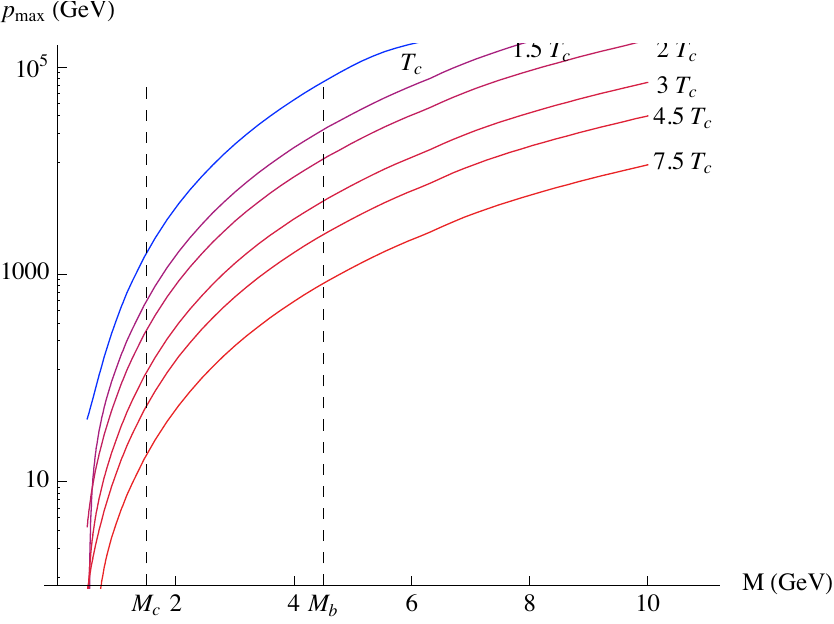}
\caption{This figure displays the upper bound on momentum $p_{max}$ (in logarithmic scale)
 beyond which the trailing string picture ceases to be valid, as a function of the quark mass, and for different temperatures.}
\label{fig pmax}
\end{center}
\end{figure}

Finally, we need to check what is the allowed range of $p$ such
that the local approximation to the Langevin equation is reliable,
as discussed in Section \ref{Sec valid}. For this to be the case,
we need the quantity $T_s/\eta_D$ to be large. For an
ultra-relativistic quark, this condition translates to equation
(\ref{bound-p1}). In our numerical solution we obtain $\l_s \sim 3
\times 10^{-2}$. Therefore we expect that, for moderate
temperatures the bound is pretty mild\footnote{There is a certain
degree of arbitrariness in the choice of normalization of $\l$.
However, changing the normalization of $\l$ would result in a
value of $\ell/\ell_s$ different from the one we are using here,
equation (\ref{ells}). The important thing is that, ones we insist
in fixing the confining string scale at a certain physical value,
the quantity $\ell/\ell_s$ scales as $\l^{-2/3}$ under an overall
scaling of $\l$ \cite{gkmn3}. Therefore the bound (\ref{bound-p1})
is independent on the overall normalization of $\l(r)$.}. Taking
the above result as a reference value for $\l_s$, and $\ell_s/\ell
\simeq 0.15$ from equation (\ref{ells}), we can rewrite the bound
(\ref{bound-p1}) more explicitly as follows:
\be\label{bound-p-2} p
\ll 1.5 M_q \left(M_q\over T\right)^2
\ee
For example, for the
charm quark ($M_q \simeq 1.5~GeV$), and close to the critical
temperature $T_c$, this translates into $p \ll 2~GeV (1.5~
GeV/T_c)^2$.

However, the situation changes dramatically as temperature
increases: from eq. (\ref{bound-p-2}) we observe that the upper
limit at temperature $T$ decreases as $(T_c/T)^2$.  A graphical
representation of the validity condition is shown in Figure \ref{fig
validity}, computed numerically from equation (\ref{bound-p}) for
both the charm and bottom quarks. The $p$-region in which the
diffusion process can be approximated by a local Langevin
equation, with constant friction and diffusion coefficients, lies
in the left side of the vertical lines (each corresponding to a
different temperature). From these plots we observe  that the
bound is satisfied for momenta up to $\sim 70~GeV$ (charm) and
more than $200~GeV$ (bottom) at $T=T_c$, but for larger
temperatures the bounds are much stronger. For example the bounds
at $T = 3 T_c$ are $p_{charm} < 10~ GeV$ and $p_{bottom} < 100~
GeV$.

What these values of $T$ correspond to in terms of actual physical
temperature of the QGP, is a subtle question, as we will discuss
more in detail in the next subsection. However an order of magnitude
estimate can be obtained by setting $T_c \approx 180~GeV$ in these
plots. This means that, for RHIC temperatures and momenta, the
local approximation remains valid. On the other hand, if the
holographic setup is to be applied to ALICE results, it is likely
that one should
 use the full non-local form of the generalized Langevin equation, and the simple parametrization of transverse momentum broadening in terms of $\hat{q}$ breaks down.

\begin{figure}[h!]\begin{center}\begin{tabular}{cc}
\includegraphics[width=7cm]{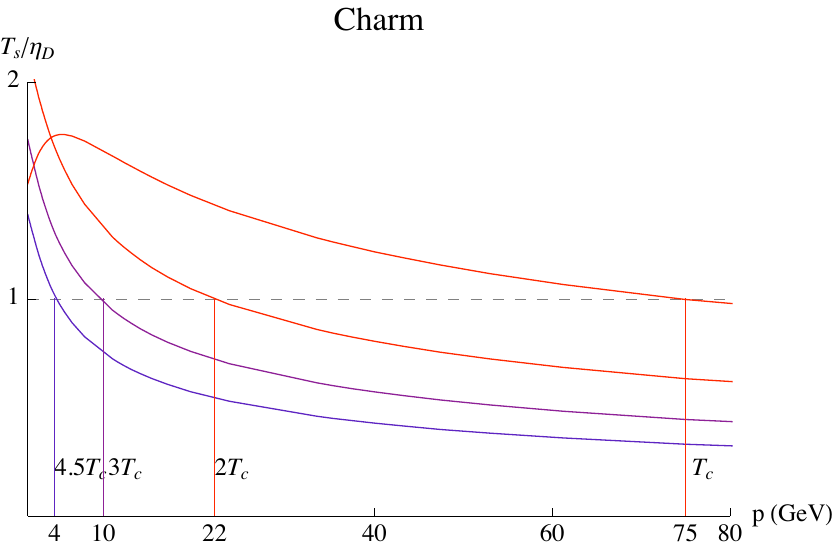} & \includegraphics[width=7cm]{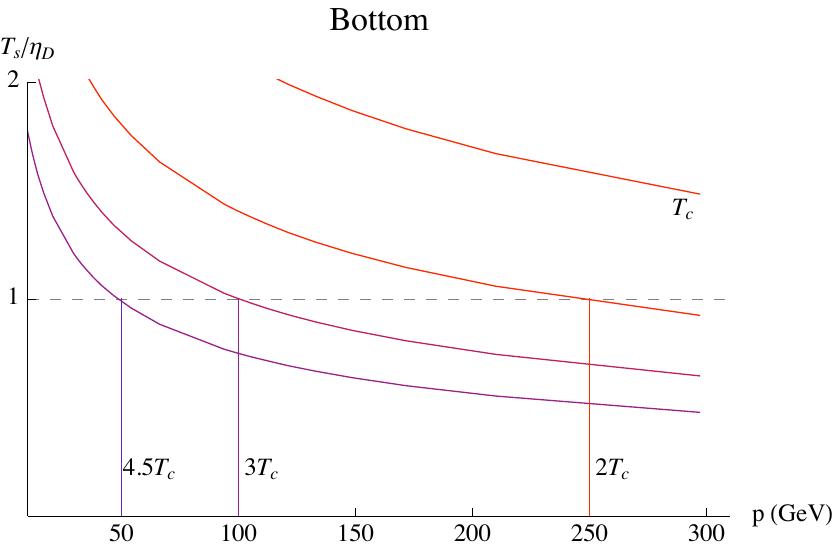} \\
(a) & (b)
\end{tabular}\end{center}
\caption{The quantity $T_s/\eta_D$ is plotted against quark momentum, for different bulk temperatures. Figures (a) and (b) refer to the charm and
bottom quark, respectively. For each temperature, the validity of the local Langevin equation constrains $p$ to the left of the corresponding vertical line, which marks the transition of $T_s/\eta_D$ across unity.}
\label{fig validity}
\end{figure}

\subsection{Comparison with heavy-ion collision observables} \label{qhat2}

Fit of RHIC data for nuclear modification factors with
hydrodynamic simulations prefer a strong jet-quenching parameter for light quarks
about $\hat{q}_{\perp} = 5$ -15 $GeV^2/fm$ (for a review of recent
results and a more references, see e.g. \cite{salgado}).

In order to compare our results to QGP observables we need to evaluate the results of the
previous section at typical temperature for QGP $T_{QGP} \approx 250 \, MeV$.

However, as discussed in detail in \cite{transport} it is not easy
to make a direct comparison, because our calculations are made
with a pure glue background (neglecting therefore the quark
contributions). It was argued recently  that quarks
contribute importantly in energy loss, beyond their enhancement of
the number of degrees of freedom. this was shown to be the case in the thermal
 $D3-D7$ system \cite{bigazzi}, and the same feature was already noted in
\cite{bigazzi2} in a non-critical  model and a in a model based on wrapped
$D5$ branes.

To proceed further  we will
 translate the physical QGP temperature to our $T$.
 To do this requires picking up a comparison scheme. In a {\em
direct} scheme one simply takes $T = T_{QGP}$.

On the other hand, one can argue that the relation between the QGP
temperature and that of the holographic model should be such that
the {\em energy densities} are the same. Energy density scales as
the number of degrees of freedom, and the holographic setup we
study is supposed to describe pure Yang Mills theory, rather than
QCD with three light flavors. Therefore, matching energy densities
leads to a holographic temperature $T$ higher than the QGP
temperature, due to the different number of degrees of freedom in
the two theories. This reasoning leads to the identification of an
alternative scheme, referred to as the {\em energy
scheme}\footnote{One can also define an {\em entropy scheme,}
where one matches entropy density rather than energy density. We
checked that the numerical results obtained in the energy and the
entropy schemes are essentially very close to each other.},where
the effective temperature $T$ is related to the real QGP
temperature $T_{QGP}$ by the implicit relation \cite{transport}:
\be\label{alt} \epsilon_{hol}(T_{energy}) \simeq 11.2\, T_{QGP}^4
\ee where $\epsilon_{hol}(T)$ is the energy density of the
holographic model.

\begin{figure}[h!]
\begin{center}
\includegraphics[width=10cm]{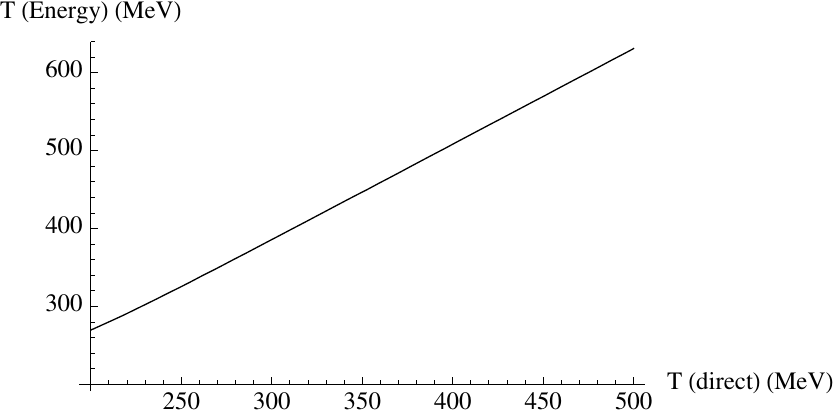}
\end{center}
\caption{Relation between the temperature in the direct and energy schemes.}
\label{fig alt}
\end{figure}

Computing the $\epsilon_{hol}(T)$ numerically one obtains the approximate
linear relation (as shown in Fig \ref{fig alt} ) between the direct scheme
and energy scheme temperatures:
\be
{T_{energy} \over MeV} = 23.7 + 1.2 {T_{dir} \over MeV}
\ee

\begin{figure}[h!]\begin{center}
\includegraphics[width=7cm]{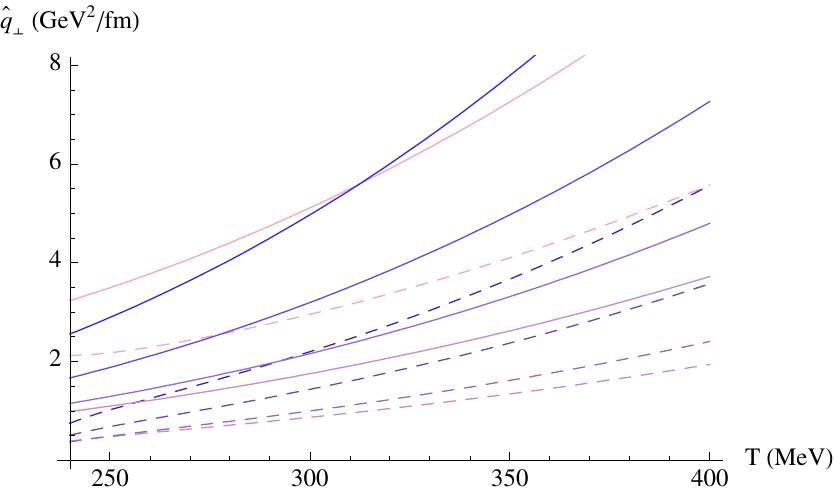} \hspace{0.8cm}
\includegraphics[width=7cm]{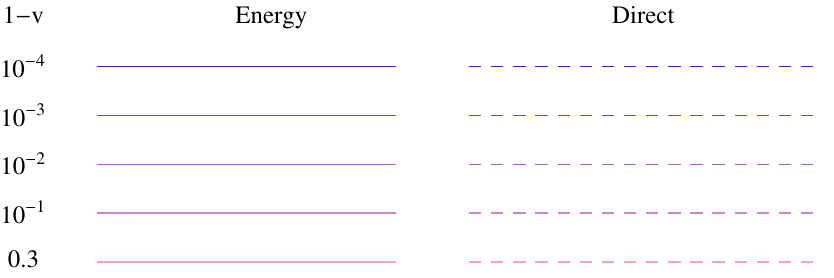}
\end{center}
\caption{The jet-quenching parameter $\hat q_\perp$ in the direct
(dashed lines) and energy (solid lines) schemes, as a function of
temperature, for different quark velocities.} \label{fig q dir
alt}
\end{figure}

The same $T_{QGP}$ corresponds to a higher temperature of the holographic model in the energy scheme,
 than in the direct scheme. Therefore, using the energy scheme to match the QGP temperature results
  in higher values for $\hat{q}$, than those obtained in the direct scheme.
   This behavior is apparent in Figure \ref{fig q dir alt}.

We are now ready to translate in physical temperatures the results
for $\hat{q}_\perp$ presented in the previous Subsection. This is
done in Figures \ref{fig q dir alt p} and \ref{fig q dir alt T},
which are analogous to Figures \ref{fig q p} and \ref{fig q T},
except that the temperature and $\hat{q_\perp}$ are displayed in
physical units. In order to express $\hat{q}$ and $T$ in
$GeV^2/fm$ and $MeV$, respectively, we have to introduce physical
energy units. The overall energy scale was fixed, as briefly
explained at the beginning of  Section 6 (and in more detail in
\cite{gkmn3}), by matching one dimensionfull quantity to its
physical value. As in \cite{gkmn3}, for this purpose we used  the
lattice value of the lowest glueball mass.

\begin{figure}[h!]\begin{center}\begin{tabular}{cc}
\includegraphics[width=8cm]{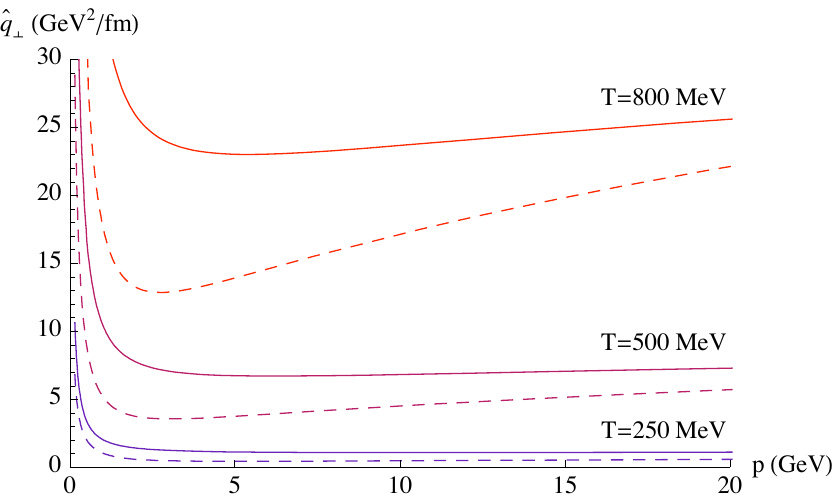} &
\includegraphics[width=8cm]{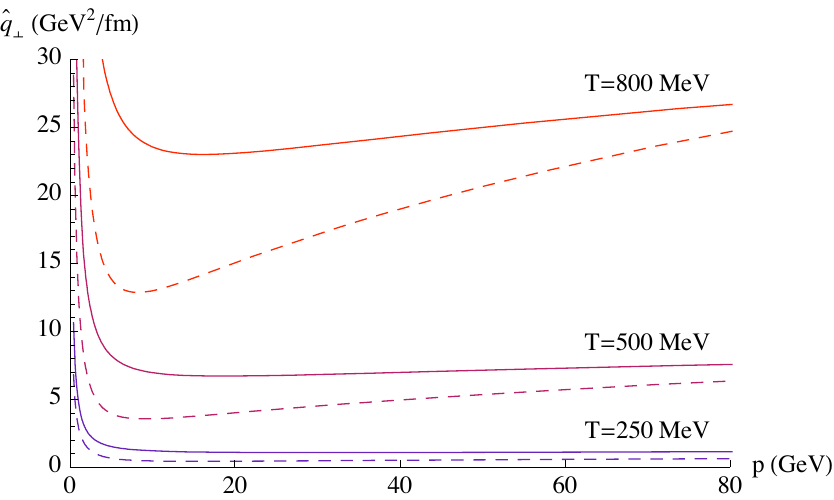} \\
$\hat{q}_\perp$ charm & $\hat{q}_\perp$ bottom
\end{tabular}\end{center}
\caption{The jet-quenching parameter $\hat q_\perp$ in the direct
(dashed lines) and alternative (solid lines) schemes, as a
function of momentum, for different physical values of
temperature.}\label{fig q dir alt p}
\end{figure}

Figure \ref{fig q dir alt p} shows $\hat{q}_{\perp}$ in both
schemes, as a function of the probe quark momentum, for charm and
bottom quarks and for various temperatures in the range relevant
for RHIC and for the ALICE experiment at LHC. We observe  that,
although the values in the energy scheme are higher, at the
temperatures relevant for RHIC ($T\approx 250 \, MeV$),
$\hat{q}_\perp$ varies in the  $1-5~GeV^2/fm$ range except at low
momenta where it is substantially higher. A full Langevin fit is necessary in order to ascertain if
these numbers fit the data.

For the highest temperature shown in Figure \ref{fig q dir alt p},
$T\sim 800~MeV$ (which is not in the range of RHIC, but may be
within the reach of ALICE), the predicted value in the energy
scheme reaches $\hat{q} \simeq 25~GeV^2/fm$.

We notice from Figure \ref{fig q dir alt p} that up to very large momenta $\hat{q}_\perp$
is effectively independent of $p$. Therefore one can safely neglect the non-linearity in the Langevin equation for large ranges of $p$. This
also allows to pick a reference momentum (say, $p \approx 10~GeV$) and study
more closely the behavior of $\hat{q}_\perp$ as a function of temperature.
This is done in Figure \ref{fig q dir alt T}, which is analogous to Figure \ref{fig q T} but with physical units for the temperature.

\begin{figure}[h!]\begin{center}\begin{tabular}{cc}
\includegraphics[width=8cm]{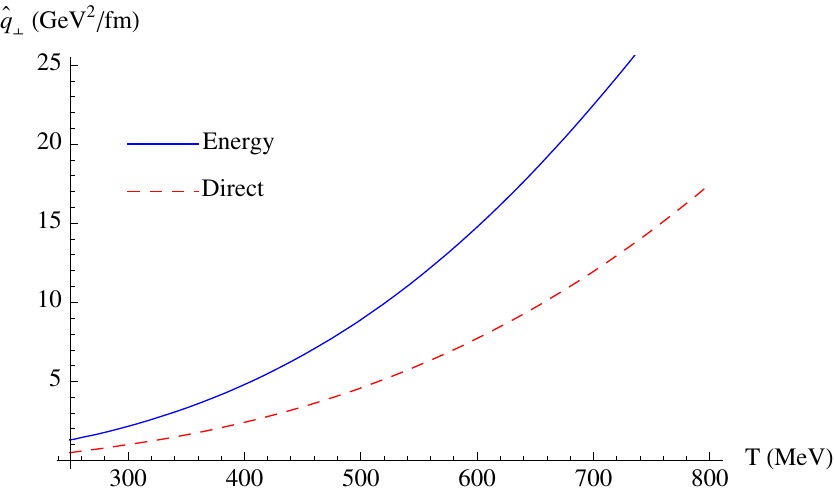} &
\includegraphics[width=8cm]{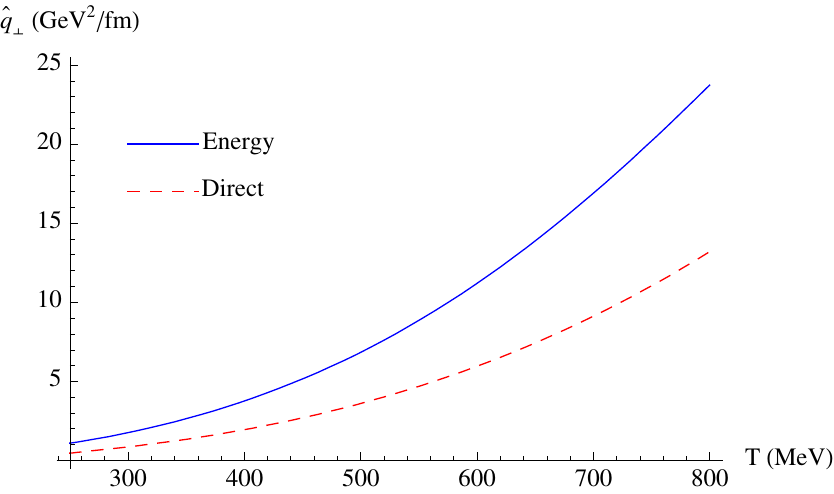} \\
\includegraphics[width=8cm]{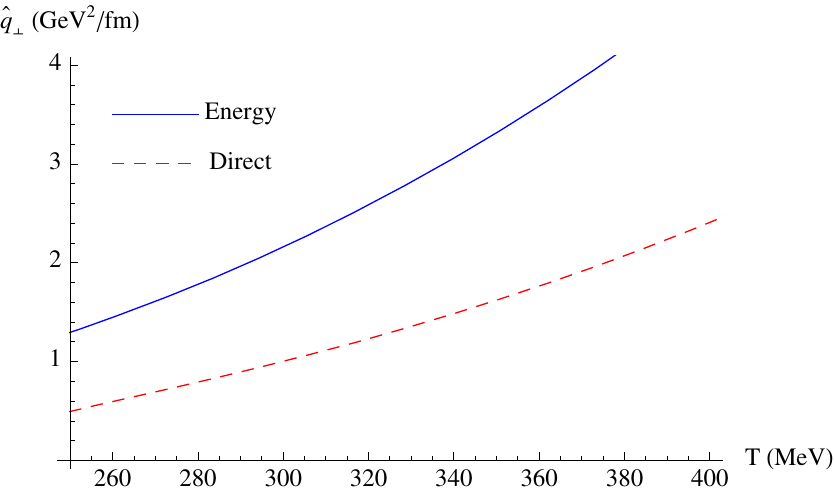} &
\includegraphics[width=8cm]{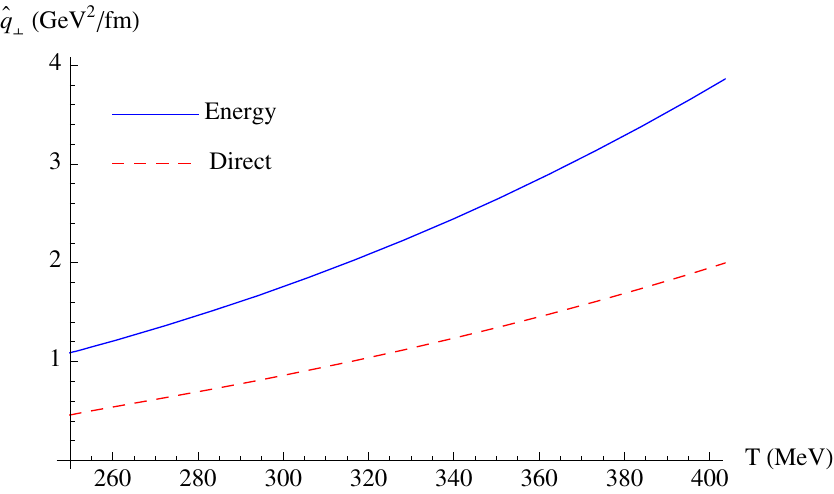} \\
Charm, $p=10 GeV$ & Bottom, $p=10 GeV$
\end{tabular}\end{center}
\caption{The jet-quenching parameter $\hat q_\perp$ in the direct (dashed lines) and
energy (solid lines) schemes, as a function of temperature, for
probe charm and bottom quarks with $p_T \approx 10\, GeV$. The bottom plots on  show the same functions on a narrower temperature range than the top plots. }
\label{fig q dir alt T}
\end{figure}

In figures \ref{fig q dir alt p} and \ref{fig q dir alt T} one
must keep in mind that the parametrization of transverse momentum
broadening by $\hat{q}_\perp$ fails for momenta that are out of
the range validity of the long time approximation to the diffusion
process. These bounds are given at the end of the previous section
as a function of $T/T_c$. In the direct scheme, a plasma
temperature of $250~MeV$ corresponds to $T \simeq T_c$ in the
holographic model\footnote{We remind the reader that the critical
temperature of the IHQCD model we are using is roughly $T_c = 247~
MeV$ \cite{gkmn3}.}, but in the energy scheme the same plasma
temperature corresponds to $T\simeq 325~MeV = 1.3 T_c$ according
to the relation (\ref{alt}). Similarly, the temperature $T_{QGP} =
800~MeV$ corresponds to $T \simeq 980~MeV \simeq 4 T_c$ in the
energy scheme. Therefore, comparing with Figure \ref{fig
validity}, we deduce that at least for the charm quark, we cannot
really trust the analysis in terms of $\hat{q}_\perp$ for the
highest temperature displayed in Figures \ref{fig q dir alt
p}-\ref{fig q dir alt T}, unless we look at momenta smaller than
$4~GeV$. On the other hand, the results shown for the bottom quark
are consistent with the long-time approximation.

Finally, we should remind the reader that our results in the UV
are not totally trustable due to the fact that the background used
may not describe properly the details of QCD.

 \addcontentsline{toc}{section}{Acknowledgements}
\paragraph{Acknowledgements}\label{ACKNOWL}

We would like to thank J. Casalderrey-Solana, A. Dainese, Y. Foka, E. Iancu, A. Kiessel, M. Panero, G. Policastro, E. Shuryak, U. Wiedemann, for useful conversations.
This work was partially supported by a European Union grant FP7-REGPOT-2008-1-CreteHEP
 Cosmo-228644. The work of LM was supported in part by MICINN and FEDER under grant FPA2008-01838 and by the Spanish Consolider-Ingenio 2010 Programme CPAN (CSD2007-00042).

\newpage
 \addcontentsline{toc}{section}{Appendices}
\appendix
\section*{Appendix}

\section{Boundary terms and scheme dependence in the propagator} \label{App GR}

The form of the retarded propagator we have obtained in Section 4,
equation (\ref{full GR}), depends on the form of the action we took as a
starting point, namely  equation (\ref{NGACTION}).
 As the correlators have UV divergences there is potential scheme dependence in their calculation that we now address.
The associated counterterms  do not modify the wave equations (\ref{H18}), but they change the value of the on-shell action and therefore the correlator.

Since the boundary terms that we add to the action must be real, the
scheme dependence  can only manifest itself in the definition of ${\rm Re}\, G_R(\o)$.
Therefore, it does not affect the physical quantities described in Sections
4 and 5, such as the diffusion constants and spectral densities.
In the case at hand, as we show below, the only scheme dependence in ${\rm Re}\, G_R (\o)$ amounts to a renormalization in the (heavy) quark mass.

 In what follows we study the divergence structure
of the action (\ref{NGACTION}), expanded to quadratic order in the fluctuations
defined in equation (\ref{H4}), around the classical trailing string solution.
To regulate the action, we cut-off the $r$-integration at $r=\e >0$,
and we study the divergences in the $\e \to 0$ limit.

Starting from equation (\ref{H2}) and (\ref{H3}), we obtain to quadratic order in the fluctuations,
\be
S_{NG} = S_0 + S_1 + S_2+\cdots .
\ee
Below, we write explicitly and discuss each term separately.
\begin{itemize}
\item {\bf Zeroth order.}\\
The zeroth order term reads simply:
\be
S_0 = -{1\over 2\pi\ell_s^2}\int dt \int_\e^{r_s} dr b^2(r) Z(r)
\ee
For small $r$, the integrand is approximately equal to
$(\ell/r)^2 \, \l^{4/3}(r) \gamma^{-1}$. Therefore the integral is dominated
by the region around $r\simeq \epsilon$, and it is given approximately by:
\be\label{div 0}
S_0^{(div)} = {\l^{4/3}(\e) \over \e} {\ell^2\over 2\pi\gamma\ell_s^2}\int dt,
\ee
 giving a $1/\e$ divergence.
\item {\bf First order.}\\
The first order term in the fluctuations is a boundary term of the form:
\be
S_1 = C \int dt\, \delta X^\parl(\e,t), \qquad C \equiv -{v\, b^2(r_s)\over 2\pi\ell_s^2}.
\ee
Since on-shell $\delta X^\parl(r=0,t)$ is finite, this term is not divergent.
\item {\bf Second order.}\\
At quadratic order the action is given by equation (\ref{H16}),
\be S_2=-{1\over
2\pi \ell_s^2}\int d\tau dr ~{1\over 2} H^{\a\b}\left[{1\over Z^2}
\partial_{\a}\dx\partial_{\b}\dx + \sum_{i=2}^3 \partial_{\a}
\delta X^i\partial_{\b}\delta X^i\right]
\label{H16-app}\ee with
\be \label{Hab-app}
 H^{\a\b} = \left( \begin{array}{cc}
-{b^4\over \sqrt{(f-v^2)(b^4f-C^2)}}~~ & ~~0\\
0~~ & \sqrt{(f-v^2)(b^4f-C^2)}\end{array}\right)\;,
\ee
It is convenient to write the fluctuation in Fourier space. The solution of the field equations, equations ~(\ref{H18}) reads, close to the boundary:
\be
\delta X^a(r,\o) \simeq C_s^a(\o) + C_v^a(\o) r^3/ \l^{4/3}(r), \qquad
a=\perp,\parl
\ee
Inserting the above expression into the action, and using the asymptotic expressions,
 \be
H^{tt} \simeq - \gamma b^2, \qquad H^{rr} \simeq b^2/ \gamma, \qquad Z^2 \simeq 1 /\gamma^2,
\ee
we observe that the only divergent term as $\e \to 0$ originates from the terms involving two time derivatives of $\delta X$:
\bea\label{div 2}
S_2^{(div)}&& = {\l^{4/3}(\e)\over \e} {\ell^2\over \gamma}\,{1\over 2}\int d\o \,\o^2 \bigg(\gamma^2 |C_s^\perp(\o)|^2 + \gamma^4 |C_s^\parl(\o)|^2 \bigg) \nonumber \\ && = {\l^{4/3}(\e)\over \e} {\ell^2\over \gamma}\,{1\over 2}\int dt\, \gamma^2 \left(\delta \dot{X}^\perp \right)^2 + \gamma^4 \left(\delta \dot{X}^\parl \right)^2
\eea
\end{itemize}
\vspace{0.5cm}

We can reabsorb both divergences, (\ref{div 0}) and (\ref{div 2}), with a single covariant boundary counterterm,
\be\label{counterterm}
S_{count} = \Delta M(\e) \int dt\, \sqrt{\dot{X}^\mu \dot{X}_\mu},
\ee
which corresponds to a renormalization of the quark mass. Indeed, expanding equation (\ref{counterterm}) to second order in $\vec{X} = \vec{v} t + \delta
\vec{X}$, we find:
\be
 S_{count} \simeq {\Delta M(\e)\over \gamma}\left\{ \int dt\, + {1\over 2}\int dt \left[\gamma^2 \left(\delta \dot{X}^\perp \right)^2 + \gamma^4 \left(\delta \dot{X}^\parl \right)^2\right]\right\}
\ee
Comparing with equations (\ref{div 0}) and (\ref{div 2}) it is clear that
the following choice of the leading divergence of $\Delta M$ cancels
both the leading and second order divergences:
\be
\Delta M^{(div)}(\e) = -{\l^{4/3}(\e) \over \e} {\ell^2 \over 2\pi \ell_s^2}\;.
\ee

{}From the discussion above, we conclude that the only boundary term that can lead to a scheme dependence of ${\rm Re}\,G_R(\o)$ can come from the finite part of the counterterm
(\ref{counterterm}), with a finite coefficient $\delta m$ completely specified by fixing the renormalized quark mass.

An independent  way to see the same effect, is as follows. According
to the first line in equation (\ref{div 2}), a finite counterterm of the form (\ref{counterterm}) with coefficient $\delta m$ would shift ${\rm Re}\,G_R(\o)$
by a term proportional to $\o^2$:
\be\label{shift}
\Delta {\rm Re\, G}^\perp(\o) ={\ell^2 \over 2\pi \ell_s^2}\, \delta m \, \gamma\, \o^2, \qquad \Delta {\rm Re\, G}^\parl(\o) = {\ell^2 \over 2\pi \ell_s^2}\,\delta m \, \gamma^3 \,\o^2
\ee
In the Langevin equation, this amounts simply to a finite shift in
the quark mass: the generalized Langevin equations (\ref{langeq}) read, in Fourier space,
\be
\o^2\,\gamma M_q\, \delta X^\perp(\o) + G_R^\perp(\o) X^\perp(\o) + \xi^\perp(\o) = 0,
\ee
\be \o^2\, \gamma^3 M_q\, \delta X^\parl(\o) + G_R^\parl(\o) X^\parl(\o) + \xi^\parl(\o) = 0.
\ee
after expanding to first order in fluctuations.

The conclusion is that the shifts (\ref{shift}) are equivalent to a finite renormalization
of the quark mass, $ M_q \to M_q + \ell^2/(2\pi \ell_s^2) \delta m$. Therefore,
once the renormalized quark mass is fixed e.g. at zero-temperature by
fixing the counterterm, there are no further ambiguities in the
two-point function.

\section{Analytic calculation of the diffusion constants}\label{details kappa}

Here, we provide a derivation of equations (\ref{kappa perp}) and (\ref{kappa par}). These equations follow from the flux (\ref{current}) which can be evaluated at any point,
in particular at the horizon. Let us define,
\be\lab{m0}
\tilde\o = \frac{\o}{4\pi T_s},
\ee
for notational convenience.

Near the horizon, the solution of the fluctuation equations (\ref{H19}) and (\ref{H20}) are of the form:
\be\lab{m1}
\d X_{\perp} = C_{\perp} (r_s-r)^{-i\tilde\o}, \qquad \d X_{\parl} = C_{\parl} (r_s-r)^{-i\tilde\o}.
\ee
In (\ref{current}) we also need the near-horizon expression for the r-r component of the world-sheet metric $H^{rr}$ equation (\ref{Hab}).
Using the definitions $C=v b(r_s)$ and $f(r_s) = v^2$ (\ref{H5}) and the definition of $T_s$ in (\ref{Ts}), we find:
\be\lab{m2}
H^{rr} \to 4\pi T_s b^2(r_s)\, (r-r_s).
\ee
 Substituting (\ref{m1}) and (\ref{m2}) in (\ref{current}) yields the flux near the horizon (and everywhere):
\be\lab{m3} \textrm{Im}\, G_R = \left\{
\begin{array}{ll}
 |C_{\perp}|^2\,b^2(r_s) \o, \\ |C_{\parl}|^2\,\frac{b^2(r_s)}{Z^2(r_s)}\,\o,
\end{array} \right.
\ee From (\ref{H10}) we also find, \be\lab{Zrs} Z(r)\to
\frac{f'(r_s)}{4\pi T_s},\qquad r\to r_s \ee Therefore the
calculation is reduced to finding the coefficients $C_{\perp}$ and
$C_{\parl}$. This can be done by matching the near-horizon
solution (\ref{m1}) for small $\o$ to the exact analytic solution
of the fluctuation equations again for small $\o$. We give details
for $\d X_{\perp}$, the other component is entirely analogous. For
small $\o$ (\ref{m1}) expands as, \be\lab{m6} \d X_{\perp} \approx
C_{\perp} - i\tilde\o C_{\perp} \log(r_s-r). \ee In the strict
$\o=0$ limit this gives \be\lab{m8} \d X_{\perp} = C_{\perp},
\qquad \o = 0.\ee On the other hand, (\ref{H19}) can be solved
exactly in the strict $\o=0$ limit: \be\lab{m7}\d X_{\perp} = C_1
+ C_2 \int_0^r \frac{dt}{\sqrt{(f-v^2)(b^4f-C^2)}}. \ee Requiring
unit norm on the boundary fixes $C_1=1$. The second term diverges
at the horizon, therefore in the strict $\o=0$ limit, one should
have $C_2=0$. Therefore one has, \be\lab{m9} \d X_{\perp} = 1, \ee
 in the strict $\o=0$ limit.
Since the solution (\ref{m9}) is valid everywhere, including the horizon, one can match it with (\ref{m8}) and obtain
\be\lab{m11}
C_{\perp}=1.
\ee

More generally, the solution at all $r$ can be written as

\bea
\d X_\perp = C_\perp(\omega) \left(r_s - r\right)^{-{i\tilde\o}}
\left[ 1 + D_{1\perp}(\omega) (r_s - r) + D_{2\perp}(\omega) (r_s - r)^2 + {\cal O}\left((r_s - r)^3\right)\right].
\nonumber\\
\eea

\noindent Expanding $C_\perp$ and $D_\perp$ around $\omega = 0$
(the regular expansion is guaranteed by the regularity of the
$\omega = 0$ solution), we get

\bea
\d X_\perp &=& C_\perp(0) \left(r_s - r\right)^{-{i\tilde\o}}
\bigg[ 1 + \left( D_{1\perp}(0) (r_s - r) + {\cal O}\left((r_s - r)^2\right) \right) + \nonumber\\
&&\left( {C'_\perp(0)/C_\perp(0)} + D'_{1\perp}(0) (r_s - r) + {\cal O}\left((r_s - r)^2\right) \right) \omega +
{\cal O}\left( \omega^2 \right)\bigg] \;.
\eea

\noindent Now, equation~\refeq{m9} implies that $D_{1\perp}(0)=0$ --- and so on for all the $D_{i\perp}(0)$ --- and $C_\perp \equiv C_\perp(0) = 1$. Hence
the solution for all values of the radial coordinate is \refeq{wave small omega} :

\bea
\d X_\perp = \left(r_s - r\right)^{-{i\tilde\o}} \left[ 1 + \tilde {C}_{1\perp}(r) \omega + {\cal O}(\omega^2) \right] \;,
\eea
with $\tilde {C}_{1\perp}(r) \simeq {C'_\perp(0)/C_\perp(0)} + D'_{1\perp}(0) (r_s - r) $, close to
the horizon.

In passing, we note that slightly away from the $\o=0$ limit one can allow for the second term in (\ref{m7}), and
expanding the integrand near the horizon, one obtains,
\be\lab{m10}\d X_{\perp} = C_1 + \frac{C_2}{b^2(r_s)4\pi T_s} \log(r_s-r). \ee
Matching this with (\ref{m6}) one can also determine $C_2$ in the small $\o$ limit:
$C_2 = -i \o b^2(r_s)$. This information is not required to calculate the diffusion coefficients.

Use of (\ref{m11}) in (\ref{m3}) and eventually in (\ref{kappa}) yields the desired result (\ref{kappa perp}) (after including the
string tension in front of the world-sheet action). The discussion for the parallel component is similar.
Solving the fluctuation equation (\ref{H20}) for $\o=0$ and by matching (\ref{m1}) one finds $C_{\parl}=1$ and using this and (\ref{Zrs}) in
(\ref{m3}) yields (\ref{kappa par}).

\section{Details of the WKB approximation}\label{AppWKB}

We follow the steps outlined in section \ref{WKB}.
It is convenient to define the dimensionless variables $x \equiv r/r_s \in (0,1)$ and $\o_s\equiv\o r_s$. The analog Schr\"odinger equation is
\be\label{schro1-app} -\y''+V_s(x)\y=0,\qquad V_s(x) =
-\frac{\o_s^2 b^4}{R^2} + \half \big(\log{\cal R}\big)'' + \frac14
\big(\log {\cal R}\big)^{'2}. \ee
where
\be\label{calR}
{\cal R} = \left\{\begin{array}{l} R \\ R/Z^2 \end{array}\right., \qquad {\cal Z} = \left\{\begin{array}{l} 1 \\ Z \end{array}\right. \qquad \begin{array}{l} \perp \\ \parl \end{array}
\ee
and the functions $R(x)$ and $Z(x)$ are:
%%%%%%%%%%%%%%%%%%%%%%%%%%%%%%
\be\lab{R-app} R = \sqrt{(f-v^2)(b^4f-C^2)}, \qquad Z = b^2\sqrt{(f-v^2)/(b^4f -C^2)} \ee
%%%%%%%%%%%%%%%%%%%%%%%%%%%%%%
We divide the range $0< x < 1$ in three regions, in each of which
we use different approximations to solve the Schr\"odinger equation.
The following discussion holds for both $\perp$ and $\parl$ fluctuations,
so we will not make any distinction from now on.

\begin{enumerate}
\item {\bf Near Boundary:} $x \ll 1$ \\
In this region we have the following asymptotics:
\be\label{RasUV}
R(x) \sim b^2(x)/\gamma , \qquad Z \sim 1/\gamma, \qquad x\ll 1,
\ee
and the Schr\"odinger potential is approximately
\be\label{vbdr-app}
V_s \simeq -\gamma^2 \omega_s^2 + \big(\log b\big)'' + \left( \log b \right)^{\prime2} ,\qquad x\to 0.
\ee

In the near-boundary region the Einstein frame scale factor becomes that
of $AdS$ space-time, and we have:
\be\label{bdmetric}
 b(x)\simeq {\ell\over r_s}{\l^{2/3}(x)\over x}
\ee

One important property of this region, is that the quantity $r \l'/\l$
is small. The reason is that the field equation for $\lambda(r)$ is \cite{ihqcd1,ihqcd2}:
\be
\l'(r) \sim {b_E(r) \over \ell} \l^2 \sim {\l^2\over r} \quad \Rightarrow \quad r{\l'\over\l } \sim \l \ll 1
\ee
where $b_E(r) \sim \ell/r$ is the Einstein frame scale factor close to the boundary.

As a consequence, all terms proportional to $r \l'/\l$ , or corresponding higher derivative terms in $\l$, can be treated, to a first approximation,
as subleading in an expansion in $\l$.

\item {\bf Near Horizon:} $ x \simeq 1$\\
In this region we have:
\be\label{Ras}
R(x) \simeq (4\pi T_s r_s)\, b^2(r_s)\, (1-x), \qquad Z \simeq {f'(r_s)\over 4\pi T_s}
\ee
leading to:
\be\label{vhor-app}
V_s(x) \simeq -\left(\tom^2 + \frac14 \right) \frac{1}{(1-x)^2},\qquad x\to 1
\ee
where $\tilde{\omega} \equiv \omega/4\pi T_s$.

\item {\bf WKB Region:} $x_{tp} \ll x < 1$ \\
This is the classically allowed region, where $V_s(x) < 0$. For
large $\omega_s$, the first term in equation (\ref{schro1-app})
dominates, except close to the turning point $x_{tp}$, where the
contributions of the other terms get large, \be V_s \simeq -
{\omega_s^2 b^4 \over R} , \qquad x_{tp} \ll x < 1 \ee Since, for
any large but finite $\omega_s$, $V_s(x=0) = +\infty$, the
turning point for large $\omega_s$ is close to the boundary and
it is found by solving the equation $V_s(x)=0$ in this limit,
i.e. using $V$ in the form (\ref{vbdr-app}). Keeping in mind that
derivatives of $\lambda(x)$ close to the boundary produce
corrections of $O(\l) \ll 1$, we find the turning point for large
$\omega_S$: \be x_{tp} = {\sqrt{2}\over \omega_s \gamma}\left(1 +
O(\l)\right) , \qquad \omega_s \gg 1. \ee
\end{enumerate}

The crucial fact is that, for large $\omega_s$, $x_{tp}\ll 1$, and regions
1 and 3 overlap. On the other hand, regions 2 and 3 overlap close to $x\sim 1$. Therefore,  the solution in the WKB region can be used to connect
the near-boundary and near-horizon asymptotics.

To find the wave-function in the large $\omega_s$ regime, we follow
the steps outlined in Section \ref{WKB}.

\begin{enumerate}

\item
Consider first the WKB region. The two independent solutions to
$-\ddot{\y} + V\y=0$ in the region $V\ll 0$ are written, in the
WKB approximation:
%%%%%%%%%%%%%%%%%%%%%%%%%
\be\lab{a1} \psi_{1}\sim
 \frac{1}{\sqrt{p}}\cos\int^{x} p, \qquad \y_2\sim \frac{1}{\sqrt{p}}\sin\int^{x} p, \qquad p(x)\equiv \sqrt{-V_s(x)} = {\omega_s b^2\over R}.
 \ee
Explicitly, the general solution has the form:
%%%%%%%%%%%%%%%%%%%%%%%%%
\be\lab{a2} \psi_{wkb}
 = C_1 \frac{\sqrt{R}}{b}\cos\left[\int_{0}^{x}\frac{\o_s b^2}{R} \right] + C_2 \frac{\sqrt{R}}{b}\sin\left[\int^{x}_{0}\frac{\o_s b^2}{R} \right] \quad x_{tp}\ll x \leq 1
 \ee
In the equation above, we made the arbitrary choice $x=0$
for the lower integration limit, in order to avoid ambiguities in the definitions of the integration constants.

%%%%%%%%%%%%%%%%%%%%%%%%%
\item Consider now the near-horizon region. There, we can solve Schr\"odinger's equation with the potential (\ref{vhor-app}).
The solution with in-falling boundary condition at the horizon is
\bea\lab{infall3}
\y_h \simeq C_h (1-x)^{-i\tom +\half},\quad \tom \equiv {\o\over 4\pi T_s} \qquad x\simeq1 ,
\eea

Since both forms (\ref{a2}) and (\ref{infall3}) are valid in the near-horizon region {\it and} for large $\omega$, we can relate the coefficients by
 evaluating (\ref{a2}) near the horizon. In order to use the near-horizon expansion of the equation (\ref{Ras}) in the integrands appearing in equation (\ref{a2}), we change the extremum $x_0$ to another point $x_1$ belonging to the horizon region. This introduces a common phase shift, $\theta = \int_{0}^{x_1} \omega_sb^2/R$, in the sine and cosine functions. Taking this into account, the near-horizon expansion of equation (\ref{a2}) reads:
%%%%%%%%%%%%%%%%%%%%%%%%%
\bea\lab{a3}
\psi_{wkb} &\simeq&
 \left(4\pi T_s r_s\right)^{1/2} (1-x)^{1/2}\big\{C_1\cos[\q - \tom\log(1-x) ] + \nonumber\\ && + C_2 \sin[\q -
 \tom\log(1-x)]\big\}, \qquad x\to 1.
 \eea
%%%%%%%%%%%%%%%%%%%%%%%%%
Comparing equations  (\ref{a3}) and (\ref{infall3}) gives the relations:
%%%%%%%%%%%%%%%%%%%%%%%%%
\be\lab{cch}
-iC_2 = C_1 = {C_h \over (4\pi T_s r_s)^{1/2}} e^{-i\theta}.
\ee

%%%%%%%%%%%%%%%%%%%%%%%%%
\item Next, we consider the boundary region, $x\ll1$. Here the potential
has the form (\ref{vbdr-app}). Since the potential diverges as
$1/x^2$, a WKB treatment is impossible all the way to $x=0$, so we
must resort to another method. The strategy we follow is that of
an expansion in the derivatives of $\l(x)$, more precisely in the
small quantity $r \l'/\l \sim O(\lambda) \ll 1$. This will allow
us to write an approximate expression for the solution, valid for
any $\omega_s$.

Using the approximation for the metric in (\ref{vbdr-app}), the second entering the near-boundary potential can be written as:
\be\label{logexp}
(\log b)'' \simeq {1\over x^2} \left(1 + {2\over 3}{x^2 \l^{''}\over\l} - {2\over 3}\left({x\l'\over\l}\right)^2 \right), \;
\left(\log b\right)' \simeq - \frac1x \left( 1 - \frac23 \frac{x\lambda'}{\lambda}\right) .
\ee
The $\l$-dependent terms in the parentheses are $O(\l)$ or $O(\l^2)$.

One may naively think that it suffices to solve Schr\"odinger's equation
keeping only the leading terms in (\ref{logexp}), and
 neglecting the $O(\l)$ corrections.
However, as we show below, these subleading terms
 {\em affect the leading term in the solution.}

Let us ignore for the moment the terms containing derivatives of $\l$ in
equation (\ref{logexp}). them, the near-boundary Schr\"odinger equation with the potential (\ref{vbdr-app}) reads:
\be
-\psi'' + {2\over x^2}\psi = \gamma^2 \o_s^2 \psi
\ee
whose general solution is:
\be\label{psiuv0a}
\psi_{UV}^0(x) = A_1 \left[\sin(\gamma \o_s x ) + {\cos(\gamma \o_s x) \over \gamma \o_s x}\right] + A_2 \left[\cos(\gamma \o_s x ) -{\sin(\gamma \o_s x) \over \gamma \o x}\right]
\ee
However, this cannot be the full story, for the following reason.
 For both small $x$ {\em and} small $\gamma \omega_s x$, we can expand this solution as:
\be\label{psiuv0b}
\psi_{UV}^0(x)\simeq \left({A_1\over \gamma \omega_s}\right) \,{1\over x} - \left({\gamma^2 \omega_s^2 A_2 \over 3}\right) \, x^2 , \qquad \gamma \omega_s x \ll1
\ee
On the other hand, in the same regime $\gamma \o_s x \to 0$, we can
ignore the constant term $(\gamma \omega_s)^2$ in the potential, and
Scrh\"odinger equation becomes:
\be
\psi'' = \frac{b''}{b} \psi
\ee
which has the exact solution:
\be\label{psiuv1}
\psi_{UV}^{1}(x) = C_{s}\, b(x) + C_v \, b(x) \int_0^x {dx'\over b^{2}(x')}
\ee
where $C_s$ and $C_v$ are integration constants corresponding to normalizable
and non-normalizable solutions.
Using the near-boundary form of the metric (\ref{bdmetric}) this expression
becomes:
\be\label{schr zero freq}
\psi_{UV}^{1}(x) \simeq {C_s \over x} \l^{2/3}(x) + C_v \, x^2 \l^{-2/3}(x), \qquad \gamma \omega_s x \ll 1
\ee
and it does not agree with the small $\gamma \o_sx$ expansion (\ref{psiuv0b})
due to the extra factors of $\l^{\pm2/3}$. From this discussion, we conclude that to find the correct behavior of the solution near the boundary, one
cannot completely ignore the terms containing $\l'$ and $\l''$ in equation (\ref{logexp}). Therefore, the true solution in the boundary region, rather than
(\ref{psiuv0a}), will read instead:
\bea \label{psiuvtrue}
\psi_{UV}&& = A_1 \psi_{source} + A_2 \psi_{vev} \nonumber \\
&& =A_1 \left[\sin(\gamma \o_s x ) + {\cos(\gamma \o_s x) \over \gamma \o_s x}\right] F_1(x,\gamma \omega_s) \\ \nonumber
&& + A_2 \left[\cos(\gamma \o_s x ) -{\sin(\gamma \o_s x) \over \gamma \o_s x}\right] F_2(x,\gamma \omega_s), \quad x\ll 1
\eea
where $F_1$ and $F_2$ are some unknown functions, with asymptotics:
\be\label{F boundary}
F_{1}(x \gamma \omega_s) \sim \l^{2/3}(x), \quad F_{2}(x \gamma \omega_s) \sim \l^{-2/3}(x),\qquad \gamma \omega_s x \ll 1.
 \ee

On the other hand, we know that the functions $F_1$ and $F_2$ must be
replaced by constants in the limit when $\l(x)$ is not changing at all. This
suggests that we can parametrize the functions $F_1$ and $F_2$ as:
\be\label{varphi}
F_{i}(x,\gamma \omega_s) = \l^{\pm2/3}(x) \left[ 1 + \varphi_{i}(x,\gamma \omega_s) \right]
\ee
where the functions $\varphi_{1,2}(x,\gamma \omega_s)$ are small compared to unity for small $x\l'/\l$ and the $+$ sign in the exponent of $\lambda$
corresponds to $i=1$, and the $-$ sign to $i=2$.

One could in principle derive differential equations for
$\varphi_{1,2}$, and solve them perturbatively. This is equivalent
solving the fluctuation equation $h'' + (\log R)' h' + \a h = 0$
with some book-keeping parameter $\a$ and where the prime denotes
derivative w.r.t. the variable $x\gamma\o_s$. The solution can be
found perturbatively in a series expansion {\em both in $\a$ and
$x$, such that $\lambda(x)$ can be regarded as a small expansion
parameter}. A simple limit of this solution is to keep only the
leading term in the latter expansion and sum up the perturbative
series in $\a$ fully and then set $\a=1$ to recover the original
fluctuation equation. This can be achieved in an iterative manner
and the answer is indeed given by (\ref{psiuvtrue}) and (\ref{F
boundary}). This method justifies the appearance of the extra
factors $F_1$ and $F_2$ in (\ref{psiuvtrue}).

However, proceeding with this method and combining it with the WKB
approximation in order to achieve the full WKB solution requires
going beyond the leading term in the perturbative series in the
expansion in $\l(x)$, which is very cumbersome. Therefore, in the
following we will follow a different strategy, which will give us
a simpler (albeit more crude) way to estimate the coefficients of
the WKB wave functions.

\item

In the limit $\gamma \omega_s \gg 1$, the UV region $x\ll 1$ overlaps with the
WKB region $x> x_{tp}$ , because $x_{tp}\ll1$. Therefore, the UV solution (\ref{psiuvtrue}) must match, for large $\gamma \omega_s$, the small-$x$
limit of the WKB solution (\ref{a2}), which
using the UV expansion of $R(x)$, equation (\ref{RasUV}), reads:
%%%%%%%%%%%%%%%%%%%%%%%%%
 \be\lab{a14}
 \psi_{wkb}\simeq \frac{C_1}{\sqrt{\gamma}}\cos(\gamma \o_s x ) + \frac{C_2}{\sqrt{\gamma}} \sin(\gamma \o_s x ), \qquad x \ll 1.
 \ee
%%%%%%%%%%%%%%%%%%%%%%%%%
In order for (\ref{psiuvtrue}) to agree with this expression, it is necessary that the functions $F_{1}(x)$ and $F_2(x)$ become approximately constant for $x\gg x_{tp}$,
where (\ref{a14}) can be trusted. Since $x_{tp} \sim 1/\omega_s\gamma$,
we conclude that:
\be\label{Fconst}
F_{i}(x) \to F_{i} = const \qquad \gamma\omega_s x \gg 1.
\ee
$i=1,2$. In order to complete the matching we must estimate these constants $F_{1}$ and $F_2$.
One way to argue is as follows: to satisfy equation (\ref{Fconst}), we need
the functions $\varphi_{1}$ and $\varphi_2$ defined in (\ref{varphi}) to have the following
property:
\be
\varphi_{i} (x)\to -1 + {F_{i} \over \l^{\pm2/3}(x)}, \qquad x > x_{tp}
\ee
again, where the $+$ is for $i=1$, while the $-$ for $i=2$. But as we argued that the functions $\varphi_{1}$ and $\varphi_2$ must stay small for
slowly varying $\l$, setting $\varphi_i \approx 0$ gives us an estimate: \be\label{F large omega} F_1 \simeq \l^{2/3}(x_0)\;, \quad F_2 \simeq
\l^{-2/3}(x_0) \;,\qquad x\gg x_{tp}
\ee
where $x_0$ is a point in the vicinity of the turning point $x_{tp}$. Since
$\l$ is slowly varying, we can take $x_0=x_{tp}$ , and the
error we make will be of the order $x_{tp}\l'/l \sim O(\l) \ll 1$.

Therefore, we match (\ref{a14}) with the large $\omega_s$ limit of (\ref{psiuvtrue}), in which the functions $F_{1}$ and $F_2$ are replaced by the
constants $\l^{\pm 2/3}(x_{tp})$:
\be\label{psiuvwkb}
\psi_{UV} \to A_1 \l_{tp}^{2/3}\sin (\gamma \o_s x) + A_2 \l_{tp}^{-2/3}\cos (\gamma \o_s x)
\ee
where we have defined $\l_{tp}=\l(x_{tp})$.
For large $\omega$ the turning point is given by:
\begin{equation}\label{turnpoint}
 x_{tp} \simeq \frac{\sqrt{2}}{\o_s\gamma}, \qquad r_{tp} \simeq \frac{\sqrt{2}}{\o
 \gamma}.
\end{equation}

Matching (\ref{psiuvwkb}) and (\ref{a14}) we obtain:
%%%%%%%%%%%%%%%%%%%%%%%%%
\be\lab{a17} \qquad C_1 = \l_{tp}^{-\frac23} A_2
{\sqrt{\gamma}}, \qquad C_2 = \l_{tp}^{\frac23}\, A_1 {\sqrt{\gamma}},
\ee
which through equation (\ref{cch}) relates $C_h$ to $A_1$.

\item

Finally, we determine $A_1$, and consequently all other coefficients, by imposing unit normalization
 of the function $\Psi(x) = {\cal R}^{-1/2}\psi(x)$ at $x=r_b/r_s$, i.e. the point where the string is attached.
For a quark with infinite mass, $r_b=0$; for finite mass $r_b=r_Q$ defined
in equation (\ref{mass cutoff}). Therefore, we must distinguish the following two situations:

\begin{itemize}
\item[$\blacktriangleright$] {\bf Infinite Quark Mass.}
 Taking into account the definition (\ref{calR}) and the asymptotics (\ref{RasUV}) and (\ref{bdmetric}), we impose:
\be\label{a18}
1=\Psi(0,\omega) = {\sqrt{\gamma} r_s \over \ell }\lim_{x\to0}\l^{-2/3} A_1 \psi_{source} = {A_1 \over \sqrt{\gamma} \ell \omega},
\ee
where in the last line we used the definition $\o_s = \o r_s$.

Using the chain of equations (\ref{a18}), (\ref{a17}) and (\ref{cch}) we can finally fix, in the large $\omega$ regime,
 the coefficient $C_h$ as:
\be
|C_h|= \ell \gamma (4\pi T_s r_s)^{1/2} \l_{tp}^{2/3}\o.
\ee
{}From this one obtains the coefficient $\Psi_h$ defined in (\ref{psihor}) using the relation
$\Psi(x) = {\cal R}^{-1/2}\psi(x)$ and (\ref{Ras}). This finally yields (\ref{psih}).

\item[$\blacktriangleright$] {\bf Finite Quark Mass.}
In an analogous way as for an infinitely massive quark, we use the form of the solution \ref{psiuvtrue} to write the normalization condition at the
cutoff $r_Q$. However, the background scale factor $b(x)$ cannot be approximated by the expression in \refeq{bdmetric} at a generic cutoff $r_Q$, but
we rather need to keep and evaluate the full ${\cal R}^{-1/2}(x) \sim b(x)/\sqrt\gamma$ entering the expression for $\Psi(x)$, without approximating it
to $\Psi(x)\sim\sqrt\gamma x \lambda(x)^{-2/3}$. In fact, the necessary condition allowing to make use of this approximation is that $\lambda(x) \ll
1$. The normalization condition then reads:
\be\label{norm finite mass}
1=\Psi(r_Q,\omega) = {\cal R}^{-1/2}(x_Q) \left[ A_1 \psi_{source}(x_Q) + A_2 \psi_{vev} (x_Q) \right] \;.
\ee
We now have to use the general form of the functions $F_1$ and $F_2$, and keep both the source and vev solutions, since at $r_Q$ the
solution $\psi_{vev}$ in not negligible. Hence, using the relations \refeq{cch} and \refeq{a17} yielding $$ A_2 = i \lambda_{tp}^{4/3} A_1, $$ the
normalization condition determines $A_1$ as
\bea
A_1 &=& \gamma \omega_s x_Q {\cal R}^{1/2}(x_Q)\big[ F_1(x_Q) \left( \cos (\gamma \omega_s x) + \gamma \omega_s x \sin (\gamma \omega_s x) \right) \nonumber\\
&& + i \lambda_{tp}^{4/3} F_2(x_Q) \left( \sin (\gamma \omega_s x) - \gamma \omega_s x \cos (\gamma \omega_s x) \right) \big]^{-1}.
\eea
Moreover the asymptotics of $F_1$ and $F_2$ must be generalized w.r.t. \refeq{F boundary}, since we need to evaluate them at $x_Q$, which is not
necessarily close enough to the boundary to allow us to use \refeq{F boundary}--- namely if $\lambda(x_Q)$ is not very small. In fact, one should keep
the exact form of the solution at zero frequency \refeq{schr zero freq}, which reads
\bea
\psi_{UV}^{1}(x) \simeq {C_s} \sqrt\gamma {\cal R}^{1/2}(x) + {C_v \over \sqrt\gamma} {\cal R}^{1/2}(x) \int^x {dx' \over {\cal R}(x')}\;.
\eea
Comparing it to \refeq{psiuvtrue} one obtains the following behavior of function $F_1$ and $F_2$:
\bea\label{F cutoff}
F_{1}(x) &\sim& \frac{r_s}{\ell}\sqrt\gamma x {\cal R}^{1/2}(x), \;
F_{2}(x) \sim \frac{\ell}{r_s} {{\cal R}^{1/2}(x) \over \sqrt\gamma x^2}
\int^x {dx' \over {\cal R}(x')},\quad \gamma \omega_s x \ll 1 \nonumber\\.
\eea
Consequently, one gets the value of the modulus square of the coefficient $C_h$:
\bea
|C_h|^2 &=& 4 \pi \gamma^2 T_s r_s^3 \l_{tp}^{4/3} {x_Q^2 {\cal R}(x_Q) \over F_1(r_Q)^2}
\bigg[ 1 + \left( \gamma \omega_s x_Q \right)^2 \nonumber\\
&& + \left( \lambda_{tp}^{8/3} {F_2(r_Q)^2 \over F_1(r_Q)^2} - 1 \right) \left( \sin (\gamma \omega_s x_Q) - \gamma \omega_s x_Q \cos (\gamma \omega_s
x_Q) \right)^2 \bigg]^{-1}. \nonumber\\
\eea

As we explained previously in this appendix, $F_1$ and $F_2$ are approximated by \refeq{F cutoff} and \refeq{F large omega}, respectively for
$\gamma\omega_s x\ll 1$ and $x \gg x_{tp} \simeq \sqrt2/\gamma \omega$. Therefore, substituting this asymptotics in $|C_h|^2$ and using equation~\refeq{IMGR}, we
arrive at the results of section~\ref{WKB}.

\end{itemize}

\end{enumerate}

Throughout the calculation we assumed that one can neglect the
$\cO(\l)$ terms compared to terms of order $\cO(1)$. This
criterion is indeed satisfied in the numerical examples we study
in this paper. In the cases where this is not satisfied, or one
needs better accuracy in the WKB approximation, than one should
work out the sub-leading corrections.

\section{Correlators in ${\cal N}=4$} \label{AppN=4}

In this Appendix we would like to collect and derive some results on the imaginary part of the retarded correlator for the ${\cal N}=4$ theory. Some
features of this quantity were discussed in \cite{gubser} (see also \cite{iancu}). More specifically, in \cite{gubser} the symmetric correlator for ${\cal
N}=4$ ---~related to ${\rm Im}~G_R$ by equation \refeq{Gsym}~--- is numerically computed and an analytic approximation is proposed. Here we show the
numeric result for ${\rm Im}~G_R$ and compare it to a linear plus cubic function. The advantage of considering ${\rm Im}~G_R$ with respect to $G_{sym}$ is
the possibility of distinguishing the corrections to the large and small frequency behavior associated to the $\coth$ factor in \refeq{Gsym} from the wave
function corrections, appearing in the coefficient $\Psi_h$ in \refeq{IMGR}. In $G_{sym}$ both kinds of corrections arise, while in ${\rm Im}~G_R$
only the wave function contributes.

Unlike the non conformal case, the ${\cal  N}=4$ correlators for the longitudinal and transverse modes only differ by a factor $\gamma^2$, due to
the fact that the wave functions satisfy the same equation for both kinds of modes. The extra $\gamma^2$ comes from the $Z^{-2}$ factor in
equation~\refeq{Gab} for the longitudinal modes, that is constant in the conformal limit: $Z \to 1/\gamma$. Moreover, the fluctuation equation for
both transverse and longitudinal modes only depends on the dimensionless variables $x\equiv r/r_s$ and $\tilde \omega \equiv {\omega\over 4\pi T_s}={\omega \sqrt{\gamma}\over 4\pi T}$:
\bea\label{fluctu conf}
\partial_x \left[ \frac{1-x^4}{x^2} \partial_x \Psi(x,\tilde\omega) \right] + \frac{(4 \tilde \omega)^2}{x^2 (1-x^4)} \Psi(x,\tilde\omega) = 0.
\eea
Following the same steps as in Section \ref{correlators}, we obtain:
\be
\Psi_R(\xi,\tilde \omega)=C(\tilde \omega)(1-x)^{-i\tilde \omega}+\cdots
\ee
\be\label{imgr-app-n=4}
{\rm Im}~G_R^\perp={2\over \pi}\sqrt{\lambda_{{\cal N}=4}}\sqrt{\gamma}(\pi T)^3\,\,\tilde \omega |C(\tilde \omega)|^2
\ee

The WKB approximation implies that the imaginary part of the retarded correlator grows as $\omega^3$
for large frequencies, in the infinite mass case. More precisely, taking the conformal limit, $b(r)\to \ell/r$ and $\ell/\ell_s\to\l_{{\cal N}=4}^{1/4}$,
we obtain
\be
C(\tilde \omega)=4\tilde\omega +{\cal O}(1)
\ee
and
\bea\label{ImGR conf large}
{\rm Im}~G_R^\perp = \gamma^{-2} {\rm Im}~G_R^{\parl} \simeq \frac{\gamma^2}{2\pi}\sqrt{\lambda_{{\cal N}=4}}\,\omega^3, \quad \mbox{for}~\omega\gg{1\over r_s}.
\eea
This result\footnote{The fact that the large-$\omega$ limit is a cubic power-law can be expected from the corresponding zero-temperature result, in which one can analytically compute this quantity, and from the consideration that the large-frequency limit should be conformal.} can be obtained by applying the WKB method of the previous subsection the the wave functions obeying equation~\refeq{fluctu conf}.

On the other hand, the result for small frequencies is well known \cite{tea,gubser,iancu}
since it provides the diffusion constants\footnote{In  reference \cite{iancu} it is
 claimed that the imaginary part of the correlator is exactly linear in $\omega$, for
 all frequencies. This is not so, as explicitly shown here.}. It is derived by
analyzing the wave function solution to \refeq{fluctu conf}, in the regime where $\omega r_s = 4 \tilde \omega$ is small. The expressions for small
frequencies read
\be
C(\tilde \omega)=1+{\cal O}(\tilde \omega)
\ee
and
\bea\label{ImGR conf small}
{\rm Im}~G_R^\perp = \gamma^{-2} {\rm Im}~G_R^{\parl} \simeq \frac{\pi\gamma}{2} \sqrt{\lambda_{{\cal N}=4}}\, T^2 \omega, \quad \mbox{for}~\omega\ll
{1\over r_s}.
\eea This result can also follow from taking the conformal limit of the diffusion constants derived in the next section. In fact, equations
\refeq{kconf1}--\refeq{kconf2} are related to \refeq{ImGR conf small} by the formula \refeq{kappa}.

In figure \ref{conformal correlators} we show the numeric result for the ${\cal N}=4$ correlator for the transverse modes compared to a linear plus
cubic polynomial approximation,
\be\label{linearcubic}
{\rm Im}~G_R^\perp \approx  c_1 \omega + c_3 \omega^3
\ee
 with $c_1$ and $c_3$ given by the small and large frequency asymptotics of the correlator, as in equations
\refeq{ImGR conf small} and \refeq{ImGR conf large}, respectively:
\bea\label{c1c3}
c_1= \frac{\pi\gamma}{2}\sqrt{\lambda_{{\cal N}=4}}\, T^2 \quad \mbox{and} \quad
c_3=\frac{\gamma^2}{2\pi}\sqrt{\lambda_{{\cal N}=4}}.
\eea

\begin{center}
\FIGURE{
\begin{tabular}{cc}
\includegraphics[width=7cm]{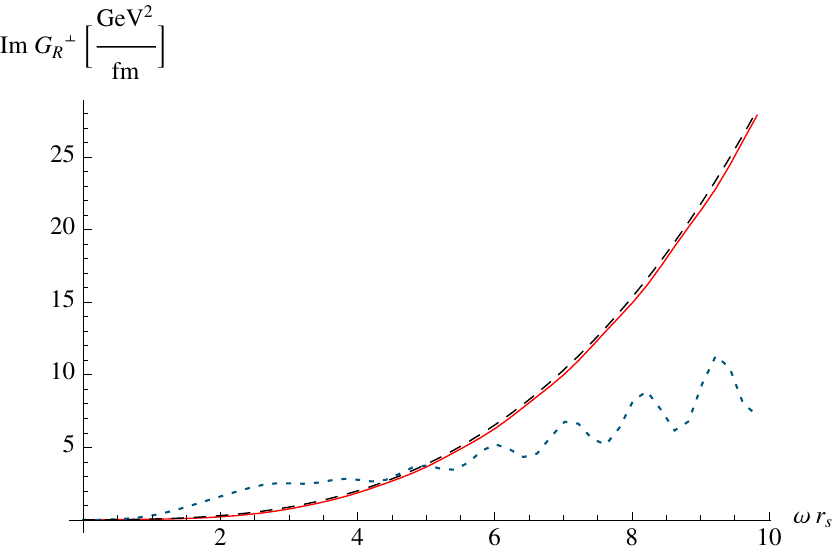}
&
\includegraphics[width=7cm]{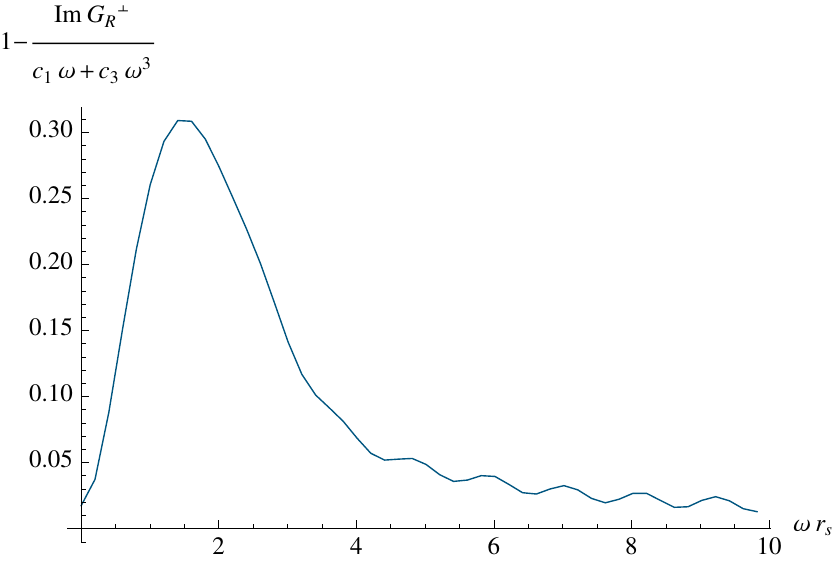}
\end{tabular}
\caption{The picture on the left shows three curves (for $\lambda_{{\cal N}=4}=5.5$ and $T=250~MeV$): i) the numeric result for ${\rm Im}~G_R$ (red), ii) the linear plus cubic function $c_1 \omega +
c_3 \omega^3$ where we use \refeq{c1c3} for $c_1$ and $c_3$ (black dashed), iii) the difference of the linear plus cubic function with respect to the
numeric result for ${\rm Im}~G_R$, multiplied by a factor of 20 (blue dotted). The right plot shows the relative difference between the linear plus
cubic function and the numeric result for the correlator.}\label{conformal correlators}}
\end{center}

The relative difference between the polynomial (\ref{linearcubic}) and the numeric result for the imaginary part of the retarded correlator vanishes, as
expected, both for small and for large frequencies. Nevertheless, the plots of figure \ref{conformal correlators} show that for $1\lesssim\omega
r_s\lesssim4$ there is a sensible difference, of the order of 10-30\%, between the two results\footnote{In Figure \ref{conformal correlators} we had to fix a value for $T$ and $\lambda_{{\cal N}=4}$ in order to plot ${\rm Im}~G_R$, but
it is important to stress that the dependence on these quantities is simply given by the overall prefactor appearing in equation (\ref{imgr-app-n=4}): the non-trivial part of the correlator  depends on $T$  only through the  combination $\tilde{\omega} = \omega/T_s$, and is independent of $\lambda_{{\cal N}=4}$.}.

In the non-conformal cases the $\omega\to \infty$ limit of the correlator is again controlled by the UV. Conformal invariance again implies an ${\cal O}(\omega^3) $ behavior although it may be corrected by logarithms.

It is also interesting to study the sub-leading corrections to $Im G_R$ in the freqency.  We first consider small $\o$. 
The leading term is determined by the Kubo's formula to be linear. On the other hand in any P-invariant  quantum field theory the imaginary part of a 
retarded correlator is guaranteed to be odd in $\o$, see for example \cite{FKT} for a recent discussion. Therefore we learn that {\em the sub-leading correction at small frequencies is cubic}. 

At high frequencies, the question is answered by extending the WKB analysis of Appendix C to sub-leading frequencies. 
First of all we note that the form of the Schrodinger potential near horizon as written in (\ref{vhor-app}). Now, 
the WKB wave-functions are obtained from this by making use of (\ref{a1}) and (\ref{a2}). The piece that is relevant for 
the current discussion is the $1/\sqrt{p}$ part that in front of the cosine and the sine terms that clearly yields a $\cO(\o^{-2}$ 
correction to the wave-functions in the large $\o$ limit. Matching the WKB solution to the near-horizon solution 
as in Appendix C, yields  $C_2= i C_1$ and one clearly obtains, 
\be\lab{nt1}
C_h \propto C_1 (1 + \cO(\o^{-2})). 
\ee
 as a correction to equation (\ref{cch}).  
 
 On the other hand, the Schrodinger equation near the boundary, in the ${\cal N}=4$ theory, can be written 
 as a function of $x\o_s$. Therefore the wave-function near the boundary is a function of $x\o_s$. This means 
 that, upon demanding unit normalization of the source term---the coefficient $A_1$ in Appendix C--- one always 
 has $A_1 \times 1/\o \sim 1 $ with {\em no correction} perturbative in $1/\o$.  Therefore matching the wave-functions 
 near the boundary and the WKB region yields $C_1 \propto A_1 \propto \o. $ Substituting in (\ref{nt1}) we learn that 
 $C_h \sim \o + \cO(\o^{-1}).$   Finally, using (\ref{IMGR}) that determines the frequency dependence as 
 $Im G_R \propto \o |C_h|^2$, we find that the subleading correction is {\em linear} in $\o$, in the high $\o$ limit.

\end{document}